\documentclass[aps,prd,superscriptaddress,nofootinbib,11pt]{revtex4-2}

\usepackage{amsfonts,amssymb,amsmath,mathtools} 
\usepackage{xcolor}
\usepackage{cancel}

\usepackage{mathrsfs}
\usepackage[colorlinks=true,citecolor=blue]{hyperref} 
\usepackage{orcidlink}

\begin{document}

\title{On the mass spectrum of Viel-dreibein gravity}

\author{Eloy Ay\'on-Beato\,\orcidlink{0000-0002-4498-3147}}
\email{eloy.ayon-beato@cinvestav.mx} 
\affiliation{Departamento de F\'{\i}sica, Cinvestav, Av.~IPN 2508, 07360, CDMX, M\'exico}

\author{Elizabeth Rodr\'{\i}guez Querts\,\orcidlink{0000-0002-0825-2945}}
\email{elizabeth@icimaf.cu}
\affiliation{Departamento de F\'{i}sica Te\'orica, Instituto de Cibern\'{e}tica Matem\'{a}tica y F\'{\i}sica (ICIMAF), \\ Calle E esq.~15 No.~309 Vedado, La Habana, 10400, Cuba}

\begin{abstract}
We study the AdS-wave configurations of Viel-dreibein gravity, a three-dimensional theory of $N$ interacting spin-2 fields, as a pretext to characterize the mass spectrum of this multigravity theory. We manage to implement the known strategy of rewriting the system as a higher-derivative theory for a single dreibein, thereby deriving the precise $2N$th-order equation governing the dynamics of the AdS-wave profile in question. The second novelty we introduce is a factorization of such an equation as the product of $N$ Klein-Gordon operators, exploiting the recent representation of the roots of any $N$th-degree polynomial in terms of the increasingly celebrated $\mathscr{A}$-hypergeometric functions. Remarkably, this allows us to determine the mass spectrum of the full theory independently of the number of involved gravities, which is essential for an explicit construction of the AdS-wave solutions. This enables us to explore the parameter space of the theory and find all its critical points (not only the Breitenlohner-Freedman bounds) where a much wider range of degeneracies in the squared masses arises. At those points, the solutions allow a zoo of new logarithmic modes leading to several $\log^{\chi-1}$ asymptotic behaviors, with $\chi\le N$, a signature of the existence of allegedly dual logarithmic conformal field theories.
\end{abstract}

\maketitle

\tableofcontents

\section{Introduction}

It is clear that Wigner's pioneering work \cite{Wigner:1931,*Wigner:1959} paved the way for the contemporary understanding that quantum states are built from the irreducible representations of the physical system symmetries. For relativistic systems, the irreps of flat spacetime Poincar\'e isometries are labeled by mass and spin (in the massless case by helicity, the spin projection along the direction of motion). For lowest spins, the relativistic equations determining the dynamics of free fields and how to incorporate interactions are the subject of any standard textbook on QFT, see e.g.~\cite{Weinberg:1995mt} and sequels. Starting from spin 2, the non-trivial subtleties involved become more common than exceptional. First, the free theory was proposed by Fierz and Pauli \cite{Fierz:1939ix} later than its unique massless self-interacting generalization, which is nothing more than General Relativity (see the foreword to \cite{Feynman:1996kb} and Refs.\ therein). In contrast, incorporating self-interactions in the massive case took many decades and significant effort \cite{Hinterbichler:2011tt,deRham:2014zqa}, as preliminary attempts were obstructed by the appearance of so-called Boulware-Deser ghosts \cite{Boulware:1972yco}. Eventually, de~Rham, Gabadadze, and Tolley achieved the long-sought result by recognizing that a ghost-free self-interaction for massive gravity can be built with the help of a background second metric \cite{deRham:2010kj}. It was later shown by Hassan and Rosen that ghosts are not reintroduced when the dynamics of the second metric is turned on, giving rise to the so-called ghost-free bigravity \cite{Hassan:2011zd}.

The next step was given by Hinterbichler and Rosen, who formulated how $N$ spin-2 fields generically interact in an arbitrary spacetime dimension $D$ via a multivielbein formalism \cite{Hinterbichler:2012cn}. The related Lagrangian starts from considering the corresponding $N$ Einstein-Cartan terms together with their respective cosmological constants for each vielbein, and incorporates new non-derivative interaction multiplicative terms coupling up to $D$ vielbeins.\footnote{See also Ref.~\cite{Li:2015iwc} for generalizations of the kinetic terms mixing different vielbeins.} It was also argued in \cite{Hinterbichler:2012cn} that such theories will contain the degrees of freedom of one massless spin-2 field, $N-1$ massive spin-2 fields, and $N-1$ scalar Boulware-Deser ghosts (see more recently Ref.~\cite{Flinckman:2025bje}). 

Subsequently, Afshar, Bergshoeff, and Merbis explored the subset of the three-dimensional version of such theories satisfying the necessary constraints to become ghost-free \cite{Afshar:2014dta}. In particular, it was found that interaction terms mixing more than two dreibein are forbidden. They also showed that the linearized Lagrangian of the ghost-free version of the theory with $N=3$ (Drei-dreibein gravity) around anti-de Sitter (AdS) spacetime is a combination of one massless and two massive Fierz-Pauli Lagrangians. There are regions in the parameter space of the model within the linear approximation, where it is possible to get both positive black hole masses and AdS central charge simultaneously, suggesting preservation of unitarity in the bulk and in the alleged dual CFT (see \cite{Afshar:2014dta} for details).

Here, we will go beyond the linear approximation in studying the dynamics of the Viel-dreibein gravity by exploring the AdS-wave configurations of the theory. To that end, we will alternatively understand Viel-dreibein gravity as a higher-derivative ($2N^{th}$-order, in this case) theory for one dreibein \cite{Hassan:2013pca}. This approach was already shown to be useful to find AdS-wave solutions for Zwei-dreibein gravity \cite{Bergshoeff:2014eca}, that is, for Viel-dreibein gravity with $N=2$ \cite{Bergshoeff:2013xma}. The related fourth-order equation for the AdS-wave profile becomes factorizable as the product of two Klein-Gordon operators, allowing one to easily find its solutions \cite{Bergshoeff:2014eca}. An interesting aspect of such solutions is the appearance of new modes (beyond the generic ones), with logarithmic asymptotic behavior at the critical points of the theory. Here, we will also analyze the different asymptotic behaviors of the solutions, and will find a more general $\log^{\chi-1}$ asymptotic decay, with $\chi\le N$. The use of AdS-waves to explore the dynamical content of generically massive gravities as an exact manifestation of their degrees of freedom beyond the perturbative level, especially in their critical sectors, is not new. It was first exhibited for Topologically Massive Gravity \cite{Deser:1981wh,*Deser:1982vy} in Refs.~\cite{Ayon-Beato:2004nrg,*Ayon-Beato:2005pnc,Ayon-Beato:2005gdo} and later extended to New Massive Gravity \cite{Bergshoeff:2009hq} in Ref.~\cite{Ayon-Beato:2009cgh}. The more complex four-dimensional configurations of ghost-free bigravity \cite{Hassan:2011zd} display a similar behavior \cite{Ayon-Beato:2018hxz}.

The first challenge we face in finding the AdS-wave configurations for Viel-dreibein gravity resides in solving the recursive relations leading to the higher-derivative equation determining the AdS-wave profile. We accomplish this goal by rewriting them in terms of new commuting operators satisfying three-term recurrence relations, since the latter are naturally expressed as the determinant of a tri-diagonal matrix \cite{*[{}] [{ Ch.~5 - Orthogonal Polynomials, pp. 240-276.}] Andrews_Askey_Roy:1999}. This ultimately allows us to find the equation as a $N$th-degree polynomial of the d'Alembertian acting on the AdS-wave profile.

The second and more demanding challenge is how to factorize the involved $2N$th-order operator again in terms of $N$ Klein-Gordon operators. This is no longer a straightforward task for $N\ge5$ since it is well-known that the Abel-Ruffini theorem, which was indeed rigorously proved by Galois after introducing the group concept \cite{*[{}] [{ Ch.~4 - Galois theory of equations, pp.\ 210-305.}] Jacobson:1985}, forbids the roots of generic higher-degree polynomials from being expressed in terms of radicals. However, this does not ban the use of non-elementary functions, and in this precisely resides the main result of this work: we succeed in the task thanks to the expressions recently obtained by Sturmfels \cite{Sturmfels:2000} for the roots of any polynomial with arbitrary coefficients in terms of $\mathscr{A}$-hypergeometric functions. This approach allows us to determine, for the first time, formal expressions for the mass spectrum of the full theory independently of the number of involved gravities, which in general is impossible by traditional methods \cite{Flinckman:2024zpb}. 

It must be said that the oldest method for finding the general roots of a polynomial of any degree is due to Birkeland, and relies on a superposition of more standard higher-order hypergeometric functions \cite{Birkeland:1927,Mayr:1936}, see the appendix in Ref.~\cite{Akbar:2003gf} and also \cite{Passare:2004} for an account of this treatment in English. However, for arbitrary coefficients as those we have to deal with here, this method is highly cumbersome since it leads to an expansion of too many standard hypergeometric functions, whereas the representation of the roots in terms of $\mathscr{A}$-hypergeometric functions is more optimal and succinct \cite{Sturmfels:2000}. 

The $\mathscr{A}$-hypergeometric functions are a generalization of the classical ones and were originally introduced by Gel'fand, Kapranov, and Zelevinsky in the early nineties \cite{Gelfand:1989,Gelfand:1990bua}. Their study has become prominent in Physics mainly due to that Feynman integrals can be interpreted as a particular class of them \cite{delaCruz:2019skx}, which facilitates their evaluation. A further striking observation is that the solutions to the constraints determining Mirror Symmetry, a duality between string compactifications on different Calabi-Yau manifolds \cite{Candelas:1990rm}, also satisfy the differential equations defining $\mathscr{A}$-hypergeometric functions \cite{Batyrev:1993}, see \cite{Stienstra:2007} for a review. A different use of the factorization of \cite{Sturmfels:2000} in terms of $\mathscr{A}$-hypergeometric functions is the construction of axisymmetric spacetimes with prescribed multipole moments \cite{Backdahl:2005uz}. Here, we are providing another interesting and highly nontrivial application.

It is also relevant to emphasize that exact configurations for a generic number $N$ of gravities are unfortunately scarce. In fact, the only examples to our knowledge are the recent higher-dimensional black hole solutions of Refs.~\cite{Wood:2024acv,*Wood:2024eol}, see also \cite{Garcia-Compean:2026cnq}. There, they manage to generalize the Kerr-Schild approach of Ref.~\cite{Ayon-Beato:2015qtt}, rooted in the stationary axisymmetric one of General Relativity \cite{Ayon-Beato:2015nvz,*Ayon-Beato:2025ahb} and allowing a transparent derivation of the spinning black holes \cite{Babichev:2014tfa,Ayon-Beato:2015qtt,Garcia-Compean:2025wkj} of ghost-free bigravity \cite{deRham:2010kj,Hassan:2011zd}, to vacuum multigravity in $D$ dimensions. Hence, the configurations presented here are new examples of exact solutions for multigravity.

The paper is organized as follows. In Sec.~\ref{sec:AdSwaves} we give a brief introduction to the AdS-wave configurations. For didactic reasons and as a warm-up to the main ingredients allowing to build AdS-wave solutions (see for example Ref.~\cite{Ayon-Beato:2009cgh}), we begin by studying the simplest non-trivial case of Drei-dreibein gravity ($N=3$), obtaining its AdS-wave solutions step by step in the whole Sec.~\ref{sec:DDG}. We start by reviewing in Sec.~\ref{sec:revDDG} the main properties of the Drei-dreibein gravity presented in \cite{Afshar:2014dta}, in particular, how to rewrite it as a higher-derivative theory \cite{Hassan:2013pca,Bergshoeff:2014eca}. This allows us to derive the precise sixth-order equation governing the AdS-wave dynamics in Sec.~\ref{sec:EqAdSw}. Thanks to this equation being factorized as the product of one massless and two massive Klein-Gordon operators, we are able to obtain the generic AdS wave solution in Sec.~\ref{sec:SolAdSw}, and exhaustively examine all the possible cases giving rise to degeneracies in the squared mass spectrum. We end by analyzing their asymptotic behavior in Sec.~\ref{sec:AdSasympt}. In Sec.~\ref{sec:VDG}, we extend the results to the general case of Viel-dreibein gravity with arbitrary $N$, for clarity we structure the section in a similar way to the previous one. Viel-dreibein gravity is reviewed in Sec.~\ref{sec:revVDG} according to Ref.~\cite{Afshar:2014dta}. The $2N$-order equation determining the AdS-wave profile is derived in Sec.~\ref{sec:EqAdSwN} in terms of a polynomial operator of the d'Alembertian. Its factorization in terms of a massless and $N-1$ massive Klein-Gordon operators is obtained in Sec.~\ref{sec:Mfact}. It is precisely here where the mass spectrum of Viel-dreibein gravity is explicitly found, by expressing the involved squared masses as $\mathscr{A}$-hypergeometric functions of the coupling constants. The generic solutions are obtained in Sec.~\ref{sec:SolAdSwN} and the critical ones in Sec.~\ref{sec:CP}, where their AdS asymptotic sectors are also discussed. Finally, in Sec.~\ref{sec:VDGeg} we illustrate how to recover from the general results for arbitrary $N$ the AdS-wave solutions of Zwei-dreibein gravity $(N=2)$ from Ref.~\cite{Bergshoeff:2014eca} and those of Drei-dreibein gravity $(N=3)$ studied in Sec.~\ref{sec:DDG}. For completeness, the new nontrivial case of Vier-dreibein gravity ($N=4$) is also explicitly considered to illustrate the power of the method. The last Sec.~\ref{sec:conclu} is devoted to our conclusions. We leave the inductive construction of the polynomial operator characterizing the AdS waves for App.~\ref{app:p_k}. In order to make our work as self-contained as possible we briefly review $\mathscr{A}$-hypergeometric functions in App.~\ref{app:AHyper}. Their recent use to represent the roots of generic polynomials \cite{Sturmfels:2000}, the essential ingredient to determine the mass spectrum of the full theory, is later revised in App.~\ref{app:Roots}.

\section{AdS-wave configurations \label{sec:AdSwaves}}

We start by briefly describing the AdS-wave configurations, which are exact gravitational waves; concretely, a special kind of those propagating in the presence of a nontrivial cosmological constant. The latter exact gravitational waves were first introduced by Garc\'{\i}a and Pleba\'nski \cite{Garcia:1981} (see also Refs.~\cite{Salazar:1983,Garcia:1983,Ozsvath:1985qn,Bicak:1999ha,*Bicak:1999hb}) who generalized the Kundt \cite{Kundt:1961} and Robinson-Trautman \cite{Robinson:1962zz} algebraically special solutions, originally derived in absence of a cosmological constant. Within the spacetimes considered in \cite{Garcia:1981}, there exists a special limit where the multiple principal null direction of the Weyl tensor becomes a Killing vector for the exact wave solutions. This forces the nontrivial cosmological constant to be necessarily negative, and one recovers a class of spacetimes thoroughly studied later by Siklos \cite{Siklos:1985}. The Siklos spacetimes can be interpreted as exact gravitational waves propagating along the AdS space \cite{Podolsky:1997ik}, reason why they were coined AdS waves in \cite{Ayon-Beato:2005gdo}. In fact, they can be viewed as a generalized Kerr-Schild transformation starting from the AdS metric 
\begin{equation}\label{eq:Kerr-Schild}
g_{ \mu\nu}=g_{ \mu\nu}^\text{AdS}-F\,k_{\mu}k_{\nu},
\end{equation}
where $k_{\mu}$ is a null geodesic vector and $F$ is a profile function satisfying the restriction of being independent of the integral parameter along $k^{\mu}$, but otherwise arbitrary
\cite{Ayon-Beato:2005gdo}.

In Poincaré coordinates the three-dimensional AdS metric is written as
\begin{equation}\label{eq:AdSPoincare}
ds^2_\text{AdS}=\dfrac{l^2}{y^2}\left(-2dudv+dy^{2} \right),
\end{equation}
where $l$ is the AdS radius related to the constant scalar curvature as $R=6\Lambda=-\frac{6}{l^2}$, furthermore $u$ and $v$ are retarded and advanced times, respectively. As is explained in the next section, the $2+1$ multigravity theories are described in terms of dreibeine. For the AdS space we shall use the following one
\begin{equation}\label{eq:AdSdreibein}
\begin{aligned}
e_\text{AdS}^{0}&=\frac{l}{\sqrt{2}y}\left(du+dv\right) ,\\
e_\text{AdS}^{1}&=\frac{l}{\sqrt{2}y} \left(-du+dv\right),\\
e_\text{AdS}^{2}&=\frac{l}{y} dy.
\end{aligned}
\end{equation}
The metric of the AdS waves
(\ref{eq:Kerr-Schild}) then becomes
\begin{equation}\label{eq:AdSwaveMetric}
ds^2=\dfrac{l^2}{y^2}\left(-F(u,y)du^{2}-2dudv+dy^{2} \right),
\end{equation}
after choosing as null geodesic field
$k^{\mu}\partial_{\mu}=(y/l)\partial_{v}$, with the advanced time playing the role of the parameter along the Killing field $\partial_{v}$. The associated dreibein is then
\begin{equation}\label{eq:AdSwave_dreibein}
\begin{aligned}
e_\text{AdSw}^{0}&=\frac{l}{y}\left( \sqrt{F}du+\frac{1}{\sqrt{F}}dv\right) ,\\ 
e_\text{AdSw}^{1}&=\frac{l}{y}\frac{1}{\sqrt{F}}dv,\\
e_\text{AdSw}^{2}&=\frac{l}{y}dy.
\end{aligned}
\end{equation}

The behavior of AdS waves has been studied for several lower-dimensional massive gravities as Topologically Massive Gravity \cite{Ayon-Beato:2004nrg,*Ayon-Beato:2005pnc,Ayon-Beato:2005gdo}, New Massive Gravity \cite{Ayon-Beato:2009cgh}, Zwei-dreibein gravity \cite{Bergshoeff:2014eca}, and even for the four-dimensional ghost-free bigravity \cite{Ayon-Beato:2018hxz}. We devote this paper to the generalization of the above results by characterizing the AdS-wave configurations of Viel-dreibein gravity \cite{Afshar:2014dta}, as a means to explicitly find the mass spectrum of such a theory. As a first step, we focus in the next section on the theory with the next level of difficulty with respect to the already studied: Drei-dreibein gravity.

\section{AdS waves in Drei-dreibein gravity \label{sec:DDG}}

As has been mentioned, AdS-wave solutions were already found for Zwei-dreibein gravity (that is, for Viel-dreibein gravity with two
dreibeine). In this section we will investigate the existence of such configurations for the next number of dreibeine, $N=3$. We begin by revising the related theory.

\subsection{Review of Drei-dreibein gravity \label{sec:revDDG}}

The dynamics of the Drei-dreibein gravity \cite{Afshar:2014dta} is given by
the Lagrangian density three-form
\begin{align}
L={}&-M_\text{P}\sum_{I=1}^{3}\left(\eta_{ab}\sigma_{I}e_{I}^{a}R_{I}^{b}+\dfrac{m^2}{6}\alpha_{I}\varepsilon_{abc}e_{I}^{a}e_{I}^{b}e_{I}^{c} \right )\notag \\
& + \dfrac{m^2}{2}M_\text{P}\varepsilon_{abc}\left( \beta_{12}e_{1}^{a}e_{1}^{b}e_{2}^{c}+\beta_{21}e_{2}^{a}e_{2}^{b}e_{1}^{c}+\beta_{13}e_{1}^{a}e_{1}^{b}e_{3}^{c} \right. \notag \\
& + \left.
 \beta_{31}e_{3}^{a}e_{3}^{b}e_{1}^{c}+ \beta_{23}e_{2}^{a}e_{2}^{b}e_{3}^{c}+\beta_{32}e_{3}^{a}e_{3}^{b}e_{2}^{c}+\beta_{123}e_{1}^{a}e_{2}^{b}e_{3}^{c}\right),
\label{eq:Lagrangian_N=3}
\end{align}
which depends on three frame fields one-forms (or dreibeine) $e_{I}^{a}=e_{I \mu}^{a} dx^{\mu}$, $I=1,2,3$. Here and in what follows, spacetime and Lorentz indices are denoted by Greek and Latin letters, respectively, and the wedge products of differential forms are implicit as standard. The first line in (\ref{eq:Lagrangian_N=3}) is a set of three copies of Einstein-Cartan Lagrangian densities. The parameters $\sigma_{I}=\pm 1$ are dimensionless, so one of them can always be set to unity without losing generality. The $\alpha_{I}$ are parameters related to cosmological constants and are also dimensionless. The second and third lines in (\ref{eq:Lagrangian_N=3}) contain the non-derivative interaction terms between two and three dreibeine, coupled by the constants $\beta_{IJ}$ and $\beta_{123}$, respectively. The Planck mass is denoted by $M_\text{P}=\frac{1}{8\pi G}$ and $m$ is a mass parameter. The related curvature and torsion two-forms are defined by the Cartan structure equations 
\begin{align}
\label{eq:Curvature}
R_{I a}&\equiv\mathit{D}_{I}\omega_{Ia}=d\omega_{I a}+\dfrac{1}{2}\varepsilon_{abc}\omega_{I}^{b}\omega_{I}^{c},\\
 \label{eq:Torsion}
T_{I a}&\equiv\mathit{D}_{I}e_{Ia}=d e_{I a}+\varepsilon_{abc}\omega_{I}^{b}e_{I}^{c},
\end{align}
where $\mathit{D}_{I}$ are the covariant derivatives with respect to the
corresponding spin connections one-forms $\omega_{I
a}=\dfrac{1}{2}\varepsilon_{abc}\omega_{I}^{bc}$.

As is explained in detail in \cite{Afshar:2014dta}, a theory with three
dreibeine should describe $2$ massive spin-2 modes, with 2 degrees of freedom each, and one massless mode, which is a pure gauge and can always be removed; the physical dimension of the phase space should then be $D_\text{PhS}^{3}=4\times 2=8$. But, after a spacetime decomposition of the fields, for each $I=1,2,3$, the spatial parts of the dreibeine and the spin connections $(e_{Ii}^{a},\omega_{Ii}^{a})$ add $12$ components to the dynamical phase space, while the time components
$(e_{It}^{a},\omega_{It}^{a})$ act as Lagrange multipliers, giving $6$
primary constraints. Out of the total $18$ primary constraints, $6$ are of
first class, and correspond to the diffeomorphism and local Lorentz
invariance of the theory. Then, $4$ additional second class constraints are required to eliminate the nonphysical degrees of freedom and get the desired
\begin{equation}\label{eq:PhysSpaceDim}
D_\text{PhS}^{3}=8=12\times 3-6\times 3-6-4.
\end{equation}
It was also shown in \cite{Afshar:2014dta} that the interaction between the three dreibeine is not allowed, that is, $\beta_{123}\neq0$ does not lead to the required secondary constraints.

For a general class of 3-dimensional Chern-Simons-like theories, to which Viel-dreibein gravity belongs, the secondary constraints can be derived from the Bianchi and Cartan identities (see \cite{Bergshoeff:2014bia} for details)
\begin{align}\label{eq:BianchiId}
\mathit{D}_{I}R_{I a}&=0,\\
 \label{eq:CartanId}
 \mathit{D}_{I}T_{I a}&=\varepsilon_{abc}R_{I}^{b}e_{I}^{c}.
\end{align}
After setting $\beta_{123}=0$, the Cartan identities (\ref{eq:CartanId}) look like
\begin{align}\nonumber
\eta_{ab}\left[ (\beta_{12}e_{1}^{c}+\beta_{21}e_{2}^{c})e_{1}^{a}e_{2}^{b}+(\beta_{13}e_{1}^{c}+\beta_{31}e_{3}^{c})e_{1}^{a}e_{3}^{b}\right]&=0,\\ \label{eq:CartanIdConstraintsN=3}
\eta_{ab}\left[ (\beta_{12}e_{1}^{c}+\beta_{21}e_{2}^{c})e_{2}^{a}e_{1}^{b}+(\beta_{23}e_{2}^{c}+\beta_{32}e_{3}^{c})e_{2}^{a}e_{3}^{b}\right]&=0,\\
 \nonumber
\eta_{ab}\left[ (\beta_{13}e_{1}^{c}+\beta_{31}e_{3}^{c})e_{3}^{a}e_{1}^{b}+(\beta_{32}e_{3}^{c}+\beta_{23}e_{2}^{c})e_{3}^{a}e_{2}^{b}\right]&=0,
\end{align}
while the Bianchi identities (\ref{eq:BianchiId}) are written as
\begin{align}\nonumber
\eta_{ab}\left[\beta_{12}\omega_{12}^{c}e_{1}^{a}e_{2}^{b}+\beta_{13}\omega_{13}^{c}e_{1}^{a}e_{3}^{b}+ e_{2}^{c} (\beta_{12}e_{1}^{a}+\beta_{21}e_{2}^{a})\omega_{12}^{b}+e_{3}^{c} (\beta_{13}e_{1}^{a}+\beta_{31}e_{3}^{a})\omega_{13}^{b}\right]&=0,\\ \label{eq:BianchiIdConstraintsN=3}
\eta_{ab}\left[\beta_{21}\omega_{21}^{c}e_{2}^{a}e_{1}^{b}+\beta_{23}\omega_{23}^{c}e_{2}^{a}e_{3}^{b}+ e_{1}^{c} (\beta_{12}e_{1}^{a}+\beta_{21}e_{2}^{a})\omega_{21}^{b}+e_{3}^{c} (\beta_{32}e_{3}^{a}+\beta_{23}e_{2}^{a})\omega_{23}^{b}\right]&=0,\\ \nonumber
\eta_{ab}\left[\beta_{31}\omega_{31}^{c}e_{3}^{a}e_{1}^{b}+\beta_{32}\omega_{32}^{c}e_{3}^{a}e_{2}^{b}+ e_{1}^{c} (\beta_{13}e_{1}^{a}+\beta_{31}e_{3}^{a})\omega_{31}^{b}+e_{2}^{c} (\beta_{23}e_{2}^{a}+\beta_{32}e_{3}^{a})\omega_{32}^{b}\right]&=0,
\end{align}
where $\omega_{IJ}=\omega_I-\omega_J$. Among the possible choices to get a ghost-free theory, we select the dreibeine $e_{1}$ and $e_{2}$ to be invertible and
\begin{equation}\label{eq:ParameterChoiseN=3}
\beta_{12}\neq0\neq\beta_{23},
\end{equation}
while all other $\beta$-parameters are set to zero. Under these assumptions, the Cartan identities (\ref{eq:CartanIdConstraintsN=3}) are reduced to two constraints
\begin{equation}\label{eq:CartanSecConstraintsN=3}
\eta_{ab}e_{1}^{a}e_{2}^{b}=0, \qquad \eta_{ab}e_{2}^{a}e_{3}^{b}=0,
\end{equation}
and the Bianchi identities (\ref{eq:BianchiIdConstraintsN=3}) give the other two needed secondary constraints
\begin{equation}\label{eq:BianchiSecConstraintsN=3}
\eta_{ab}e_{1}^{a}\omega_{12}^{b}=0, \qquad
\eta_{ab}e_{2}^{a}\omega_{23}^{b}=0,
\end{equation}
to reduce the number of degrees of freedom of the model to $4$ (that is, $D_\text{PhS}^3=8$), which accounts for the two helicity states of each massive graviton. It is worth noting that the conditions (\ref{eq:ParameterChoiseN=3}) turn the model into a so called line theory, according to the theory graph (for a detailed explanation see, for instance, \cite{Scargill:2014wya}). While conditions (\ref{eq:CartanSecConstraintsN=3}), being stronger than the symmetric vielbein conditions $e_{I \mu}^{a}e_{J \nu}^{b} \eta_{ab}=e_{I \nu}^{a}e_{J \mu}^{b} \eta_{ab}$, ensure compliance with the latter for every two interacting vielbeine (dreibeine, in our case). The symmetric vielbein condition is closely related to the ghost-free character of the theory and the absence of the cycle type interactions (see \cite{deRham:2015cha,Goon:2014paa,Deffayet:2012zc}).

The equations of motion for a given parameter choice, where only interactions between $e_1,e_2$, and between $e_2,e_3$ remain, are
\begin{align}\label{eq:EqsMotion_e 1}
\sigma_{1}R_{1 a}&=\dfrac{m^2}{2}\varepsilon_{abc}\left(-\alpha_{1}e_{1}^{b}e_{1}^{c}+2\beta_{12}e_{1}^{b}e_{2}^{c} \right) ,\\ \label{eq:EqsMotion_e2}
\sigma_{2}R_{2 a}&=\dfrac{m^2}{2}\varepsilon_{abc}\left(-\alpha_{2}e_{2}^{b}e_{2}^{c}+\beta_{12}e_{1}^{b}e_{1}^{c}+2\beta_{23}e_{2}^{b}e_{3}^{c} \right) ,\\ \label{eq:EqsMotion_e3}
\sigma_{3}R_{3 a}&=\dfrac{m^2}{2}\varepsilon_{abc}\left(-\alpha_{3}e_{3}^{b}e_{3}^{c}+\beta_{23}e_{2}^{b}e_{2}^{c} \right),
\end{align}
and the torsion equations
\begin{equation} \label{eq:EqsMotion_TI N=3}
T_{I a}=0.
\end{equation}

It is known that bimetric theories can also be interpreted as higher-derivative theories \cite{Hassan:2013pca}. In particular, in
\cite{Bergshoeff:2014eca} it was explained how to go from the two dreibeine approach to the higher-derivative one in the Zwei-dreibein gravity case. Here we will show how to do it for Drei-dreibein gravity $(N=3)$. We first remark that we can use the inverse of the dreibein $e_{\nu}^{c}$ in two-form equations of the type
\begin{equation} \label{eq:Invertible2-form}
 F_{\mu \nu a}= \tau \varepsilon_{abc}y_{\mu}^{b}e_{\nu}^{c },
\end{equation}
to isolate the accompanying one-form in the right-hand side as
\begin{subequations}\label{eq:Inverted2-form}
\begin{align} 
 y_{\beta}^{l}&= \frac{1}{\tau}e_{e}^{\alpha}e_{d}^{\mu}e_{f}^{\nu}\varepsilon^{edf} F_{\mu \nu a} P^{al}_{\beta \alpha},\\
 P^{al}_{I\beta \alpha} &= e_{I\beta}^{a} e_{I\alpha}^{l}-\frac{1}{2} e_{I\alpha}^{a} e_{I\beta}^{l}.
\end{align}
\end{subequations}
We then use the property
\begin{equation}\label{eq:Property}
\frac{1}{2}e_{Ie}^{\alpha}e_{Id}^{\mu}e_{If}^{\nu}\varepsilon^{edf}R_{I\mu \nu a} P^{al}_{I\beta \alpha}= S^{l}_{I\beta},
\end{equation}
where $S_{I\mu}^{a}=S_{I\mu \nu}e_{I}^{\nu a}$ and $S_{I\mu \nu}=R_{I\mu
\nu}-\dfrac{1}{4}g_{I\mu \nu}R_{I}$ is the Schouten tensor for $e_I$, to obtain from Eqs.~(\ref{eq:EqsMotion_e 1}) and (\ref{eq:EqsMotion_e2}) the following expressions for $e_2$ and $e_3$
\begin{align} \label{eq:EqsMotion_Invert_e1N=3}
e_{2 \mu}^{a}&=\dfrac{\alpha_1}{2\beta_{12}}e_{1 \mu}^{a}+\dfrac{\sigma_1}{m^2\beta_{12}}S_{1\mu}^{a},\\
\label{eq:EqsMotion_Invert_e2}
e_{3 \mu}^{a}&=\dfrac{\alpha_2}{2\beta_{23}}e_{2 \mu}^{a}+\dfrac{\sigma_2}{m^2\beta_{23}}S_{2\mu}^{a}-\dfrac{\beta_{12}}{2\beta_{23}}H_{21 \mu}^{a},
\end{align}
where we use the notation
\begin{subequations}\label{eq:MixedMetric}
\begin{align}
 H_{IJ\mu}^{a}&=2(N_{IJ b}^{a}-\dfrac{1}{4}\delta_{b}^{a}N_{IJ c}^{c})e_{I \nu}^{b}, \\ 
 N_{IJ a}^{b}&=L_{IJ a}^{b}L_{IJ c}^{c}-L_{IJ a}^{c}L_{IJ c}^{b},\\ 
 L_{IJ a}^{b}&=e_{I a}^{\mu}e_{J \mu}^{b}.
\end{align}
\end{subequations}

Once $e_2$ is given in terms of $e_1$ up to its second derivatives in (\ref{eq:EqsMotion_Invert_e1N=3}), $e_3$ can also be written in terms of $e_1$ and its higher derivatives by using (\ref{eq:EqsMotion_Invert_e1N=3}) into (\ref{eq:EqsMotion_Invert_e2}). Finally, Eq.~(\ref{eq:EqsMotion_e3}) can be thought as a further higher-derivative equation of motion for $e_{1}$. We will explicitly show this procedure for the AdS waves in the next subsection.

\subsection{AdS-waves equation \label{sec:EqAdSw}}

Since AdS waves describe exact gravitational waves propagating along AdS spacetime, we need to first find possible restrictions in the parameter space of the theory needed to admit the AdS vacuum as a solution. We first substitute the proportionality ansatz
\begin{equation}
 e_{1}=\gamma_1 e_\text{AdS},\label{eq:AdSvac_e1}
\end{equation}
for the AdS dreibein (\ref{eq:AdSdreibein}) into Eqs.~(\ref{eq:EqsMotion_Invert_e1N=3}) and (\ref{eq:EqsMotion_Invert_e2}) to obtain
\begin{equation}\label{eq:AdSvac_eI N=3}
 e_{I}=\gamma_I e_\text{AdS}, \qquad I=2,3,
\end{equation}
where
\begin{subequations}\label{eq:vac_parametersN=3}
\begin{align}
 -\frac{\sigma_{1}}{l^{2}m^{2}}&=-\alpha_{1} \gamma_{1}^{2} +2\beta_{12}\gamma_{1}\gamma_{2},\\ 
 -\frac{\sigma_{2}}{l^{2}m^{2}}&=-\alpha_{2} \gamma_{2}^{2} +2\beta_{23}\gamma_{2}\gamma_{3}+\beta_{12}\gamma_{1}^{2},\\ 
 -\frac{\sigma_{3}}{l^{2}m^{2}}&=-\alpha_{3} \gamma_{3}^{2}+\beta_{23}\gamma_{2}^{2}.
\end{align}
\end{subequations}
The conditions (\ref{eq:vac_parametersN=3}) define the proportionality factors $\gamma_{2}$ and $\gamma_{3}$ together with the AdS radius $l$ in terms of the coupling constants of the theory, while $\gamma_{1}$ is put by hand and it is usually set to unity. The above will be the proportionality factors we use in what follows.

We are ready to propose the study ansatz 
\begin{equation}
 e_{1}=\gamma_1 e_\text{AdSw},\label{eq:AdSwave_e1}
\end{equation}
as proportional to the AdS-wave dreibein (\ref{eq:AdSwave_dreibein}) to characterize the AdS-wave configurations of Drei-dreibein gravity. After using the Cartan structure equations (\ref{eq:Curvature}) and (\ref{eq:Torsion}) to find $\omega_1$ and $R_1$, we get from (\ref{eq:EqsMotion_Invert_e1N=3})
\begin{subequations}\label{eq:AdSwave_e2}
\begin{align}\nonumber
e_{2}^{0}&=\gamma_{2}\left(
\frac{l}{y}\frac{h_{2}-F}{2\sqrt{F}}du + e_\text{AdSw}^{0} \right) , \\ 
e_{2}^{1}&=\gamma_{2}\left(
\frac{l}{y}\frac{h_{2}-F}{2\sqrt{F}}du + e_\text{AdSw}^{1} \right) ,\\ \nonumber
e_{2}^{2}&=\gamma_2 e_\text{AdSw}^{2},
\end{align}
where
\begin{align}
h_{2}&=\frac{1}{a_{12}}\left( -\frac{\sigma_{1}\gamma_{1}}{m^{2}}\square +a_{12}\right) F, \label{eq:RecursiveRel h_1N=3} \\ 
a_{12}&=\beta_{12} \gamma_{1}^{2}\gamma_{2},
\label{eq:a12N=3}
\end{align}
\end{subequations}
and the box is the d'Alembertian operator on the AdS-wave background (\ref{eq:AdSwaveMetric}) for profiles independent of the advanced time 
\begin{equation}\label{eq:box_operator}
 \square=\frac{1}{l^{2}}\left( y^2 \partial_{y }^{2}-y \partial_{y }\right).
\end{equation}
We then obtain $e_3$ by replacing $e_1$ and $e_2$ given by (\ref{eq:AdSwave_e1}) and (\ref{eq:AdSwave_e2}) into (\ref{eq:EqsMotion_Invert_e2})
\begin{subequations}\label{eq:AdSwave_e3}
\begin{align}\nonumber
e_{3}^{0}&=\gamma_{3}\left(
\frac{l}{y}\frac{h_{3}-F}{2\sqrt{F}}du + e_\text{AdSw}^{0} \right) , \\ 
e_{3}^{1}&=\gamma_{3}\left(
\frac{l}{y}\frac{h_{3}-F}{2\sqrt{F}}du + e_\text{AdSw}^{1} \right) ,\\ \nonumber
e_{3}^{2}&=\gamma_3 e_\text{AdSw}^{2},
\end{align}
where
\begin{align} 
h_{3}&=\frac{1}{a_{23}}\left(-\frac{\sigma_{2}\gamma_{2}}{m^{2}}\square +a_{23}+a_{12}\right) h_{2}-a_{12}F \nonumber \\
&=\left[ \frac{1}{a_{12}a_{23}}\left(-\frac{\sigma_{2}\gamma_{2}}{m^{2}}\square +a_{23}+a_{12}\right) \left( -\frac{\sigma_{1}\gamma_{1}}{m^{2}}\square +a_{12}\right)-a_{12}\right] F, \label{eq:RecursiveRel h_2N=3} \\
a_{23}&=\beta_{23} \gamma_{2}^{2}\gamma_{3}.
\label{eq:a23N=3}
\end{align}
\end{subequations}

The equation of motion for $e_3$ (\ref{eq:EqsMotion_e3}) leads to a single
differential equation
\begin{equation}\label{eq:EqMotion h_3}
\left( -\frac{\sigma_{3}\gamma_{3}}{m^{2}}\square + a_{23}\right) h_{3} - a_{23}h_{2} = 0,
\end{equation}
which, after using (\ref{eq:RecursiveRel h_1N=3}) and (\ref{eq:RecursiveRel
h_2N=3}), can be written as a sixth-order Euler differential equation for the wave profile
\begin{subequations}\label{eq:EqMotion h_3F}
\begin{equation}
\left[ \tilde{p}_{3}\left( \frac{\square}{m^2}\right)^3+\tilde{p}_{2}\left( \frac{\square}{m^2}\right)^2+\tilde{p}_{1}\left( \frac{\square}{m^2}\right)\right] F=0,
\end{equation}
\begin{align}\nonumber
\tilde{p}_{3}&=-\frac{\sigma_1\gamma_1\sigma_2\gamma_2\sigma_3\gamma_3}{a_{12}a_{23}},\\ \label{eq:p_i}
\tilde{p}_{2}&=\frac{\sigma_1\gamma_1\sigma_2\gamma_2 a_{23}+\sigma_1\gamma_1\sigma_3\gamma_3 (a_{12}+a_{23})+\sigma_2\gamma_2\sigma_3\gamma_3 a_{12}}{a_{12}a_{23}},\\\nonumber
\tilde{p}_{1}&=-(\sigma_1\gamma_1+\sigma_2\gamma_2+\sigma_3\gamma_3).
\end{align}
\end{subequations}

The resulting Eq.~(\ref{eq:EqMotion h_3F}) is finally the one characterizing the AdS waves of Drei-dreibein gravity and can be factorized as
\begin{equation}\label{eq:HigherOrder H_3 Factor}
\square\left(\square-M_{-}^{2} \right)\left(\square-M_{+}^{2} \right)F=0,
\end{equation}
where $M_{\pm}$ are in general two non-zero effective masses 
\begin{equation}\label{eq:Mpm}
 M_{\pm}^2=\frac{m^{2}}{2\tilde{p}_{3}}\left(-\tilde{p}_{2} \pm \sqrt{\tilde{p}_{2}^2-4\tilde{p}_{1}\tilde{p}_{3}}\right),
\end{equation}
which are given in terms of the coupling constants of the theory by means of the polynomial coefficients (\ref{eq:p_i}), defined through the constants \eqref{eq:a12N=3} and \eqref{eq:a23N=3}, in addition to the proportionality factors \eqref{eq:vac_parametersN=3}.

\subsection{AdS-waves solutions \label{sec:SolAdSw}}

From factorization (\ref{eq:HigherOrder H_3 Factor}) it is easy to see that the profile function $F$ will behave as a superposition of modes, each one satisfying Klein-Gordon equations $\left(\square-\mu^2 \right)F=0$ with masses $\mu=0, M_{\pm}$, respectively. In fact, by employing the standard substitution $F=y^{n}$ and the local d'Alembertian expression \eqref{eq:box_operator} to solve the Euler differential equation (\ref{eq:HigherOrder H_3 Factor}) we get the sixth order in $n$ characteristic polynomial
\begin{equation}\label{eq:CharPolynN=3}
n(n-2)\left[n(n-2)-M_{-}^{2} l^2 \right]\left[n(n-2)-M_{+}^{2} l^2\right]=0.
\end{equation}
Accordingly, the generic solution for the wave profile is
\begin{subequations}\label{eq:F GenericSolutionN=3}
\begin{equation}
F(u,y)=F_{-}(u,y)+ F_{+}(u,y),
\end{equation}

\begin{equation}\label{eq:F+-}
F_{\pm}(u,y) = 
\begin{cases}
 \displaystyle
 F_{\pm}^{+}(u)\left(\frac{y}{l} \right)^{1+l\sqrt{M_\pm^2-M_\text{BF}^2}} +F_{\pm}^{-}(u)\left(\frac{y}{l} \right)^{1-l\sqrt{M_\pm^2-M_\text{BF}^2}}, & 
 M_{\pm}^2 > M_\text{BF}^2, \\ \\
 \displaystyle
 \frac{y}{l}\left[ F_{1\pm}(u)\ln \left(\frac{y}{l} \right)+F_{0\pm}(u)\right], & 
 M_{\pm}^2 = M_\text{BF}^2.
\end{cases}
\end{equation}
Here, $F_{\pm}^{\pm}$, $F_{1\pm}$, and $F_{0\pm}$ are arbitrary integration functions of the retarded time $u$. The second line in \eqref{eq:F+-} corresponds to the cases where only one of the two masses saturates the $2+1$-dimensional Breitenlohner-Freedman bound \cite{Breitenlohner:1982jf,Mezincescu:1984ev}
\begin{equation}\label{eq:BF}
M_\text{BF}^2=-\frac1{l^2},
\end{equation}
\end{subequations}
the lowest mass square allowing a stable behavior on AdS, and where it is known that logarithmic AdS decays appear \cite{Ayon-Beato:2009cgh,Ayon-Beato:2018hxz}. Notice that no homogeneous ($n=0$) or quadratic ($n=2$) dependence of the wave-front coordinate $y$ do appear in the solution (\ref{eq:F GenericSolutionN=3}) since they always can be eliminated by a coordinate transformation; both represent the pure-gauge modes of the massless sector in $2+1$ dimension (for a detailed discussion on this topic see \cite{Ayon-Beato:2005gdo}, also \cite{Ayon-Beato:2015xsz} and \cite{Ayon-Beato:2018hxz} for more extended analysis on the residual symmetries of AdS waves). So, we are left with four modes, two for each non-zero mass $M_-$ and $M_+$, corresponding to the two helicity states of a massive graviton in $2+1$ dimensions. The solution (\ref{eq:F GenericSolutionN=3}) is valid in a region of the parameters space of the theory where $M_{\pm}^{2}\neq0$ and $M_{+}^{2}\neq M_{-}^{2}$, that is, in the region where the spectrum of the squared masses is nondegenerate.

Besides the generic solution (\ref{eq:F GenericSolutionN=3}) we need to consider the possibility of having degeneracies in the squared mass spectrum. When such multiplicities occur in the roots of the characteristic polynomial \eqref{eq:CharPolynN=3} the theory becomes of the critical gravity kind
\cite{Bergshoeff:2012ev} and new logarithmic modes, additional to the Breitenlohner-Freedman ones, must be added in order to span the whole space of linearly independent solutions of the exact modes equation (\ref{eq:HigherOrder H_3 Factor}). The maximum degeneracy achieved in \eqref{eq:CharPolynN=3} is when the theory becomes completely massless $M_{-}=0=M_{+}$, but the corresponding modes are no longer pure gauge due to the multiplicity. This threefold multiplicity arises when the condition
\begin{equation}\label{eq:3multiplicity_cond}
\tilde{p}_{1}=\tilde{p}_{2}=0,
\end{equation}
holds in \eqref{eq:Mpm} for the coefficients defined in \eqref{eq:p_i}. The points satisfying (\ref{eq:3multiplicity_cond}) are called tri-critical points, and form a three-dimensional surface in the five-dimensional parameter space of the theory, once we fix the values of $\sigma_I$ and $\gamma_1$. The wave profile function is then
\begin{equation}\label{eq:F 3crit}
F(u,y) = \ln \left(\frac{y}{l} \right)\left[ F_{12}(u) \left(\frac{y}{l} \right)^2+F_{10}(u)\right] +
 \ln^2 \left(\frac{y}{l} \right)\left[ F_{22}(u) \left(\frac{y}{l} \right)^2+F_{20}(u)\right].
\end{equation}

The following degeneracy is when
\begin{equation}\label{eq:2multiplicity_cond zero mass}
\tilde{p}_{1}=0, \qquad \tilde{p}_{2}\neq0,
\end{equation}
where the roots of \eqref{eq:CharPolynN=3} have a double multiplicity since $M_+=0$ and $M_-\neq0$, or $M_-=0$ and $M_+\neq0$, depending on the sign of the nontrivial coefficient $\tilde{p}_2/\lvert\tilde{p}_2\rvert=\pm1$. The condition (\ref{eq:2multiplicity_cond zero mass})
defines a hypersurface in the parameter space, for which the profile is given by
\begin{equation}\label{eq:F 2crit zero mass}
F(u,y)=\ln \left(\frac{y}{l} \right)\left[ F_{12}(u) \left(\frac{y}{l} \right)^2+F_{10}(u)\right] + 
F_{\mp}(u,y),
\end{equation}
where $F_{\mp}$ are given in (\ref{eq:F GenericSolutionN=3}), and the upper (lower) sign applies when $\tilde{p}_2$ is positive (negative). 

Another two-fold multiplicity in \eqref{eq:CharPolynN=3} is achieved when both masses \eqref{eq:Mpm} collapse to a single nontrivial value, $M_+=M_-=M\neq0$, which occurs for the parameters hypersurface 
\begin{equation}\label{eq:2multiplicity_cond non zero mass}
\tilde{p}_{2}-4\tilde{p}_{1}\tilde{p}_{3}=0.
\end{equation}
The corresponding solution for the wave profile is
\begin{subequations}\label{eq:F 2crit non zero mass}
\begin{equation}
F(u,y)=\hat{F}_{-}(u,y)+ \hat{F}_{+}(u,y),
\end{equation}
\begin{equation}
\hat{F}_{\pm}(u,y) = \left( \frac{y}{l}\right)^{1\pm l\sqrt{M^2-M_\text{BF}^2}}\left[ \hat{F}_{1\pm}(u)\ln \left(\frac{y}{l} \right)+\hat{F}_{0\pm}(u)\right],
\end{equation}
\end{subequations}
for $M^2 > M_\text{BF}^2$, or
\begin{equation}\label{eq:F 2crit Ml^2eq-1}
F(u,y) = \frac{y}{l}\left[ F_{31}(u)\ln^3 \left(\frac{y}{l} \right)+ F_{21}(u)\ln^2 \left(\frac{y}{l} \right)+ F_{11}(u)\ln \left(\frac{y}{l} \right)+ F_{11}(u)\right],
\end{equation}
when the single mass saturates the Breitenlohner-Freedman bound \eqref{eq:BF}, $M^2 = M_\text{BF}^2$.

The above exhausts the set of AdS-wave solutions of Drei-dreibein gravity for a mass spectrum respecting the Breitenlohner-Freedman bound. It is worth emphasizing that in addition to them, there also exist solutions to \eqref{eq:HigherOrder H_3 Factor} below the Breitenlohner-Freedman bound \eqref{eq:BF}. They describe oscillatory behaviors on the front-wave coordinate $y$, see \cite{Ayon-Beato:2009cgh}. However, we don't address them here since they are less physically interesting.

\subsection{Asymptotically AdS sector \label{sec:AdSasympt}}
 
At this point, it is important to analyze the asymptotic behavior of the AdS-wave configurations given in the previous subsection, and to discuss which of them are further asymptotically AdS. First, let's remind that the asymptotically AdS behavior is traditionally understood according to Brown-Henneaux boundary conditions \cite{Brown:1986nw}, where for a metric $g_{ \mu\nu}=g_{ \mu\nu}^\text{AdS}+h_{ \mu\nu}$ the deviation $h_{ \mu\nu}$ from the AdS one $g_{ \mu\nu}^\text{AdS}$ must fall off as
\begin{equation}\label{eq:Brown-Henneaux BC}
h_{uu}\sim h_{uv}\sim h_{vv}\sim h_{yy}\sim \mathscr{O} (1), \qquad 
h_{uy}\sim h_{vy}\sim \mathscr{O} (y),
\end{equation}
in the limit to the conformal boundary $y\rightarrow 0$. For the AdS-wave ansatz \eqref{eq:AdSwaveMetric} it simply means that the wave profile must obey $F(u,y) \sim \mathscr{O}(y^{2})$ in this limit. It is obvious that the generic solution (\ref{eq:F GenericSolutionN=3}) is asymptotically AdS if $M_{-}^{2}>0$, $M_{+}^{2}>0$, and $F_{\pm}^{-}=0$.

On the other hand, at critical points the Brown-Henneaux boundary conditions (\ref{eq:Brown-Henneaux BC}) need to be relaxed to admit weaker $\log$ and $\log^2$ AdS behaviors. In fact, the solutions (\ref{eq:F 2crit zero mass}) and (\ref{eq:F 2crit non zero mass}) (valid when $M_+^2=0,M_-^2\neq0$ or $M_-^2=0,M_+^2\neq0$, and $M_+^2=M_-^2=M^2\neq0$, respectively) are compatible with $\ln$ boundary conditions
\begin{equation}\label{eq:log BC}
h_{uv}\sim h_{yy}\sim \mathscr{O}(1), \qquad h_{uu}\sim h_{vv}\sim \mathscr{O} (\ln(y)), \qquad h_{uy}\sim h_{vy}\sim \mathscr{O} (y\ln(y)),
\end{equation}
These boundary conditions also arise when studying AdS waves in other three-dimensional massive gravity theories, like Topological Massive Gravity \cite{Ayon-Beato:2004nrg,*Ayon-Beato:2005pnc,Ayon-Beato:2005gdo}, New Massive Gravity \cite{Ayon-Beato:2009cgh}, and Zwei-dreibein Gravity \cite{Bergshoeff:2014eca}. They are typical of higher-derivative gravity theories up to fourth order, and even appear in the semiclassical corrections of their black holes \cite{Chernicoff:2024dll}.

An even weaker $\ln^2$ decay, characteristic of sixth-derivative gravity theories (see \cite{Bergshoeff:2012ev}, \cite{Setare:2013fza}, \cite{Setare:2014zea})
\begin{equation}\label{eq:log^2 BC}
h_{uv}\sim h_{yy}\sim \mathscr{O}(1) \qquad 
h_{uu}\sim h_{vv}\sim \mathscr{O} (\ln^2(y)), \qquad 
h_{uy}\sim h_{vy}\sim \mathscr{O} (y\ln^2(y)),
\end{equation}
appears at the tri-critical point, when the AdS-wave configuration is described by (\ref{eq:F 3crit}).

\section{AdS waves in Viel-dreibein gravity \label{sec:VDG}}

After studying the AdS-wave configurations of Drei-dreibein gravity and exhibiting the general procedure in a concrete example, in this section we generalize the results to Viel-dreibein gravity by considering an arbitrary number $N$ of dreibeine. We first briefly review the involved theory.

\subsection{Review of Viel-dreibein gravity \label{sec:revVDG}}

The Lagrangian density three-form of the Viel-Dreibein gravity \cite{Afshar:2014dta} is written in terms of $N$ dreibeine as
\begin{align}
L={}&-M_\text{P}\sum_{I=1}^{N}\left(\eta_{ab}\sigma_{I}e_{I}^{a}R_{I}^{b}+\dfrac{m^2}{6}\varepsilon_{abc}\alpha_{I}e_{I}^{a}e_{I}^{b}e_{I}^{c} \right ) \notag \\
& + \frac{m^2}{2}M_\text{P}\varepsilon_{abc}\left( \sum_{I\neq J}^{N}\beta_{IJ}e_{I}^{a}e_{I}^{b}e_{J}^{c}
+ \sum_{I<J<K}^{N}\beta_{IJK}e_{I}^{a}e_{J}^{b}e_{K}^{c}
\right), \label{eq:Lagrangian_N}
\end{align}
Now we have $N$ sign parameters $\sigma_{I}=\pm 1$, and $N$ cosmological constants $\alpha_{I}$. The coupling constants $\beta_{IJ}$ and $\beta_{IJK}$
characterize the interaction between two and three dreibeine, respectively. As in the case of Drei-dreibein theory, $m$ is a mass parameter, $M_\text{P}$ is the Planck mass, and the curvature and torsion two-forms are given by Cartan structure equations (\ref{eq:Curvature}) and (\ref{eq:Torsion}).

The theory with $N$ dreibeine shall describe $N-1$ massive spin-2 modes, and the physical dimension of the phase space should then be $D_\text{PhS}^N=4(N-1)$. But now the spatial parts of the $N$ dreibeine and $N$ spin connections give a total of $12N$ components, while the time components acting as Lagrange multipliers lead now to $6N$ primary constraints, out of which again six are first class related to invariance under diffeomorphism and local Lorentz transformations \cite{Afshar:2014dta}. Then, we need $2(N-1)$ extra second class constraints to get
\begin{equation}\label{eq:PhysSpaceDimN}
D_\text{PhS}^N=4(N-1)=12N-6N-6-2(N-1).
\end{equation}

After setting all the $\beta_{IJK}=0$, because the corresponding terms in the Lagrangian density do not lead to secondary constraints, as we already saw for $N=3$, the Cartan identities (\ref{eq:CartanId}) read
\begin{equation}\label{eq:CartanIdConstraints}
\sum_{\substack{ J=1 \\ J\neq I}}^{N}
\eta_{ab}\left( \beta_{IJ}e_{I}^{c}+\beta_{JI}e_{J}^{c}\right)e_{I}^{a}e_{J}^{b}=0, \qquad I=1,\ldots,N,
\end{equation}
and the Bianchi identities (\ref{eq:BianchiId}) are
\begin{equation}
 \label{eq:BianchiIdConstraints}
\sum_{\substack{J=1 \\ J\neq I}}^{N}
\eta_{ab}\left[ \beta_{IJ}\omega_{IJ}^{c}e_{I}^{a}e_{J}^{b}+e_{J}^{c}\left( \beta_{IJ}e_{I}^{a}+\beta_{JI}e_{J}^{a}\right)\omega_{IJ}^{b}\right]=0, \qquad I=1,\ldots,N,
\end{equation}
where again $\omega_{IJ}=\omega_{I}-\omega_{J}$. We assume the first $N-1$ dreibeine $e_{I}$, $I=1,\ldots,N-1$, to be invertible, additionally that
\begin{equation}\label{eq:ParameterChoise}
\beta_{I,I+1}\neq 0, \qquad I=1,\ldots,N-1,
\end{equation}
and $\beta_{IJ}=0$, $\forall J\neq I+1$. After making this choice, the Cartan identities (\ref{eq:CartanIdConstraints}) lead to
\begin{equation}\label{eq:CartanSecConstraints}
\eta_{ab}e_{I}^{a}e_{I+1}^{b}=0, \qquad I=1,\ldots,N-1,
\end{equation}
and from the Bianchi identities (\ref{eq:BianchiIdConstraints}) we get
\begin{equation}\label{eq:BianchiSecConstraints}
\eta_{ab}e_{I}^{a}\omega_{I,I+1}^{b}=0, \qquad I=1,\ldots,N-1,
\end{equation}
giving the right number of extra constraints: $2(N-1)$.

The equations of motion become
\begin{align}\label{eq:EqsMotion_e1}
\sigma_{1}R_{1 a}&=\dfrac{m^2}{2}\varepsilon_{abc}\left(-\alpha_{1}e_{1}^{b}e_{1}^{c}+2\beta_{12}e_{1}^{b}e_{2}^{c} \right) ,\\ \label{eq:EqsMotion_eI}
\sigma_{I}R_{I a}&=\dfrac{m^2}{2}\varepsilon_{abc}\left(-\alpha_{I}e_{I}^{b}e_{I}^{c}+\beta_{I-1,I}e_{I-1}^{b}e_{I-1}^{c}+2\beta_{I,I+1}e_{I}^{b}e_{I+1}^{c} \right) , \qquad I=2,\ldots,N-1,\\ \label{eq:EqsMotion_eN}
\sigma_{N}R_{N a}&=\dfrac{m^2}{2}\varepsilon_{abc}\left(-\alpha_{N}e_{N}^{b}e_{N}^{c}+\beta_{N-1,N}e_{N-1}^{b}e_{N-1}^{c} \right),
\end{align}
along with $N$ torsion equations (\ref{eq:Torsion}). 

Since each dreibein $e_I$, $I=2,\ldots,N-1$, only interacts with the previous $e_{I-1}$ and the following $e_{I+1}$ ones, the equations (\ref{eq:EqsMotion_eI}) are exact copies of (\ref{eq:EqsMotion_e2}). This allows us to apply the same prescription used in the previous section to consider Viel-dreibein gravity as a higher-derivative theory for the single dreibein $e_{1}$. In fact, we use the same property (\ref{eq:Property}) to rewrite the equations of motion (\ref{eq:EqsMotion_e1}) and (\ref{eq:EqsMotion_eI}) as
\begin{align} \label{eq:EqsMotion_Invert_e1}
e_{2 \mu}^{a}&=\frac{\alpha_1}{2\beta_{12}}e_{1 \mu}^{a}+\frac{\sigma_1}{m^2\beta_{12}}S_{1\mu}^{a},\\
\label{eq:EqsMotion_Invert_eI}
e_{I+1 \mu}^{a}&=\frac{\alpha_{I}}{2\beta_{I,I+1}}e_{I \mu}^{a}+\frac{\sigma_{I}}{m^2\beta_{I,I+1}}S_{I \mu}^{a}-\frac{\beta_{I-1,I}}{2\beta_{I,I+1}}H_{I,I-1 \mu}^{a}, \qquad I=2,\ldots,N-1.
\end{align}
Finally, (\ref{eq:EqsMotion_Invert_e1}) and (\ref{eq:EqsMotion_Invert_eI}) enable us to transform the equation of motion for $e_{N}$
(\ref{eq:EqsMotion_eN}) into a $2N^{th}$-derivative equation of motion for $e_{1}$, just as we did in the previous section.

\subsection{AdS-waves equation \label{sec:EqAdSwN}}

Our analysis begins once more with understanding how AdS spacetime is embedded within Viel-dreibein gravity. By proposing the same proportionality ansatz to the AdS dreibein \eqref{eq:AdSdreibein} for the first dreibein (\ref{eq:AdSvac_e1}), the higher dreibeine (\ref{eq:EqsMotion_Invert_e1}) and (\ref{eq:EqsMotion_Invert_eI}) become again proportional to AdS according to
\begin{equation}
 e_{I}=\gamma_I e_\text{AdS},\label{eq:AdSvac_eI}
\end{equation}
where the proportionality factors $\gamma_{I}$ for $I=2,\ldots,N$, and the AdS radius $l$ are related to the coupling constants of the theory through 
\begin{subequations}\label{eq:vac_parameters}
\begin{align}
 -\frac{\sigma_{1}}{l^{2}m^{2}}&=-\alpha_{1} \gamma_{1}^{2} +2\beta_{12}\gamma_{1}\gamma_{2},\\ 
 -\frac{\sigma_{I}}{l^{2}m^{2}}&=-\alpha_{I} \gamma_{I}^{2} +2\beta_{I,I+1}\gamma_{I}\gamma_{I+1}+\beta_{I-1,I}\gamma_{I-1}^{2},\qquad I=2,\ldots,N-1,\\
 -\frac{\sigma_{N}}{l^{2}m^{2}}&=-\alpha_{N} \gamma_{N}^{2}+\beta_{N-1,N}\gamma_{N-1}^{2}.
\end{align}
\end{subequations}
These are the proportionality factors used in the forthcoming expressions, where $\gamma_1$ is again chosen arbitrarily and can be set to unity, for example.

To explore the AdS-wave configurations of Viel-dreibein gravity, we start again with the proportionality ansatz to the AdS wave \eqref{eq:AdSwave_dreibein} for the initial dreibein (\ref{eq:AdSwave_e1}). Then the next dreibeine are obtained from (\ref{eq:EqsMotion_Invert_e1}) and (\ref{eq:EqsMotion_Invert_eI}) giving 
\begin{subequations}\label{eq:AdSwave_eI}
\begin{align}\nonumber
e_{I}^{0}&=\gamma_{I}\left(
\frac{l}{y}\frac{h_{I}-F}{2\sqrt{F}}du + e_\text{AdSw}^0 \right), \\ 
e_{I}^{1}&=\gamma_{I}\left(
\frac{l}{y}\frac{h_{I}-F}{2\sqrt{F}}du + e_\text{AdSw}^1 \right), \qquad I=1,\ldots,N, \\ \nonumber
e_{I}^{2}&=\gamma_I e_\text{AdSw}^2,
\end{align}
where $h_1=F$ and the remaining functions $h_{I}$ satisfy the recursive relations
\begin{align}\label{eq:RecursiveRel h_1}
a_{12}h_{2}&=\Upsilon_{1}(h_{1}), & & \\ \label{eq:RecursiveRel h_I}
a_{I,I+1}h_{I+1}&=\Upsilon_{I}(h_{I})-a_{I-1,I}h_{I-1}, & I&=2,\ldots,N-1, \\
\label{eq:aII+1}
a_{I,I+1}&=\beta_{I,I+1} \gamma_{I}^{2}\gamma_{I+1}, & I&=1,\ldots,N-1,
\end{align}
defined by the commuting linear differential operators
\begin{align}
\label{eq:def_Upsilon1}
\Upsilon_{1}&= -\frac{\sigma_{1}\gamma_{1}}{m^{2}}\square +a_{12}, && \\
\label{eq:def_UpsilonI}
\Upsilon_{I}&= -\frac{\sigma_{I}\gamma_{I}}{m^{2}}\square +a_{I,I+1}+a_{I-1,I}, & I&=2,\ldots,N-1.
\end{align}
\end{subequations}

Finally, it is easy to show that the last equation of motion (\ref{eq:EqsMotion_eN}) yields
\begin{subequations}\label{eq:Rel h_N}
\begin{equation}
\Upsilon_{N}(h_{N})-a_{N-1,N}h_{N-1}=0,
\end{equation}
with linear differential operator
\begin{equation}
\label{eq:def_UpsilonN}
\Upsilon_{N}= -\frac{\sigma_{N}\gamma_{N}}{m^{2}}\square + a_{N-1,N}.
\end{equation}
\end{subequations}
Our goal is to write Eq.~(\ref{eq:Rel h_N}) in terms of the wave profile $F=h_{1}$. To that end, we will transform the set of second-order recursive relations (\ref{eq:RecursiveRel h_1}) and (\ref{eq:RecursiveRel h_I}) into an equivalent system. We start by applying the operator $\Upsilon_{1}$ defined in \eqref{eq:def_Upsilon1} to Eq.~(\ref{eq:RecursiveRel h_I}) with $I=2$. Then we use (\ref{eq:RecursiveRel h_1}) to eliminate $h_{1}$ from the obtained equation, after which the resulting equation and (\ref{eq:RecursiveRel h_1}) are rewritten as
\begin{align}\label{eq:RecursiveRel h_1_2}
a_{12}\mathscr{H}_{0}(h_{2})-\mathscr{H}_{1}(h_{1})&=0, & \mathscr{H}_{0}&=\mathbf{I},\qquad \mathscr{H}_{1}=\Upsilon_{1},\\ \label{eq:RecursiveRel h_2_2}
a_{23}\mathscr{H}_{1}(h_{3})-\mathscr{H}_{2}(h_{2})&=0, & \mathscr{H}_{2}&=\mathscr{H}_{1}\Upsilon_{2}-a_{12}^{2}\mathscr{H}_{0}.
\end{align}
Proceeding in a similar way with Eq.~(\ref{eq:RecursiveRel h_I}) for $I=3$ and using (\ref{eq:RecursiveRel h_2_2}) to eliminate $h_{2}$, we obtain
\begin{align}\label{eq:RecursiveRel h_3_2}
a_{34}\mathscr{H}_{2}(h_{4})-\mathscr{H}_{3}(h_{3})&=0, & \mathscr{H}_{3}&=\mathscr{H}_{2}\Upsilon_{3}-a_{23}^{2}\mathscr{H}_{1}.
\end{align}
After repeating the same algorithm with all the equations (\ref{eq:RecursiveRel h_I}) we finally can write the resulting system in a compact form
\begin{subequations}\label{eq:RecursiveRel h_I_2}
\begin{equation}
a_{I,I+1}\mathscr{H}_{I-1}(h_{I+1})-\mathscr{H}_{I}(h_{I})=0, \qquad I=1,\ldots,N-1,
\end{equation}
where the new commuting operators are recursively defined as
\begin{equation}\label{eq:RecursiveRel H_I_2}
\mathscr{H}_{I}=\mathscr{H}_{I-1}\Upsilon_{I}-a_{I-1,I}^{2}\mathscr{H}_{I-2}, \qquad I=2,\ldots,N.
\end{equation}
\end{subequations}
We can use the new operators to solve the recursive relations \eqref{eq:RecursiveRel h_I} as
\begin{equation}\label{eq:ClosedRelh_N}
h_{I}=\prod_{K=1}^{I-1}\dfrac{1}{a_{K,K+1}}\mathscr{H}_{I-1}(h_{1}), \qquad I=1,\ldots,N,
\end{equation}
from which Eq.~(\ref{eq:Rel h_N}) finally leads to a $2N^{th}$ order differential equation for the wave profile
\begin{equation}\label{eq:HigherOrder H_N}
\mathscr{H}_{N}(h_{1})=\left( \mathscr{H}_{N-1}\Upsilon_{N}-a_{N-1,N}^{2}\mathscr{H}_{N-2}\right) (F)=0.
\end{equation}

The problem now is reduced to finding the explicit form of the differential operator $\mathscr{H}_{N}$. Notice that objects defined by three-term recurrence relations like (\ref{eq:RecursiveRel H_I_2}) are equivalently determined from the determinant of a tri-diagonal matrix, see Ref.~\cite{*[{}] [{ Ch.~5 - Orthogonal Polynomials, pp. 240-276.}] Andrews_Askey_Roy:1999} for the case of orthogonal polynomials. In particular, the highest-order operator in \eqref{eq:HigherOrder H_N} is equivalent to the determinant 
\begin{equation}\label{eq:HigherOrder H_N Det}
\mathscr{H}_{N}=
\begin{vmatrix}
\Upsilon_{1} & a_{12} &0& &0&0&0\\
a_{12} & \Upsilon_{2} & a_{23} & \cdots &0&0&0\\
0& a_{23} & \Upsilon_{3} & &0&0&0\\
&\vdots & & \ddots & & \vdots & \\
0&0&0& & \Upsilon_{N-2} & a_{N-2,N-1} &0\\
0&0&0& \cdots & a_{N-2,N-1} & \Upsilon_{N-1} & a_{N-1,N}\\
0&0&0& &0& a_{N-1,N} & \Upsilon_{N}
\end{vmatrix}.
\end{equation}
In turn, determinant (\ref{eq:HigherOrder H_N Det}) can be evaluated after some algebra giving the explicit form of the differential operator as a polynomial of the d'Alembertian operator
\begin{subequations}\label{eq:HigherOrder H_N Polynomial}
\begin{equation}
\mathscr{H}_{N}= \sum_{k=0}^{N}p_{k}\left( \frac{\square}{m^{2}}\right) ^{k},
\end{equation}
with coefficients given by the following expressions, which are inductively proved in App.~\ref{app:p_k},
\begin{equation}\label{eq:p_k}
\begin{aligned}
p_{N}={}&(-1)^{N} \sigma_{1}\gamma_{1} \cdots \sigma_{N}\gamma_{N},\\ 
p_{N-1}={}&(-1)^{N-1}
\smashoperator[l]{
\sum_{\substack{
 I_{1}<\cdots<I_{N-1}\\
 J\neq I_{k}\forall k}}}
\sigma_{I_1}\gamma_{I_{1}}\cdots \sigma_{ I_{N-1}}\gamma_{ I_{N-1}} (a_{J-1,J}+a_{J,J+1}),\\ 
\vdotswithin{=} \\ 
p_{k}={}&(-1)^{k}
\smashoperator[l]{
\sum_{\substack{
 I_{1}<\cdots<I_{k}\\
 J_{1}<\cdots<J_{N-k}\\
 J_{r}\neq I_{q}\forall r,q\\ 
 s_{q}=-1,0}}} 
\sigma_{I_1}\gamma_{I_1}\cdots \sigma_{I_k}\gamma_{I_k} 
a_{J_{1}+s_{1},J_{1}+s_{1}+1}\cdots 
a_{J_{N-k}+s_{N-k},J_{N-k}+s_{N-k}+1}\\ 
& \qquad\qquad\qquad \times \Delta(J_{1}+s_{1},\ldots,J_{N-k}+s_{N-k}),\\ 
\vdotswithin{=} \\ 
p_{1}={}& -\left(\sum_{k=1}^{N}\sigma_{k}\gamma_{k}\right) a_{12}\cdots a_{N-1,N},\\ 
p_{0}={}& a_{12}\cdots a_{N,N+1}=0,
\end{aligned}
\end{equation}
where $a_{01}$ and $a_{N,N+1}$ must be set to zero, and we denote
\begin{equation}\label{eq:Delta()}
\Delta(J_{1}+s_{1},\ldots,J_{n}+s_{n})\equiv
(1-\delta_{J_{1}+s_{1},J_{2}+s_{2}})\cdots 
(1-\delta_{J_{n-1}+s_{n-1},J_{n}+s_{n}}).
\end{equation}
\end{subequations}

\subsection{Mass spectrum from factorization \label{sec:Mfact}}

The obtained equation \eqref{eq:HigherOrder H_N Polynomial} is a higher-order Euler one, and to find its solutions it is essential to factorize it as a product of Klein-Gordon operators $(\square-\mu^{2})$ acting on the profile
\begin{equation}\label{eq:HigherOrder H_N Factor}
\square\left(\square-M_{2}^{2} \right)\left(\square-M_{3}^{2} \right)\cdots \left(\square-M_{N}^{2} \right)F=0, \qquad M_{k}^{2}=\theta_{k}m^{2}, 
\end{equation}
where $M_{k}^2$ are the squared effective masses, proportional to the roots $\theta_{k}$ of the polynomial $\mathscr{H}_{N}$. This is precisely our next step, which will provide for the first time the full mass spectrum of Viel-dreibein gravity.

It is well-known that the Abel-Ruffini theorem \cite{*[{}] [{ Ch.~4 - Galois theory of equations, pp.\ 210-305.}] Jacobson:1985} states that there is no general algebraic solution (that is, solution in radicals) to polynomial equations of degree five or higher with arbitrary coefficients as \eqref{eq:p_k}. However, as has been proved by Sturmfels not long ago \cite{Sturmfels:2000}, the roots in \eqref{eq:HigherOrder H_N Factor} can always be succinctly expressed in terms of the generic polynomial coefficients \eqref{eq:p_k} by means of $\mathscr{A}$-hypergeometric functions
\begin{equation}\label{eq:H_N roots}
\begin{aligned}
\theta_{1}&=-\left[\frac{p_{0}}{p_{1}} \right] =
0,\\ 
\theta_{k}&=-\left[\frac{p_{k-1}}{p_{k}} \right] +\left[\frac{p_{k-2}}{p_{k-1}} \right], \qquad k=2,\ldots N,
\end{aligned}
\end{equation}
which allow the following series representation
\begin{subequations}\label{eq:H_N roots1}
\begin{equation}
\left[\frac{p_{k-1}}{p_{k}}\right]=
\sum_{i_{1},\ldots,i_{N}\geq 0}
\frac{(-1)^{i_{k}}}{i_{k-1}+1}
\begin{pmatrix}
i_{k}\\
i_{1} \cdots i_{k-1} i_{k+1} \cdots i_{N}
\end{pmatrix}
\frac{p_{1}^{i_{1}}\cdots p_{k-1}^{i_{k-1}+1}
p_{k+1}^{i_{k+1}}\cdots p_{N}^{i_{N}}}{p_{k}^{i_{k}+1}},
\end{equation}
\begin{equation}
\begin{pmatrix}
i_{k}\\
i_{1} \cdots i_{k-1} i_{k+1} \cdots i_{N}
\end{pmatrix}
\equiv\frac{i_{k}!}{i_{1}!\cdots i_{k-1}!i_{k+1}! \cdots i_{N}!},
\end{equation}
where additionally, the nonnegative integers $i_{1},\ldots,i_{N}$ satisfy the relations
\begin{equation}\label{eq:H_N roots2}
\begin{aligned}
i_{1}+\cdots+i_{k-1}-i_{k}+i_{k+1}+\cdots +i_{N}&=0,\\ 
1i_{1}+\cdots +(k-1)i_{k-1}-ki_{k}+(k+1)i_{k+1}+ \cdots + N i_{N}&=0.
\end{aligned}
\end{equation}
\end{subequations}
See App.~\ref{app:Roots} for details and notations on the Sturmfels representation for the roots. We also intend to provide a brief but gentle introduction to $\mathscr{A}$-hypergeometric functions in App.~\ref{app:AHyper}, given the recent interest in the subject in the Physics literature, see \cite{delaCruz:2019skx,Batyrev:1993,Stienstra:2007,Backdahl:2005uz}.

\subsection{AdS-wave solutions \label{sec:SolAdSwN}}

By using the substitution $F=y^{n}$ into Euler equation (\ref{eq:HigherOrder H_N Factor}), together with expression \eqref{eq:box_operator} for the d'Alembertian, drive us to the $2N^{th}$-order factorized characteristic polynomial
\begin{equation}\label{eq:CharacteristicPolynomial}
n(n-2)\left[n(n-2)-M_{2}^{2}l^{2}\right]
\left[n(n-2)-M_{3}^{2}l^{2}\right]\cdots
\left[n(n-2)-M_{N}^{2}l^{2}\right]=0.
\end{equation}
Hence, after eliminating the pure-gauge massless trivial behaviors, the generic solution for the AdS-wave profile in Viel-dreibein gravity is
\begin{subequations}\label{eq:F GenericSolution}
\begin{equation}
F(u,y)=\sum_{k=2}^{N} F_{k}(u,y),
\end{equation}

\begin{equation}
F_{k}(u,y) = 
\begin{cases}
\displaystyle
F_{k}^{+}(u)\left(\frac{y}{l}\right)^{1+l\sqrt{M_k^2-M_\text{BF}^2}} +F_{k}^{-}(u)\left(\frac{y}{l}\right)^{1-l\sqrt{M_k^2-M_\text{BF}^2}}, & M_{k}^2 > M_\text{BF}^2,\\ \\
\displaystyle
\frac{y}{l}\left[ F_{1k}(u)\ln\left(\frac{y}{l}\right)
+F_{0k}(u)\right], & M_{k}^2 = M_\text{BF}^2,
\end{cases}
\end{equation}
\end{subequations}
where $F_{k}^{\pm}$, $F_{1k}$, and $F_{0k}$ are integration functions depending arbitrarily on the retarded time $u$. We remind that the Breitenlohner-Freedman mass $M_\text{BF}^2$ is given in $2+1$ dimensions by \eqref{eq:BF}, and the solutions (\ref{eq:F GenericSolution}) are those respecting the bound in the absence of multiplicities for the roots of the characteristic polynomial \eqref{eq:CharacteristicPolynomial}, i.e., whenever
$M_{k}^{2}\neq M_{j}^{2}$ and $M_{k}^{2}\neq 0$, where $k\neq j$ and
$k,j=2,\ldots,N$. 

According to the analysis of Sec.~\ref{sec:AdSasympt}, these generic solutions will satisfy Brown-Henneaux boundary conditions (\ref{eq:Brown-Henneaux BC}) and consequently become asymptotically AdS spacetimes if additionally $M_{k}^{2}>0$ and $F_{k}^{-}=0$, for $k=2,\ldots,N$.

\subsection{Critical points \label{sec:CP}}

The critical points are defined as those where the roots of the characteristic polynomial \eqref{eq:CharacteristicPolynomial} acquire multiplicities. In such cases, the proper factorization of the polynomial is given by
\begin{equation}\label{eq:HigherOrder H_N Factor Degenerate}
\left[ n(n-2)\right]^{\lambda_{1}}
\left[ n(n-2)-M_{2}^{2}l^{2}\right]^{\lambda_{2}}\cdots 
\left[ n(n-2)-M_{j}^{2}l^{2}\right]^{\lambda_{j}}=0, \qquad \sum_{k=1}^{j}\lambda_{k}=N,
\end{equation}
where the integers $\lambda_{k}\geq1$ are the square mass multiplicities. At critical points, we must supplement the generic power-law modes with new logarithmic ones, and the resulting wave profile is given by {\small
\begin{subequations}\label{eq:F DegenerateSolution}
\begin{equation}
F(u,y) = \sum_{\mathsf{m}=1}^{\lambda_{1}-1}\left[ F_{1\mathsf{m}}^{+}(u) \left(\frac{y}{l}\right)^{2} + F_{1\mathsf{m}}^{-}(u)\right] 
\ln^\mathsf{m}\!\left( \frac{y}{l}\right)
+\sum_{k=2}^{j} F_{k}(u,y),
\end{equation}

\begin{equation} 
F_{k}(u,y) = 
\begin{cases}
 \displaystyle
 \sum_{\mathsf{m}=0}^{\lambda_{k}-1} 
 \left[ F_{k\mathsf{m}}^{+}(u)\left(\frac{y}{l}\right)^{1+l\sqrt{M_k^2-M_\text{BF}^2}}
 +F_{k\mathsf{m}}^{-}(u)\left(\frac{y}{l}\right)^{1-l\sqrt{M_k^2-M_\text{BF}^2}}\right]
 \ln^\mathsf{m}\!\left(\frac{y}{l}\right)
 , & M_{k}^2 > M_\text{BF}^2, \\ \\
 \displaystyle
 \frac{y}{l}\sum_{\mathsf{m}=0}^{2\lambda_{k}-1} F_{k\mathsf{m}}(u)
 \ln^\mathsf{m}\!\left(\frac{y}{l}\right)
 , & 
 M_{k}^2 = M_\text{BF}^2.
\end{cases}
\end{equation}
\end{subequations}}%

In the critical solutions (\ref{eq:F DegenerateSolution}), we can distinguish sectors with different asymptotic behaviors, weaker than the one determined by the Brown-Henneaux boundary conditions (\ref{eq:Brown-Henneaux BC}). Indeed, the modes
\begin{equation}
 F_{k\mathsf{m}}^{+}(u)\left(\frac{y}{l}\right)^{1+l\sqrt{M_k^2-M_\text{BF}^2}}\ln^\mathsf{m}\!\left(\frac{y}{l}\right),
\end{equation}
have a $\ln^\mathsf{m}$ asymptotic behavior, provided $M_{k}^{2}>0$. That is, for these modes the corresponding next-to-leading terms in the metric fall off as
\begin{equation}\label{eq:log_m BC}
h_{uv}\sim h_{yy}\sim \mathscr{O}(1) \qquad 
h_{uu}\sim h_{vv}\sim \mathscr{O}(\ln^\mathsf{m}(y)), \qquad 
h_{uy}\sim h_{vy}\sim \mathscr{O}(y\ln^\mathsf{m}(y)).
\end{equation}
Boundary conditions (\ref{eq:log_m BC}) were mentioned in \cite{Gurses:2015zia} and are a generalization of the $\ln^{2}$ ones discussed in \cite{Bergshoeff:2012ev}. The whole solution (\ref{eq:F DegenerateSolution}) then turns out to be compatible with the weakened $\ln^{\chi-1}$ AdS asymptotic behavior, determined by the greatest root multiplicity $\chi=\max(\lambda_{k})$. All these boundary conditions are signature of the existence of allegedly dual logarithmic conformal field theories \cite{Bergshoeff:2012ev,Grumiller:2008qz,Liu:2009bk,*Liu:2009kc}. 

This completes the generic description of AdS-waves in Viel-dreibein gravity, made possible by the formal expression found for its mass spectrum in Sec.~\ref{sec:Mfact}. In the next section we obtain explicit results for the lowest numbers of gravities.

\section{Explicit examples \label{sec:VDGeg}}

In this section we illustrate how to apply the previously described method to find the AdS-wave solutions for Viel-dreibein gravity to the simplest versions of the theory, corresponding to $N=2,3,4$.

\subsection{Zwei-dreibein gravity: \texorpdfstring{$N=2$}{}}

The Lagrangian of Viel-dreibein gravity (\ref{eq:Lagrangian_N}) for $N=2$ coincides with the Lagrangian of Zwei-dreibein gravity studied in \cite{Bergshoeff:2013xma}, after choosing $\sigma_2=1$ without losing generality as was already remarked. In this case the differential operator (\ref{eq:HigherOrder H_N Polynomial}) becomes
\begin{subequations}\label{eq:HigherOrder H_2 Polynomial}
\begin{equation}
\mathscr{H}_{2}=\mathscr{H}_{N}\Big \vert_{N=2}= p_{2}\left( \frac{\square}{m^{2}}\right) ^{2}+p_{1}\left( \frac{\square}{m^{2}}\right),
\end{equation}
where the polynomial coefficients, according to the prescriptions (\ref{eq:p_k}), are
\begin{equation}\label{eq:p_2p_1}
\begin{aligned}
p_{2}&= (-1)^{N} \sigma_{1}\gamma_{1} \cdots \sigma_{N}\gamma_{N} \Big \vert_{N=2}=\sigma_{1}\gamma_{1} \sigma_{2}\gamma_{2},\\ 
p_{1}&= -\left( \sum_{k=1}^{N}\sigma_{k}\gamma_{k}\right) a_{12}\cdots a_{N-1,N}\Big \vert_{N=2}=-(\sigma_{1}\gamma_{1} +\sigma_{2}\gamma_{2}) a_{12}.
\end{aligned}
\end{equation}
\end{subequations}
In any case, the differential polynomial can be trivially factorized, leading to the fourth-order Euler differential equation for the wave profile
\begin{equation}\label{eq:HigherOrder H_2 Factor}
\square \left(\square-M_{2}^{2} \right)F=0,
\end{equation}
with the effective mass given by
\begin{equation}\label{eq:M_2N2}
M_{2}^{2}=-m^2\frac{ p_{1}}{p_{2}}=m^2 \beta_{12} \gamma_{1}\frac{\sigma_{1}\gamma_{1}+\sigma_{2}\gamma_{2}}{\sigma_{1}\sigma_{2}},
\end{equation}
where we have used the definition of $a_{12}$ given in \eqref{eq:aII+1}, or more precisely (\ref{eq:a12N=3}). We want to remark that the same result is obtained from the general expressions (\ref{eq:HigherOrder H_N
Factor}) and (\ref{eq:H_N roots}), since
\begin{equation}
\left[\frac{p_{1}}{p_{2}} \right]=
\sum_{\substack{i_{1},i_{2}\geq 0\\
 i_1-i_2=0\\
 i_1-2i_2=0 }}
\frac{(-1)^{i_{2}}}{i_{1}+1}
\begin{pmatrix}
i_{2} \\ i_{1}
\end{pmatrix}
 \frac{p_{1}^{i_{1}+1}}{p_{2}^{i_{2}+1}}
=\frac{p_1}{p_2},
\end{equation}
resulting in $M_2^2=m^2 \theta_2=-m^2\frac{p_1}{p_2}$.

We write the generic solution for the AdS-wave profile, valid for $M_{2}\neq 0$, from (\ref{eq:F GenericSolution}) as
\begin{equation}\label{eq:F_2}
F(u,y)=F_{2}(u,y)= 
\begin{cases}
\displaystyle
F_{2}^{+}(u)\left(\frac{y}{l}\right)
^{1+l\sqrt{M_2^2-M_\text{BF}^2}} +
F_{2}^{-}(u)\left(\frac{y}{l}\right)
^{1-l\sqrt{M_2^2-M_\text{BF}^2}}, 
& M_{2}^2 > M_\text{BF}^2\\ \\
\displaystyle
\frac{y}{l}\left[F_{12}(u)\ln\left(\frac{y}{l} \right)+F_{02}(u)\right],
& M_{2}^2 = M_\text{BF}^2,
\end{cases}
\end{equation}
while at the critical point $M_{2}=0$, the solution (\ref{eq:F DegenerateSolution}) reduces to
\begin{equation}\label{eq:F DegenerateSolution2}
F(u,y) = 
\left[F_{11}^{+}(u)\left(\frac{y}{l}\right)^{2} + F_{11}^{-}(u)\right] 
\ln\left(\frac{y}{l}\right).
\end{equation}
These results totally agree with those reported in \cite{Bergshoeff:2014eca}, if we fix
$\sigma_2=\gamma_1=1$, as may be easily verified.

\subsection{Drei-dreibein gravity: \texorpdfstring{$N=3$}{}}

When $N=3$, we have the Drei-dreibein gravity case studied for the first time in Sec.~\ref{sec:DDG}. Here, we will recover those results from the general case with arbitrary $N$. We start by evaluating the differential operator (\ref{eq:HigherOrder H_N Polynomial}) for $N=3$
\begin{subequations}\label{eq:HigherOrder H_N_3 Polynomial}
\begin{equation}
\mathscr{H}_{3}=\mathscr{H}_{N}\Big \vert_{N=3}= p_{3}\left( \frac{\square}{m^{2}}\right) ^{3}+p_{2}\left( \frac{\square}{m^{2}}\right) ^{2}+p_{1}\left( \frac{\square}{m^{2}}\right),
\end{equation}
where the polynomial coefficients are given by (\ref{eq:p_k})
\begin{equation}\label{eq:p_3p_2p_1}
\begin{aligned}
p_{3}&= (-1)^{N} \sigma_{1}\gamma_{1}\cdots \sigma_{N}\gamma_{N} \Big\vert_{N=3}=-\sigma_{1}\gamma_{1} \sigma_{2}\gamma_{2}\sigma_{3}\gamma_{3},\\ 
p_{2}&=(-1)^{N-1}\smashoperator[l]{
\sum_{\substack{
 I_{1}<\cdots<I_{N-1}\\
 J\neq I_{k}\forall k}}} 
\sigma_{I_1}\gamma_{I_{1}} \cdots 
\sigma_{ I_{N-1}}\gamma_{ I_{N-1}} 
(a_{J-1,J}+a_{J,J+1})\Big \vert_{N=3}\\ 
 &= \sigma_{1}\gamma_{1} \sigma_{2}\gamma_{2}a_{23}+\sigma_{1}\gamma_{1} \sigma_{3}\gamma_{3}(a_{12}+a_{23})+\sigma_{2}\gamma_{2} \sigma_{3}\gamma_{3}a_{12},\\ 
p_{1}&= -\left( \sum_{k=1}^{N}\sigma_{k}\gamma_{k}\right) a_{12}\cdots a_{N-1,N}\Big \vert_{N=3}=-(\sigma_{1}\gamma_{1} +\sigma_{2}\gamma_{2} +\sigma_{3}\gamma_{3}) a_{12}a_{23},
\end{aligned}
\end{equation}
\end{subequations}
and we have used that $a_{01}=0=a_{34}$. Then we write factorization (\ref{eq:HigherOrder H_N Factor}) as
\begin{equation}\label{eq:HigherOrder H_N_3 Factor}
\square\left(\square-M_{2}^{2} \right)\left(\square-M_{3}^{2} \right)F=0,
\end{equation}
where the squared masses 
\begin{equation}\label{eq:M_23}
\begin{aligned} 
M_{2}^{2}&= -m^2\left[\frac{p_{1}}{p_{2}} \right],\\ 
M_{3}^{2}&= -m^2\left( \left[\frac{p_{2}}{p_{3}} \right] -\left[\frac{p_{1}}{p_{2}} \right]\right) ,
\end{aligned}
\end{equation}
are given by $\mathscr{A}$-hypergeometric functions allowing the representation (\ref{eq:H_N
roots1})
\begin{subequations}
\begin{align} 
\nonumber
\left[\frac{p_{1}}{p_{2}} \right]&=
\sum_{\substack{
 i_{1},i_{2},i_{3}\geq 0\\
 i_1-i_2+i_3=0\\
 i_1-2i_2+3i_3=0 }} 
\frac{(-1)^{i_{2}}}{i_{1}+1}
\begin{pmatrix}
i_{2}\\
i_{1} i_{3}
\end{pmatrix}
\frac{p_{1}^{i_{1}+1}p_{3}^{i_{3}}}{p_{2}^{i_{2}+1}}\\  \nonumber
&=\frac{p_2}{p_3}\sum_{i=0}^{\infty} \frac{2^i (2i-1)!!}{(i+1)!}\left( \frac{p_1 p_3}{p_2^2}\right) ^{i+1}\\ 
\label{p21}
&=\frac{p_2}{2p_3}\left[1+ \sum_{i=0}^{\infty} \frac{(2i-3)!!}{i!}\left( \frac{2p_1 p_3}{p_2^2}\right) ^{i}\right], \\ \label{p23}
\left[\frac{p_{2}}{p_{3}} \right]&=
\sum_{\substack{
 i_{1},i_{2},i_{3}\geq 0\\
 i_1+i_2-i_3=0\\
 i_1+2i_2-3i_3=0 }}
\frac{(-1)^{i_{3}}}{i_{2}+1}
\begin{pmatrix}
i_{3}\\
i_{1} i_{2}
\end{pmatrix}
\frac{p_{1}^{i_{1}}p_{2}^{i_{2}+1}}{p_{3}^{i_{3}+1}} =\frac{p_{2}}{p_{3}}.
\end{align}
\end{subequations}
We can use the Taylor expansion
\begin{equation}\nonumber
\sqrt{1-2x}=-\sum_{i=0}^{\infty} \frac{(2i-3)!!}{i!}x^{i},
\end{equation}
to write masses (\ref{eq:M_23}) as
\begin{equation}\label{M_23radical}
\begin{aligned}
M_{2}^{2}&=-m^2\frac{p_2}{2p_3}
\left(1-\sqrt{1-\frac{4p_1p_3}{p_2^2}}\right),\\ 
M_{3}^{2}&=-m^2\frac{p_2}{2p_3}
\left(1+\sqrt{1-\frac{4p_1p_3}{p_2^2}}\right).
\end{aligned}
\end{equation}
The factorization (\ref{eq:HigherOrder H_N_3 Factor}) with the masses (\ref{M_23radical}), emerging both from the general formulas of previous section, completely match the analogue ones (\ref{eq:HigherOrder H_3 Factor}) and (\ref{eq:Mpm}) from Sec.~\ref{sec:DDG}. In fact, it can be easily verified that $M_2^2=M_{+}^2$ and $M_3^2=M_{-}^2$, since $p_i=a_{12}a_{23}\tilde{p}_{i}$, $i=1,2,3$. It is also evident that the general solutions (\ref{eq:F GenericSolution}) and (\ref{eq:F DegenerateSolution}) reproduce (\ref{eq:F GenericSolutionN=3}), (\ref{eq:F 3crit}), (\ref{eq:F 2crit zero mass}), (\ref{eq:F 2crit non zero mass}), and (\ref{eq:F 2crit Ml^2eq-1}) once the former are evaluated at $N=3$.

\subsection{Vier-dreibein gravity: \texorpdfstring{$N=4$}{}}

The last example we detail corresponds to the version of Viel-dreibein gravity with $N=4$. As in the previous cases, we first evaluate the differential operator (\ref{eq:HigherOrder H_N Polynomial})
\begin{subequations}\label{eq:HigherOrder H_N_4 Polynomial}
\begin{equation}
\mathscr{H}_{4}=\mathscr{H}_{N}\Big \vert_{N=4}= p_{4}\left( \frac{\square}{m^{2}}\right) ^{4}+p_{3}\left( \frac{\square}{m^{2}}\right) ^{3}+p_{2}\left( \frac{\square}{m^{2}}\right) ^{2}+p_{1}\left( \frac{\square}{m^{2}}\right),
\end{equation}
where the expressions (\ref{eq:p_k}) again determine the polynomial coefficients as
\begin{equation}\label{eq:p_4p_3p_2p_1}
\begin{aligned}
p_{4}={}& (-1)^{N} \sigma_{1}\gamma_{1} \cdots \sigma_{N}\gamma_{N} \Big \vert_{N=4} 
=\sigma_{1}\gamma_{1}\sigma_{2}\gamma_{2}
\sigma_{3}\gamma_{3}\sigma_{4}\gamma_{4},\\ 
p_{3}={}& (-1)^{N-1}
\smashoperator[l]{
\sum_{\substack{
 I_{1}<\cdots<I_{N-1}\\
 J\neq I_{k}\forall k}}}
\sigma_{I_1}\gamma_{I_{1}} \cdots 
\sigma_{I_{N-1}}\gamma_{I_{N-1}} 
(a_{J-1,J}+a_{J,J+1})\Big\vert_{N=4}\\ 
={}& - [\, \sigma_{1}\gamma_{1} \sigma_{2}\gamma_{2}\sigma_{3}\gamma_{3}a_{34}
+ \sigma_{1}\gamma_{1} \sigma_{2}\gamma_{2}\sigma_{4}\gamma_{4}(a_{23}+a_{34}) + \sigma_{1}\gamma_{1} \sigma_{3}\gamma_{3}\sigma_{4}\gamma_{4}(a_{12}+a_{23}) \\ 
& \quad + \sigma_{2}\gamma_{2}\sigma_{3}\gamma_{3}
\sigma_{4}\gamma_{4}a_{12} \,], \\ 
p_{2}={}& (-1)^{2}
\smashoperator[l]{
\sum_{\substack{
 I_{1}<I_{2}\\
 J_{1}<\cdots<J_{N-2}\\
 J_{r}\neq I_{q}\forall r,q\\ 
 s_{q}=-1,0}}}
\sigma_{I_1}\gamma_{I_1}\sigma_{I_2}\gamma_{I_2} a_{J_{1}+s_{1},J_{1}+s_{1}+1} \cdots 
a_{J_{N-k}+s_{N-k},J_{N-k}+s_{N-k}+1}\\ 
& \qquad\qquad\qquad \times 
\Delta(J_{1}+s_{1},\ldots,J_{N-k}+s_{N-k}) 
\Big\vert_{N=4}\\ 
={}& \sigma_{1}\gamma_{1} \sigma_{2}\gamma_{2}a_{23}a_{34}+ \sigma_{1}\gamma_{1} \sigma_{3}\gamma_{3}(a_{12}+a_{23})a_{34} + \sigma_{1}\gamma_{1} \sigma_{4}\gamma_{4}(a_{12}+a_{34})a_{23} 
+ \sigma_{2}\gamma_{2}\sigma_{3}\gamma_{3}a_{12}a_{34} \\ 
& + \sigma_{2}\gamma_{2} \sigma_{4}\gamma_{4}a_{12}(a_{23}+a_{34})+
 \sigma_{3}\gamma_{3} \sigma_{4}\gamma_{4}a_{12}a_{23} ,\\ 
p_{1}={}& -\left( \sum_{k=1}^{N}\sigma_{k}\gamma_{k}\right) a_{12}\cdots a_{N-1,N}\Big \vert_{N=4} =
-(\sigma_{1}\gamma_{1} +\sigma_{2}\gamma_{2} +\sigma_{3}\gamma_{3} +\sigma_{4}\gamma_{4}) a_{12}a_{23}a_{34},
\end{aligned}
\end{equation}
\end{subequations}
after taking $a_{01}=0=a_{45}$. The differential equation for the wave profile $\mathscr{H}_{4}(F)=0$ becomes factorized to
\begin{equation}\label{eq:HigherOrder H_N_4 Factor}
\square\left(\square-M_{2}^{2} \right)\left(\square-M_{3}^{2} \right)\left(\square-M_{4}^{2} \right)F=0,
\end{equation}
with squared masses represented by \eqref{eq:H_N roots} as
\begin{equation}\label{eq:M4_234}
\begin{aligned} 
M_{2}^{2}&= -m^2\left[\frac{p_{1}}{p_{2}} \right],\\ 
M_{3}^{2}&= -m^2\left( \left[\frac{p_{2}}{p_{3}} \right] -\left[\frac{p_{1}}{p_{2}} \right]\right) ,\\ 
M_{4}^{2}&= -m^2\left( \left[\frac{p_{3}}{p_{4}} \right] -\left[\frac{p_{2}}{p_{3}} \right]\right) ,
\end{aligned}
\end{equation}
in terms of the $\mathscr{A}$-hypergeometric functions (\ref{eq:H_N roots1}) that for $N=4$ become
\begin{subequations}\label{eq:p4_12_23_34}
\begin{align} 
\left[\frac{p_{1}}{p_{2}}\right]
&= \sum_{\substack{
 i_{1},i_{2},i_{3},i_{4}\geq 0\\
 i_1-i_2+i_3+i_4=0\\
 i_1-2i_2+3i_3+4i_4=0 }}
\frac{(-1)^{i_{2}}}{i_{1}+1}
\begin{pmatrix}
i_{2}\\
i_{1} i_{3} i_{4}
\end{pmatrix}
\frac{p_{1}^{i_{1}+1}p_{3}^{i_{3}}p_{4}^{i_{4}}}{p_{2}^{i_{2}+1}} \nonumber \\ 
&= \sum_{i_3,i_4\geq 0} \frac{(-1)^{i_{4}}(2i_3+3i_4)!}{(i_3+2i_4+1)!i_3!i_4!}\frac{p_1^{i_3+2i_4+1} p_3^{i_3}p_4^{i_4}}{p_2^{2i_3+3i_4+1}}, \label{eq:p4_12} \\
\left[\frac{p_{2}}{p_{3}} \right] 
&= \sum_{\substack{
 i_{1},i_{2},i_{3},i_{4}\geq 0\\
 i_1+i_2-i_3+i_4=0\\
 i_1+2i_2-3i_3+4i_4=0 }}
\frac{(-1)^{i_{3}}}{i_{2}+1}
\begin{pmatrix}
i_{3} \\
i_{1} i_{2} i_{4}
\end{pmatrix}
\frac{p_{1}^{i_{1}}p_{2}^{i_{2}+1}p_{4}^{i_{4}}}{p_{3}^{i_{3}+1}} \nonumber \\
&= \sum_{\substack{
 i_{1},i_{2}\geq 0 }}
\frac{(-1)^{i_{1}}(3i_1+2i_2)!}{i_1!(i_2+1)!(2i_1+i_2)!}\frac{p_1^{i_1} p_2^{i_2+1}p_4^{2i_1+i_2}}{p_3^{3i_1+2i_2+1}}, \label{eq:p4_23} \\ \label{eq:p4_34}
\left[\frac{p_{3}}{p_{4}} \right] 
&= \sum_{\substack{
 i_{1},i_{2},i_{3},i_{4}\geq 0\\
 i_1+i_2+i_3-i_4=0\\
 i_1+2i_2+3i_3-4i_4=0 }}
\frac{(-1)^{i_{4}}}{i_{3}+1}
\begin{pmatrix}
i_{4} \\
i_{1} i_{2} i_{3}
\end{pmatrix}
 \frac{p_{1}^{i_{1}}p_{2}^{i_{2}}p_{3}^{i_{3}+1}}{p_{4}^{i_{4}+1}} = \frac{p_{3}}{p_{4}}.
\end{align}
\end{subequations}

Of course, the series (\ref{eq:p4_12_23_34}) are linked to the Cardano solutions \cite{Zucker:2008}
\begin{equation}\label{eq:Cardano}
\begin{aligned} 
\tilde{\theta}_{0}(p_1,p_2,p_3,p_4)&= u_{+} +u_{-} -\frac{p_3}{3p_4},\\ 
\tilde{\theta}_{+}(p_1,p_2,p_3,p_4)&= e^{\frac{2}{3} \pi i }u_{+} + e^{-\frac{2}{3} \pi i}u_{-} -\frac{p_3}{3p_4},\\
\tilde{\theta}_{-}(p_1,p_2,p_3,p_4)&= e^{-\frac{2}{3} \pi i }u_{+} +e^{\frac{2}{3} \pi i }u_{-}-\frac{p_3}{3p_4},
\end{aligned}
\end{equation}
of the cubic equation $p_4\tilde{\theta}^3+p_3\tilde{\theta}^2+p_2\tilde{\theta}+p_1=0$, where
\begin{subequations}\label{eq:upmCardano}
\begin{equation}
u_{\pm} =\left(-Q\pm\sqrt{Q^2+P^3} \right)^{\frac{1}{3}},
\end{equation}
\begin{equation}
P= -\frac{p_3^2}{9p_4^2}+\frac{p_2}{3p_4}, \qquad 
Q= \frac{p_3^3}{27p_4^3}-\frac{p_3p_2}{6p_4^2}+\frac{p_1}{2p_4},
\end{equation}
\end{subequations}
are any of the three possible pairs $(u_+, u_-)$ of cubic roots satisfying the relation $u_+ u_-=-P$. To see this, it is enough to check that the Taylor expansions of the Cardano solutions
$\tilde{\theta}_{0}(1,1,\bar{p}_3,\bar{p}_4)$ and
$\tilde{\theta}_{+}(\breve{p}_1,\breve{p}_2,1,1)$ around
$\bar{p}_3=0=\bar{p}_4$ and $\breve{p}_1=0=\breve{p}_2$, respectively, are 
\begin{subequations}\label{eq:TaylorSeriesCardanoSol_p_12_34}
\begin{align}
\tilde{\theta}_{0}(1,1,\bar{p}_3,\bar{p}_4)&=-\smashoperator[l]{\sum_{i_3,i_4=0}^{\infty}} \frac{(-1)^{2i_3+3i_4}(2i_3+3i_4)!}{i_3!i_4!(i_3+2i_4+1)!}\bar{p}_3^{i_3}\bar{p}_4^{i_4},\\ 
\tilde{\theta}_{+}(\breve{p}_1,\breve{p}_2,1,1)&=-1+\smashoperator[l]{
\sum_{\substack{
 i_1,i_2+1=0\\
 i_1+i_2\geq0}}^{\infty}}
\frac{(-1)^{3i_1+2i_2}(3i_1+2i_2)!}{i_1!(i_2+1)!(2i_1+i_2)!}\breve{p}_1^{i_1}\breve{p}_2^{i_2+1}.
\end{align}
\end{subequations}
Then one can use the twofold homogeneity property of the Cardano solutions
\cite{Passare:2004}
\begin{equation}
\lambda_1^{-1}\tilde{\theta}_{0,\pm}(p_1,p_2,p_3,p_4)=\tilde{\theta}_{0,\pm}(\lambda_0 p_1,\lambda_0\lambda_1 p_2,\lambda_0\lambda_1^2p_3,\lambda_0\lambda_1^3p_4),
\end{equation}
to get
\begin{subequations}
\begin{align}\nonumber
\tilde{\theta}_{0}(p_1,p_2,p_3,p_4)
&=\frac{p_1}{p_2}
\tilde{\theta}_{0}\left(1,1,\frac{p_1p_3}{p_2^2},\frac{p_1^2p_4}{p_2^3}\right)\\ \nonumber
&=-\frac{p_1}{p_2}
\sum_{i_3,i_4=0}^{\infty}
\frac{(-1)^{2i_3+3i_4}(2i_3+3i_4)!}{i_3!i_4!(i_3+2i_4+1)!}\left( \frac{p_1p_3}{p_2^2}\right)^{i_3}\left( \frac{p_1^2p_4}{p_2^3}\right)^{i_4}\\ \label{eq:Taylor_HypergeomSeriesCardanoSol p_12}
&= - \left[\frac{p_{1}}{p_{2}}\right],\\ \nonumber
\tilde{\theta}_{+}(p_1,p_2,p_3,p_4)
&= \frac{p_3}{p_4}
\tilde{\theta}_{+}\left(\frac{p_1p_4^2}{p_3^3},\frac{p_2p_4}{p_3^2},1,1\right)\\ \nonumber
&=\frac{p_3}{p_4}
\left[-1+\smashoperator[l]{
\sum_{\substack{
 i_1,i_2+1=0\\
 i_1+i_2\geq0}}^{\infty}}
\frac{(-1)^{3i_1+2i_2}(3i_1+2i_2)!}
{i_1!(i_2+1)!(2i_1+i_2)!}
\left(\frac{p_1p_4^2}{p_3^3}\right)^{i_1}
\left(\frac{p_2p_4}{p_3^2}\right)^{i_2+1}\right] \\ 
&= - \frac{p_3}{p_4} +\left[\frac{p_{2}}{p_{3}} \right].
\label{eq:Taylor_HypergeomSeriesCardanoSol p_34}
\end{align}
Additionally, it follows from (\ref{eq:Cardano}) that
\begin{align}\nonumber
\tilde{\theta}_{-}(p_1,p_2,p_3,p_4)
&= -\tilde{\theta}_{0}(p_1,p_2,p_3,p_4)-\tilde{\theta}_{+}(p_1,p_2,p_3,p_4) - \frac{p_3}{p_4}\\ \label{eq:Taylor_HypergeomSeriesCardanoSol p_23}
&= -\left[\frac{p_{2}}{p_{3}} \right]+\left[\frac{p_{1}}{p_{2}} \right].
\end{align}
\end{subequations}
That is, the masses \eqref{eq:M4_234} consequently allows the more conventional Cardano representation 
\begin{equation}\label{eq:Mass_CardanoSol}
M_{2}^{2}= m^2\tilde{\theta}_{0}, \qquad
M_{3}^{2}= m^2\tilde{\theta}_{+}, \qquad
M_{4}^{2}= m^2\tilde{\theta}_{-}.
\end{equation}

In terms of the above masses, the wave profile function $F$ is given by (\ref{eq:F GenericSolution}) for $N=4$ in the generic case. On the other hand, at the critical points we have \eqref{eq:F DegenerateSolution} where the mass squared multiplicities must satisfy $\displaystyle\sum_{k=1}^{j}\lambda_{k}=4$, giving a maximum degeneracy $\max(\lambda_{1})=4$ for the massless modes, or $\max(\lambda_{k})=3$ for the massive ones ($k\geq2$). As a consequence, we may have sectors with $\ln$, $\ln^2$, or $\ln^3$ asymptotic behaviors, provided the squared masses \eqref{eq:Mass_CardanoSol} are positive.

\section{Conclusions \label{sec:conclu}}

In this work, we thoroughly examined the AdS-wave configurations of the ($2+1$)-dimensional theory of $N$ interacting spin-2 fields known as Viel-dreibein gravity, as a means to understand its mass spectrum. To provide a brief context, the AdS waves---exact solutions for gravitational waves propagating on an AdS spacetime of radius $l$ and characterized in Poincaré coordinates by a profile function $F(u,y)$, depending on the retarded time $u$ and the front-wave coordinate $y$---has been a very effective tool to explore the nontrivial spectrum of massive gravities and learn about how their behavior depart from standard gravity. Such exploration started with Topologically Massive Gravity \cite{Deser:1981wh,*Deser:1982vy}, even before the relevance of critical points in massive gravities was understood \cite{Ayon-Beato:2004nrg,*Ayon-Beato:2005pnc,Ayon-Beato:2005gdo}. This explains why, after the later discovery of New Massive Gravity \cite{Bergshoeff:2009hq}, this was precisely one of the first tasks to be completed for the then-new theory \cite{Ayon-Beato:2009cgh}. With the subsequent advent of $N=2$ Zwei-dreibein gravity \cite{Bergshoeff:2013xma}, the harder characterization of its AdS waves was also accomplished soon after \cite{Bergshoeff:2014eca}, by means of the higher-derivative perspective \cite{Hassan:2013pca}. However, the latter is much more difficult to implement for $N\ge3$ and as a result the problem has remained open until now.

As a warm-up allowing us to clearly exhibit the proposed approach, we start exploring the AdS-wave solutions for the simplest open case: a ghost-free version of $N=3$ Drei-dreibein gravity that according to its graph becomes a line theory \cite{Afshar:2014dta, Scargill:2014wya}, since only interactions coupled by the constants $\beta_{12}$ and $\beta_{23}$ remain. As consequence, the field equations for the lower dreibeine are algebraically linear in the immediately higher dreibein, allowing to isolate the latter in terms of up to second-order expressions for the immediately below one. Successively inserting the results in the remaining equations when the first dreibein describes an AdS-wave ansatz, is how we successfully derive a sixth-order equation for the wave profile. The related sixth-order operator is a third-degree polynomial in the d'Alembertian, in its version acting on advanced-time symmetric profiles, that becomes the operator of an Euler ordinary equation. Such a polynomial is easily factorized in terms of Klein-Gordon operators, one massless and two others massive. Consequently, the straightforward determination of the mass spectrum in this case allows to find in detail the full space of solutions. The arising generic solution is given by a superposition of the two modes corresponding to each nontrivial mass, since the massless ones are pure-gauge in $2+1$ dimensions. As expected, the solution acquires a logarithmic mode when at least one of the two squared masses saturates the slightly negative Breitenlohner-Freedman bound, marking the limit of stability behavior on AdS. Additionally, at the points in the parameter space of the theory corresponding to nonnegative squared masses, we got solutions with new logarithmic modes when some masses coincide. They admit $\ln$ and/or $\ln^2$ AdS asymptotic decays. These boundary conditions have been conjectured to be related to the existence of dual logarithmic conformal field theories, in the context of the AdS/CFT correspondence \cite{Bergshoeff:2012ev,Grumiller:2008qz,Liu:2009bk,*Liu:2009kc}.

We then face the highly nontrivial task of generalizing the results to the Viel-dreibein gravity case, with an arbitrary number $N$ of interacting dreibeine. The first difficulty to overcome is the implementation of the previously described procedure leading to a single higher-derivative equation for the wave profile. We manage to rewrite the involved equations as a three-term iterative system that is successfully solved appealing to their properties. As consequence, a $2N$-order Euler equation is produced for the AdS-wave profile, that is encoded in a $N$-degree polynomial operator of the d'Alembertian. The main obstacle now is how to factorize this general polynomial as a product of Klein-Gordon operators in order to read the mass spectrum of Viel-dreibein gravity. We accomplish this appealing to the relatively recent Sturmfels succinct characterization of the roots of a generic polynomial with arbitrary coefficients \cite{Sturmfels:2000}, obtaining concrete expressions for the masses in terms of the coupling constants of the theory given through $\mathscr{A}$-hypergeometric functions. This factorization is again the key allowing us to obtain the AdS-wave solutions as the superposition of the two modes corresponding to $N-1$ massive Klein-Gordon equations. Once more, new logarithmic modes appear at the critical points leading to $\ln^{\chi-1}$ asymptotic behavior, being $\chi\le N$.

Regarding the perspectives of this work, the approach successfully exhibited here can be extended to the potentially more phenomenological four-dimensional multigravity theories. First, the more involved AdS waves of ghost-free bigravity \cite{Hassan:2011zd} were exhaustively studied in \cite{Ayon-Beato:2018hxz} by considering a biproportionality ansatz and recognizing the resulting decoupled massless and massive Klein-Gordon excitations on AdS as integrable Euler-Darboux equations. A more comprehensive strategy would again consist in recasting the theory as higher-derivative equations for a single vierbein \cite{Hassan:2013pca}, chosen as an AdS-wave ansatz. The hope is that the higher-order equation emerging for the wave profile becomes a polynomial in the Euler-Darboux operators and can eventually be factorized. This then can be implemented to any $N>2$ as we do here by exploiting the $\mathscr{A}$-hypergeometric factorization of generic polynomials, incidentally providing a mass spectrum for multigravities also in $D=4$. The solution profiles would be a superposition of those corresponding to each Euler-Darboux operator of the factorization associated with a given mass, which have already been characterized in \cite{Ayon-Beato:2018hxz}.

Another future direction is to adapt part of the latter strategy to the study of black holes in multigravity. The current status of this problem in the more challenging spinning case is to exploit their stationary axisymmetric Kerr-Schild representation \cite{Ayon-Beato:2015nvz,*Ayon-Beato:2025ahb}, as was originally proposed in \cite{Ayon-Beato:2015qtt}. This justifies the proportionality ans\"atze to Kerr-Schild transformations for bigravity of \cite{Ayon-Beato:2015qtt}, extended later to multigravities in \cite{Wood:2024acv,*Wood:2024eol} and \cite{Garcia-Compean:2026cnq}. The drawback of such an approach is that only the spinning black holes of General Relativity are recovered, at most allowing multiple values for their masses \cite{Babichev:2014tfa,Ayon-Beato:2015qtt,Wood:2024acv,*Wood:2024eol,Garcia-Compean:2025wkj,Garcia-Compean:2026cnq}. Resorting again to a higher-derivative system for a single vielbein \cite{Hassan:2013pca,Bergshoeff:2014eca}, rooted now in a stationary axisymmetric Kerr-Schild ansatz \cite{Ayon-Beato:2015nvz,*Ayon-Beato:2025ahb}, must lead to a higher-order linear equation for the Kerr-Schild profile, eventually solved by the factorization method exhibited here. In bigravity for example, it is expected that the infinite series of the alternative higher-derivative approach of \cite{Gording:2018not} becomes truncated by the nilpotent properties provided by the Kerr-Schild ansatz first identified in \cite{Ayon-Beato:2015qtt}, and later exploited in \cite{Ayon-Beato:2018hxz,Wood:2024acv,*Wood:2024eol,Garcia-Compean:2025wkj,Garcia-Compean:2026cnq}. However, this is not the end of the story since the enriched multigravity dynamics can provide additional nonlinear constraints. This subtle linear-nonlinear interplay is just the source of admissible departures from standard gravity. New Massive Gravity \cite{Bergshoeff:2009hq} is the iconic example where this has been explicitly proved \cite{Ayon-Beato:2014wla}, explaining how the Kerr-Schild ansatz can lead to new black holes not allowed by standard gravity \cite{Bergshoeff:2009aq,Oliva:2009ip,Ayon-Beato:2009rgu}. Obviously, the chances are much better for higher $D$ and $N$.

\begin{acknowledgments}
We would like to thank Leonardo de la Cruz, Daniel Flores-Alfonso, and Julio A.~M\'endez for useful discussions and suggestions. This research was partially funded by Conahcyt grant A1-S-11548.
\end{acknowledgments}

\appendix

\section{Polynomial form of the differential operator \texorpdfstring{$\mathscr{H}_{N}$}{HN} \label{app:p_k}}

In this appendix, we will prove by complete induction that the prescription (\ref{eq:p_k}) gives the right expressions for the coefficients $p_k$ of the polynomial form \eqref{eq:HigherOrder H_N Polynomial} of the operator $\mathscr{H}_{N}$. The induction proceeds as usual by explicitly verifying first the results for $N=2$ and $N=3$ cases, starting from the tridiagonal matrix structure of the differential operator $\mathscr{H}_{N}$ given by (\ref{eq:HigherOrder H_N Det}). Then, assuming the validity up to $N$ dreibeins, we show that the general formula (\ref{eq:p_k}) also holds for $N^{\prime}=N+1$.

For $N=2$
\begin{equation}\label{eq:H_2 Det}
\mathscr{H}_{2} =
\begin{vmatrix}
-\frac{\sigma_{1}\gamma_{1}}{m^{2}}\square +a_{12}, & a_{12} \\
a_{12} & -\frac{\sigma_{2}\gamma_{2}}{m^{2}}\square +a_{2,3}+a_{1,2}
\end{vmatrix}, \qquad \text{with} \qquad a_{2,3}=0.
\end{equation}
Direct calculation leads to 
\begin{equation}\label{eq:H_2p_k}
\mathscr{H}_{2}= \sigma_{1}\gamma_{1} \sigma_{2}\gamma_{2}\left( \frac{\square}{m^{2}}\right) ^{2}-(\sigma_{1}\gamma_{1} +\sigma_{2}\gamma_{2}) a_{12} \left( \frac{\square}{m^{2}}\right).
\end{equation}
This result coincides with that obtained from the general prescription for the coefficients $p_k$ when $N=2$, as can be seen by comparing (\ref{eq:H_2p_k}) with (\ref{eq:HigherOrder H_2 Polynomial}).

If we repeat the procedure for $N=3$, from
\begin{equation}\label{eq:H_3 Det}
\mathscr{H}_{3} =
\begin{vmatrix}
-\frac{\sigma_{1}\gamma_{1}}{m^{2}}\square +a_{12} & a_{12} & 0 \\
a_{12} & -\frac{\sigma_{2}\gamma_{2}}{m^{2}}\square +a_{2,3}+a_{1,2} & a_{2,3} \\
0 & a_{23} & -\frac{\sigma_{3}\gamma_{3}}{m^{2}}\square +a_{3,4}+a_{2,3} 
\end{vmatrix}, \qquad \text{with} \qquad a_{3,4}=0,
\end{equation}
we directly get 
\begin{align}
\mathscr{H}_{3} &= 
-\ \sigma_{1}\gamma_{1}
\sigma_{2}\gamma_{2}\sigma_{3}\gamma_{3}
\left( \frac{\square}{m^{2}}\right) ^{3}+\left[\sigma_{1}\gamma_{1} \sigma_{2}\gamma_{2}a_{23}+\sigma_{1}\gamma_{1} \sigma_{3}\gamma_{3}(a_{12}+a_{23})\right. \notag \\ 
&\phantom{=}\ + \left.\sigma_{2}\gamma_{2} \sigma_{3}\gamma_{3}a_{12} \right] \left( \frac{\square}{m^{2}}\right) ^{2} -(\sigma_{1}\gamma_{1} +\sigma_{2}\gamma_{2} + \sigma_{3}\gamma_{3}) a_{12}a_{23}\left( \frac{\square}{m^{2}}\right). \label{eq:H_3p_k}
\end{align}
From the above expression (\ref{eq:H_3p_k}) and (\ref{eq:HigherOrder H_N_3 Polynomial}) it can be seen that the general prescription also holds true for $N=3$.

Now, let's assume that for $N=K$ the differential operator given by (\ref{eq:HigherOrder H_N Det}) with $a_{K,K+1}=0$, satisfying $\mathscr{H}_{K}(F)=0$ can be written as
\begin{subequations}\label{eq:HigherOrder H_N_K Polynomial}
\begin{equation}
\mathscr{H}_{K}= \sum_{k=0}^{K}p_{k}(K)\left( \frac{\square}{m^{2}}\right) ^{k},
\end{equation}
where
\begin{equation}\label{eq:p_k_N}
\begin{aligned}
p_{K}(K)={}&(-1)^{K} \sigma_{1}\gamma_{1} \cdots \sigma_{K}\gamma_{K},\\ 
p_{K-1}(K)={}&(-1)^{K-1}
\smashoperator[l]{
\sum_{\substack{
 I_{1}<\cdots<I_{K-1}\\
 J\neq I_{k}\forall k}}}
\sigma_{I_1}\gamma_{I_{1}}\cdots \sigma_{ I_{K-1}}\gamma_{ I_{K-1}} (a_{J-1,J}+a_{J,J+1}),\\ 
\vdotswithin{=} \\ 
p_{k}(K)={}&(-1)^{k}
\smashoperator[l]{
\sum_{\substack{
 I_{1}<\cdots<I_{k}\\
 J_{1}<\cdots<J_{K-k}\\
 J_{r}\neq I_{q}\forall r,q\\ 
 s_{q}=-1,0}}} 
\sigma_{I_1}\gamma_{I_1}\cdots \sigma_{I_k}\gamma_{I_k} 
a_{J_{1}+s_{1},J_{1}+s_{1}+1}\cdots 
a_{J_{K-k}+s_{K-k},J_{K-k}+s_{K-k}+1}\\ 
& \qquad\qquad\qquad \times \Delta(J_{1}+s_{1},\ldots,J_{K-k}+s_{K-k}),\\ 
\vdotswithin{=} \\ 
p_{1}(K)={}& -\left(\sum_{k=1}^{K}\sigma_{k}\gamma_{k}\right) a_{12}\cdots a_{K-1,K},\\ 
p_{0}(K)={}& a_{12}\cdots a_{K,K+1}=0,
\end{aligned}
\end{equation}
\end{subequations}
and we use definition \eqref{eq:Delta()}. Then, we must evaluate 
\eqref{eq:HigherOrder H_N} for $N=K+1$ and $a_{K+1,K+2}=0$,
\begin{equation}\label{eq:HigherOrder H_K+1 Det}
\mathscr{H}_{K+1}=
\begin{vmatrix}
\Upsilon_{1} & a_{12} &0& &0&0&0\\
a_{12} & \Upsilon_{2} & a_{23} & \cdots &0&0&0\\
0& a_{23} & \Upsilon_{3} & &0&0&0\\
&\vdots & & \ddots & & \vdots & \\
0&0&0& & \Upsilon_{K-1} & a_{K-1,K}&0\\
0&0&0& \cdots & a_{K-1,K} & \Upsilon_{K} & a_{K,K+1}\\
0&0&0& &0& a_{K,K+1} & \Upsilon_{K+1}
\end{vmatrix}.
\end{equation}
Expanding the previous determinant by minors, we see that
\begin{align} \nonumber 
\mathscr{H}_{K+1} 
&= \Upsilon_{K+1} \mathscr{H}_{K} - a_{K,K+1}^2 \mathscr{H}_{K-1} \\ \nonumber 
&= \left( - \frac{\sigma_{K+1}\gamma_{K+1}}{m^{2}}\square +a_{K+1,K+2}+a_{K,K+1} \right) \sum_{k=0}^{K}p_{k}(K)\left( \frac{\square}{m^{2}}\right) ^{k} \\ \nonumber 
&\phantom{=}\ - a_{K,K+1}^2 \sum_{k=0}^{K-1}p_{k}(K-1)\left( \frac{\square}{m^{2}}\right)^{k}  \\ \label{eq:Induction_H_K+1}
&= -\: \sigma_{K+1}\gamma_{K+1} p_{K}(K)\left( \frac{\square}{m^{2}}\right) ^{K+1}+ \\ \nonumber 
&\phantom{=}\ +\left[ -\sigma_{K+1}\gamma_{K+1} p_{K-1}(K) +(a_{K+1,K+2}+a_{K,K+1})p_{K}(K) \right] \left( \frac{\square}{m^{2}}\right)^{K} \\ \nonumber 
&\phantom{=}\ + \sum_{k=1}^{K-1} \left[ -\sigma_{K+1}\gamma_{K+1} p_{k-1}(K) + (a_{K+1,K+2}+a_{K,K+1})p_{k}(K) - a_{K,K+1}^2 p_{k}(K-1) \right] \left(\frac{\square}{m^{2}}\right)^{k} \\ \nonumber 
&\phantom{=}\ +(a_{K+1,K+2}+a_{K,K+1})p_{0}(K)- a_{K,K+1}^2 p_{0}(K-1).
\end{align}
From the previous sum, it is easy to realize that
\begin{subequations}\label{eq:HigherOrder H_N_K+1 Polynomial}
\begin{equation}
\mathscr{H}_{K+1}= \sum_{k=0}^{K+1}p_{k}(K+1)\left( \frac{\square}{m^{2}}\right) ^{k},
\end{equation}
where {\small 
\begin{equation}\label{eq:p_k_K+1}
\begin{aligned}
p_{K+1}(K+1)={}& -\sigma_{K+1}\gamma_{K+1} p_{K}(K) \\
={}& 
(-1)^{K+1} \sigma_{1}\gamma_{1} \cdots \sigma_{K+1}\gamma_{K+1},\\ 
p_{K}(K+1)={}& -\sigma_{K+1}\gamma_{K+1} p_{K-1}(K) +(a_{K+1,K+2}+a_{K,K+1})p_{K}(K) \\
 ={}&(-1)^{K}
\smashoperator[l]{
\sum_{\substack{
 I_{1}<\cdots<I_{K}\\
 J\neq I_{k}\forall k}}}
\sigma_{I_1}\gamma_{I_{1}}\cdots \sigma_{ I_{K}}\gamma_{ I_{K}} (a_{J-1,J}+a_{J,J+1}),\\ 
\vdotswithin{=} \\ 
p_{k}(K+1)={}&(-1)^{k}
\smashoperator[l]{
\sum_{\substack{
 I_{1}<\cdots<I_{k}\\
 J_{1}<\cdots<J_{K+1-k}\\
 J_{r}\neq I_{q}\forall r,q\\
 s_{q}=-1,0}}} 
\sigma_{I_1}\gamma_{I_1}\cdots \sigma_{I_k}\gamma_{I_k} 
a_{J_{1}+s_{1},J_{1}+s_{1}+1}\cdots 
a_{J_{K+1-k}+s_{K+1-k},J_{K+1-k}+s_{K+1-k}+1}\\ 
& \qquad\qquad\qquad \times \Delta(J_{1}+s_{1},\ldots,J_{K+1-k}+s_{K+1-k}),\\ 
\vdotswithin{=} \\ 
p_{1}(K+1)={}& -\left(\sum_{k=1}^{K}\sigma_{k}\gamma_{k}\right) a_{12}\cdots a_{K-1,K}a_{K,K+1},\\ 
p_{0}(K+1)={}& a_{12}\cdots a_{K,K+1}a_{K+1,K+2}=0.
\end{aligned}
\end{equation}}%
\end{subequations}

Therefore, the $N=K$ expressions (\ref{eq:p_k_N}) also hold for $N=K+1$, completing the inductive step. Since the result is true for $K=2, 3$ and the inductive step has been established, we conclude the validity of expansion \eqref{eq:HigherOrder H_N Polynomial}.

\section{\texorpdfstring{$\mathscr{A}$}{A}-Hypergeometric Functions \label{app:AHyper}}

The theory of $\mathscr{A}$-hypergeometric functions, also known as \emph{Gel'fand--Kapranov--Zelevinsky} (GKZ) hypergeometric functions, was developed in the late 1980s as a unifying framework for a broad class of special functions \cite{Gelfand:1989,Gelfand:1990bua}. The aim of this theory is to generalize classical hypergeometric functions to several variables in a way that is naturally adapted to problems arising in algebraic geometry, combinatorics, and mathematical physics \cite{Beukers:2014}.

Classical hypergeometric functions are characterized as solutions of linear differential equations with regular singularities. For instance, Gauss' hypergeometric function ${}_2F_1$ satisfies a second-order differential equation with three regular singular points. Although several multivariable generalizations exist (such as the Appell and Lauricella functions \cite{Beukers:2011}), these examples belong to specific families and do not capture the full range of structures that arise in applications.

The GKZ construction provides a systematic extension. Instead of starting from a particular differential equation one begins with a combinatorial object, that is, a configuration of integer vectors, which encodes the structure of the problem. A system of linear PDE is then constructed from these data, whose solutions are the $\mathscr{A}$-hypergeometric functions. Concretely, they arise whenever two types of constraints are imposed on a multivalued function: algebraic (binomial) relations among monomials, often called toric constraints, and homogeneity relations specifying the scaling behavior under a torus action. The resulting differential system is holonomic and admits explicit power series solutions known as the $\Gamma$-series. In what follows, we provide a brief outline to the subject.

Let $\mathscr{A}$ be an integer matrix whose columns are vectors in $\mathbb{Z}^r$, 
\begin{equation}
\mathscr{A} =
\begin{pmatrix}
\vec{a}_1 & \vec{a}_2 & \cdots & \vec{a}_N
\end{pmatrix}.
\end{equation}
The set of vectors $\vec{a}_i$ is assumed to $\mathbb{Z}$-span the lattice \(\mathbb{Z}^r\), and the $r$-th component of all vectors is set equal to $1$, that is, $\vec{a}_i=(a_{1,i},a_{2,i},\cdots,a_{r-1,i},1)$. Other ingredients are the kernel of $\mathscr{A}$, which is the integer lattice defined as
\begin{equation}
L \;=\; \ker_{\mathbb{Z}}(\mathscr{A}) \;=\; 
\{\vec{\ell}\in\mathbb{Z}^N : \mathscr{A}  \vec{\ell}=0\}, 
\end{equation}
and a parameter vector $\vec{\beta}$ with components that are usually chosen to be rational numbers, $\vec{\beta}\in\mathbb{Q}^r$, but in general may be complex, $\vec{\beta}\in\mathbb{C}^r$.

The $\mathscr{A}$-hypergeometric system $H_\mathscr{A}(\beta)$ is a system of linear PDE for functions of the variables $\vec{v}=(v_1,\dotsc,v_N)$ \cite{Beukers:2011}. It is generated by two families of operators, the first ones are the so-called \emph{toric operators} defined for each element of the kernel $\vec{\ell}\in L$ as
\begin{equation}\label{eq:ToricOp}
\Box_{\vec{\ell}}
=
\prod_{\ell_j>0}\partial_j^{\ell_j}
-
\prod_{\ell_j<0}\partial_j^{-\ell_j},
\end{equation}
where $\partial_j=\frac{\partial}{\partial v_j}$. 
The second ones are the \emph{Euler operators} for each row $i=1,\dotsc,r$ of $\mathscr{A}$ and the corresponding parameter 
\begin{equation}\label{eq:EulerOp}
E_i =
\sum_{j=1}^{N} a_{ij} v_j \partial_j - \beta_i .
\end{equation}
A function $\Phi(\vec{v})$ satisfying
\begin{equation}
\Box_{\vec{\ell}}\,(\Phi) = 0,
\qquad
E_i(\Phi) = 0,
\end{equation}
for all $\vec{\ell}\in L$ and $i=1,\dotsc,r$ is called an $\mathscr{A}$-\emph{hypergeometric function} with parameter vector $\vec{\beta}$.

A canonical family of formal solutions $\Phi_{\vec{\gamma}}(\vec{v})$ can be obtained as follows. Choose $\vec{\gamma}\in\mathbb{C}^N$ satisfying the compatibility condition
\begin{equation}\label{eq:Ag=b}
\mathscr{A}\vec{\gamma} = \vec{\beta},
\end{equation}
and consider the lattice-sum 
\begin{equation}\label{eq:LattSum}
\Phi_{\vec{\gamma}}(\vec{v}) 
= 
\sum_{\vec{\ell}\in L}
\frac{\vec{v}^{\;\vec{\gamma}+\vec{\ell}}}
{\Gamma(\vec{\gamma}+\vec{\ell}+\mathbf{1})}
\equiv 
\sum_{\vec{\ell}\in L} \prod_{j=1}^N
\frac{v_j^{\,\gamma_j+\ell_j}}
{\Gamma(\gamma_j+\ell_j+1)}.
\end{equation}
A careful evaluation over the kernel $L$ shows that this series is annihilated by the toric operators $\Box_{\vec{\ell}}$. Furthermore, it is straightforward to show that condition \eqref{eq:Ag=b} ensures the Euler equations are also satisfied. Different choices of $\vec{\gamma}$ produce different local branches and convergence is not automatic, so in general these are formal solutions see \cite{Sturmfels:2000} and \cite{Beukers:2011} for details. However, complete sets of series solutions to the GKZ system are built for appropriate choices \cite{Gelfand:1989} under a so-called ``non-resonant'' condition, which we do not revise here but that warrants the GKZ system is not reducible to a simpler differential one \cite{Beukers:2011}. Note that the \(\Gamma\)-denominator furnishes the factorial/Pochhammer structure that is present in classical hypergeometric coefficients.

\section{Roots as solutions to a GKZ system \label{app:Roots}}

As a concrete application of the general construction described in App.~\ref{app:AHyper}, consider the characterization of the roots of any polynomial equation
\begin{equation}\label{eq:polynomial_eq}
f(x)=p_0+p_1x+p_2x^2+\cdots+p_nx^n=0 .
\end{equation}
Each root is necessarily a function of the coefficients, $\theta_j=\theta_j(p_0,p_1,p_2,\dotsc,p_n)$, where  $f(\theta_j)=0$, $j=1,\dotsc,n$. A classical result is that such functions satisfy a system of PDE, as was first observed by Mayr \cite{Mayr:1936}. This system was later reinterpreted within the GKZ framework by Sturmfels \cite{Sturmfels:2000} as follows. The data of the configuration of integer points
$\{0,1,2,\dots,n\}\subset\mathbb{Z}$ is encoded within the GKZ formalism in the matrix
\begin{equation}
\mathscr{A}=
\begin{pmatrix}
0 & 1 & 2 & \cdots & n \\
1 & 1 & 1 & \cdots & 1
\end{pmatrix}, 
\end{equation}
whose kernel $L$ is the lattice of vectors $\vec{\ell}=(\ell_0,\ell_1,\ell_2,\dotsc,\ell_n)\in \mathbb{Z}^{n+1}$, with integer components satisfying the constraints 
\begin{subequations}\label{eq:Kconstr}
\begin{align}
0&=\ell_1 + 2\ell_2 +\dotsb+ n\ell_n, \\
0&=\ell_0 + \ell_1 + \ell_2 +\dotsb+ \ell_n.
\end{align}
\end{subequations}
Applying the general GKZ construction to this specific matrix, with the vector of parameters fixed at $\vec{\beta}=(-1,0)$, yields the following differential equations for a function $\Phi(\vec{p})$ of the coefficients $\vec{p}=(p_0,p_1,p_2,\dotsc,p_n)$. First, the relations associated to the corresponding toric operators \eqref{eq:ToricOp} can be generated from the simpler ones \cite{Stienstra:2007}
\begin{equation}\label{eq:Toric_DE}
\frac{\partial^2 \Phi}{\partial p_i\,\partial p_j} =
\frac{\partial^2 \Phi}{\partial p_k\,\partial p_l},
\qquad \text{whenever} \qquad i+j=k+l.
\end{equation}
Second, from \eqref{eq:EulerOp} the Euler relations become
\begin{subequations}\label{eq:Euler_DEs}
\begin{align} 
\sum_{i=0}^{n} i\,p_i \frac{\partial \Phi}{\partial p_i}
&= -\Phi , \label{eq:Euler_DE1}\\
\sum_{i=0}^{n} p_i \frac{\partial \Phi}{\partial p_i}
&= 0 . \label{eq:Euler_DE2}
\end{align}
\end{subequations}
This is precisely the PDE system identified by Mayr as characterizing the roots \cite{Mayr:1936}, which allowed Sturmfels \cite{Sturmfels:2000} to conclude they are $\mathscr{A}$-hypergeometric functions of the coefficients present in their polynomial equation \eqref{eq:polynomial_eq}.

The representation of the roots as $\mathscr{A}$-hypergeometric series must be done with care as is emphasized by Sturmfels in \cite{Sturmfels:2000}, since the related Euler subsystem \eqref{eq:Euler_DEs} does not satisfy the non-resonance condition required by GKZ to construct complete sets of series solutions \cite{Gelfand:1989}. However, the proper series obtained by Sturmfels can be obtained from the GKZ ones \eqref{eq:LattSum} by a nontrivial limit \cite{Stienstra:2007,Beukers:2011,Beukers:2014}. We start with a suitable choice for $\vec{\gamma}$ in \eqref{eq:Ag=b}, as $\vec{\gamma}^{(j)}=(0,0,\cdots,0,1,-1,0,\cdots,0)$ where the unit negative value appears in the $j$th component, that is, a vector with components $\gamma^{(j)}_{i}=\delta_{i,j-1}-\delta_{i,j}$. There are $n$ vectors $\vec{\gamma}^{(j)}$ of that type for $j=1,\dotsc,n$. A superposition of the formal series solutions \eqref{eq:LattSum} corresponding to all of them remains a solution due to the linear and homogeneous character of the GKZ system
\begin{align} \label{eq:LattSumAp1}
\Phi_{\vec{\gamma}}(\vec{p})
&=\sum_{j=1}^n C^{(j)} \Phi_{\vec{\gamma}^{(j)}}(\vec{p})
=\sum_{j=1}^n C^{(j)} \sum_{\vec{\ell}\in L} \displaystyle \prod_{i=0}^{n}\frac{p_i^{\gamma_i+\ell_i}}{\Gamma(\gamma_i+\ell_i+1)} \nonumber\\ 
&=\sum_{j=1}^n C^{(j)} 
\sum_{\substack{
\ell_0,\ell_1,\dotsc,\ell_n \in \mathbb{Z}\\
\ell_1 + 2\ell_2 +\dotsb+ n\ell_n=0\\
\ell_0 + \ell_1 + \ell_2 +\dotsb+ \ell_n=0 
}} 
\frac{p_0^{\ell_0}}{\Gamma(\ell_0+1)}
\frac{p_1^{\ell_1}}{\Gamma(\ell_1+1)}
\dotsm
\frac{p_{j-2}^{\ell_{j-2}}}{\Gamma(\ell_{j-2}+1)}
\frac{p_{j-1}^{\ell_{j-1}+1}}{\Gamma(\ell_{j-1}+2)} \nonumber \\ 
& \qquad \qquad \qquad \qquad \qquad \qquad \times 
\frac{p_{j}^{\ell_{j}-1}}{\Gamma(\ell_{j})}
\frac{p_{j+1}^{\ell_{j+1}}}{\Gamma(\ell_{j+1}+1)}
\dotsm
\frac{p_n^{\ell_n}}{\Gamma(\ell_n+1)},
\end{align} 
where the $C^{(j)}$ are arbitrary constants. We now slightly modify this superposition by introducing a parameter $0<\epsilon \ll  1$ in the $j$-th component of the vectors $\gamma^{(j)}_{i}$, changing them to $\gamma^{(j)}_{i}(\epsilon)=\delta_{i,j-1}-\delta_{i,j}(1 - \epsilon)$. The superposition \eqref{eq:LattSumAp1} turns into a one-parameter family 
\begin{align} \label{eq:LattSumAp2}
\Phi_{\vec{\gamma}(\epsilon)}(\vec{p})
&=\sum_{j=1}^n C^{(j)}(\epsilon)
\sum_{\substack{
\ell_0,\ell_1,\dotsc,\ell_n \in \mathbb{Z}\\
\ell_1 + 2\ell_2 +\dotsb+ n\ell_n=0\\
\ell_0 + \ell_1 + \ell_2 +\dotsb+ \ell_n=0 
}} 
\frac{p_0^{\ell_0}}{\Gamma(\ell_0+1)}
\frac{p_1^{\ell_1}}{\Gamma(\ell_1+1)}
\dotsm
\frac{p_{j-2}^{\ell_{j-2}}}{\Gamma(\ell_{j-2}+1)}
\frac{p_{j-1}^{\ell_{j-1}+1}}{\Gamma(\ell_{j-1}+2)} \nonumber \\ 
& \qquad \qquad \qquad \qquad \qquad \qquad \quad \:
\times 
\frac{p_{j}^{\ell_{j}-1+\epsilon}}{\Gamma(\ell_{j}+\epsilon)}
\frac{p_{j+1}^{\ell_{j+1}}}{\Gamma(\ell_{j+1}+1)}
\dotsm
\frac{p_n^{\ell_n}}{\Gamma(\ell_n+1)},
\end{align} 
where we are also incorporating an arbitrary dependence on the parameter in the constants. Applying the reflection formula $\Gamma(z)\Gamma(1-z)=\frac{\pi}{\sin(\pi z)}$ to the factor $\Gamma(\ell_j+\epsilon)$, and subsequently using the identity $\sin(\pi(\ell_j+\epsilon))=(-1)^{\ell_j}\sin(\pi\epsilon)$ for $\ell_j\in \mathbb{Z}$, we get the common factor $\frac{\sin(\pi\epsilon)}\pi$. Since the latter is independent of the summation indices, it can be absorbed within the arbitrary dependence of the constants by the redefinition $C^{(j)}(\epsilon) \mapsto \tilde{C}^{(j)}(\epsilon)=\frac{\sin(\pi\epsilon)}\pi C^{(j)}(\epsilon)$. The nontrivial limit consists in assuming that the renormalized constants are nonvanishing and well-behaved when we return to $\epsilon\rightarrow0$, bringing the superposition to
\begin{align} 
\Phi_{\vec{\gamma}}(\vec{p}) & = \sum_{j=1}^n \tilde{C}^{(j)} 
\sum_{\substack{
\ell_0,\ell_1,\dotsc,\ell_n \in \mathbb{Z}\\
\ell_1 + 2\ell_2 +\cdots+ n\ell_n=0\\
\ell_0 + \ell_1 + \ell_2 +\cdots+ \ell_n=0 }} (-1)^{l_j}\frac{p_0^{\ell_0}}{\Gamma(\ell_0+1)}
\frac{p_1^{\ell_1}}{\Gamma(\ell_1+1)}
 \dotsm
\frac{p_{j-2}^{\ell_{j-2}}}{\Gamma(\ell_{j-2}+1)}
\frac{p_{j-1}^{\ell_{j-1}+1}}{\Gamma(\ell_{j-1}+2)}
\nonumber \\
& \qquad \qquad \qquad \qquad \qquad \qquad \! \times \frac{\Gamma(1- \ell_j)}{p_{j}^{- \ell_j+1}}
\frac{p_{j+1}^{\ell_{j+1}}}{\Gamma(\ell_{j+1}+1)}
 \dotsm
\frac{p_n^{\ell_n}}{\Gamma(\ell_n+1)}. \label{eq:monomial_series}
\end{align}
Noticing that $\Gamma(z)$ has simple poles at $z=-\ell$, $\forall \ell \in \mathbb{N^+}$, we can see that the sum in (\ref{eq:monomial_series}) must be restricted to nonnegative integers $\ell_i\neq l_j$, for which additionally $\Gamma(\ell+1)=\ell!$. Furthermore, after relabeling $l_j \mapsto -l_j$ the kernel constraints on the integers \eqref{eq:Kconstr} are only satisfied if the new $l_j$ is also a nonnegative integer. Then the superposition becomes
\begin{equation}\label{eq:superp_monomial}
\Phi_{\vec{\gamma}}(\vec{p}) = \sum_{j=1}^n \tilde{C}^{(j)} \left[\frac{ p_{j-1}}{p_{j}} \right],
\end{equation}
where the superposed series are precisely those proposed by Sturmfels \cite{Sturmfels:2000}
\begin{subequations}\label{eq:monomial_series_final}
\begin{align} 
\left[\frac{ p_{j-1}}{p_{j}} \right]&
\equiv \sum_{\ell_0,\ell_1,\dotsc,\ell_n \geq 0} (-1)^{l_j} 
 \frac{ \ell_j!}{\ell_0!\ell_1!\dotsm \ell_{j-2}!(\ell_{j-1}+1)!\ell_{j+1}!\dotsm \ell_n! }
\nonumber \\ 
& \qquad \qquad \qquad \! \times 
\frac{p _0^{\ell_0} p_1^{\ell_1} \dotsm p_{j-2}^{\ell_{j-2}} p_{j-1} ^{\ell_{j-1}+1} p_{j+1}^{\ell_{j+1}} \dotsm p_n^{\ell_n} }{p_{j}^{ \ell_j+1} },\label{eq:monomial_series_finala}
\end{align}
with all the involved nonnegative integers belonging now to the lattice defined by the relations
\begin{align}
0 &= \ell_1 + 2\ell_2 +\cdots+(j-1)\ell_{j-1} -j \ell_j+(j+1)\ell_{j-1}+\cdots+ n\ell_n, \\
0 &= \ell_0 + \ell_1 + \ell_2 +\cdots+\ell_{j-1} - \ell_j+\ell_{j-1}+\cdots+ \ell_n.
\end{align}
\end{subequations}
The notation used for these multi-series comes from the fact that the leading term of \eqref{eq:monomial_series_finala} is the monomial $\frac{ p_{j-1}}{p_{j}}$.

The consistency of the result \eqref{eq:superp_monomial} of the nontrivial limit was proved by Sturmfels \cite{Sturmfels:2000}. Concretely, in appropriate regions of parameters space the complex-valued functions satisfying the GKZ linear system \eqref{eq:Toric_DE}-\eqref{eq:Euler_DE2} form a vector space of dimension $n$. There, he identifies convergent linearly independent elements as superpositions \eqref{eq:superp_monomial} of the series \eqref{eq:monomial_series_final} that for precise elections of the coefficients $\tilde{C}^{(j)}$, depending on the chosen triangulation of the set of integer points $\{0,1,\dots,n\}$, are just the roots of the polynomial equation \eqref{eq:polynomial_eq}. Among all the triangulations, the \emph{finest} one subdividing the interval into $n$ segments of unit length is special. For this triangulation the $n$ roots become the following succinct superposition of $\mathscr{A}$-hypergeometric series with integer coefficients \cite{Sturmfels:2000}
\begin{equation}\label{eq:Xj_monomial}
\theta_j =
-\left[\frac{p_{j-1}}{p_j}\right]
+
\left[\frac{p_{j-2}}{p_{j-1}}\right],
\qquad j=1,\dotsc,n ,
\end{equation}
where it is understood that $p_{-1}/p_0=0$.

In conclusion, the $\mathscr{A}$-hypergeometric framework provides a powerful analytic tool for solving polynomial equations beyond radicals. The method is constructive, based on series expansions, and it connects directly with familiar techniques in theoretical physics, such as perturbation series and generating functions. By working with the finest triangulation, one obtains simple and transparent expressions for the $n$ roots, suitable for both theoretical analysis and symbolic computation. This is the one we choose to represent the mass spectrum of Viel-dreibein gravity in Sec.~\ref{sec:Mfact}.

\bibliography{references}

\begin{thebibliography}{83}%
\makeatletter
\providecommand \@ifxundefined [1]{%
 \@ifx{#1\undefined}
}%
\providecommand \@ifnum [1]{%
 \ifnum #1\expandafter \@firstoftwo
 \else \expandafter \@secondoftwo
 \fi
}%
\providecommand \@ifx [1]{%
 \ifx #1\expandafter \@firstoftwo
 \else \expandafter \@secondoftwo
 \fi
}%
\providecommand \natexlab [1]{#1}%
\providecommand \enquote  [1]{``#1''}%
\providecommand \bibnamefont  [1]{#1}%
\providecommand \bibfnamefont [1]{#1}%
\providecommand \citenamefont [1]{#1}%
\providecommand \href@noop [0]{\@secondoftwo}%
\providecommand \href [0]{\begingroup \@sanitize@url \@href}%
\providecommand \@href[1]{\@@startlink{#1}\@@href}%
\providecommand \@@href[1]{\endgroup#1\@@endlink}%
\providecommand \@sanitize@url [0]{\catcode `\\12\catcode `\$12\catcode `\&12\catcode `\#12\catcode `\^12\catcode `\_12\catcode `\%12\relax}%
\providecommand \@@startlink[1]{}%
\providecommand \@@endlink[0]{}%
\providecommand \url  [0]{\begingroup\@sanitize@url \@url }%
\providecommand \@url [1]{\endgroup\@href {#1}{\urlprefix }}%
\providecommand \urlprefix  [0]{URL }%
\providecommand \Eprint [0]{\href }%
\providecommand \doibase [0]{https://doi.org/}%
\providecommand \selectlanguage [0]{\@gobble}%
\providecommand \bibinfo  [0]{\@secondoftwo}%
\providecommand \bibfield  [0]{\@secondoftwo}%
\providecommand \translation [1]{[#1]}%
\providecommand \BibitemOpen [0]{}%
\providecommand \bibitemStop [0]{}%
\providecommand \bibitemNoStop [0]{.\EOS\space}%
\providecommand \EOS [0]{\spacefactor3000\relax}%
\providecommand \BibitemShut  [1]{\csname bibitem#1\endcsname}%
\let\auto@bib@innerbib\@empty
\bibitem [{\citenamefont {Wigner}(1931)}]{Wigner:1931}%
  \BibitemOpen
  \bibfield  {author} {\bibinfo {author} {\bibfnamefont {E.~P.}\ \bibnamefont {Wigner}},\ }\href {https://doi.org/10.1007/978-3-663-02555-9} {\emph {\bibinfo {title} {Gruppentheorie und ihre anwendung auf die quantenmechanik der atomspektren}}}\ (\bibinfo  {publisher} {F. Vieweg \& Sohn Akt.-Ges},\ \bibinfo {year} {1931})\BibitemShut {NoStop}%
\bibitem [{\citenamefont {Wigner}(1959)}]{Wigner:1959}%
  \BibitemOpen
  \bibfield  {author} {\bibinfo {author} {\bibfnamefont {E.~P.}\ \bibnamefont {Wigner}},\ }\href {https://www.sciencedirect.com/bookseries/pure-and-applied-physics/vol/5/suppl/C} {\emph {\bibinfo {title} {English translation}}}\ (\bibinfo  {publisher} {Academic Press, Inc.},\ \bibinfo {year} {1959})\BibitemShut {NoStop}%
\bibitem [{\citenamefont {Weinberg}(2005)}]{Weinberg:1995mt}%
  \BibitemOpen
  \bibfield  {author} {\bibinfo {author} {\bibfnamefont {S.}~\bibnamefont {Weinberg}},\ }\href {https://doi.org/10.1017/CBO9781139644167} {\emph {\bibinfo {title} {{The Quantum theory of fields. Vol. 1: Foundations}}}}\ (\bibinfo  {publisher} {Cambridge University Press},\ \bibinfo {year} {2005})\BibitemShut {NoStop}%
\bibitem [{\citenamefont {Fierz}\ and\ \citenamefont {Pauli}(1939)}]{Fierz:1939ix}%
  \BibitemOpen
  \bibfield  {author} {\bibinfo {author} {\bibfnamefont {M.}~\bibnamefont {Fierz}}\ and\ \bibinfo {author} {\bibfnamefont {W.}~\bibnamefont {Pauli}},\ }\href {https://doi.org/10.1098/rspa.1939.0140} {\bibfield  {journal} {\bibinfo  {journal} {Proc. Roy. Soc. Lond. A}\ }\textbf {\bibinfo {volume} {173}},\ \bibinfo {pages} {211} (\bibinfo {year} {1939})}\BibitemShut {NoStop}%
\bibitem [{\citenamefont {Feynman}(1996)}]{Feynman:1996kb}%
  \BibitemOpen
  \bibfield  {author} {\bibinfo {author} {\bibfnamefont {R.~P.}\ \bibnamefont {Feynman}},\ }\href {https://doi.org/10.1201/9780429502859} {\emph {\bibinfo {title} {{Feynman lectures on gravitation}}}},\ edited by\ \bibinfo {editor} {\bibfnamefont {F.~B.}\ \bibnamefont {Morinigo}}, \bibinfo {editor} {\bibfnamefont {W.~G.}\ \bibnamefont {Wagner}},\ and\ \bibinfo {editor} {\bibfnamefont {B.}~\bibnamefont {Hatfield}}\ (\bibinfo  {publisher} {CRC Press},\ \bibinfo {year} {1996})\BibitemShut {NoStop}%
\bibitem [{\citenamefont {Hinterbichler}(2012)}]{Hinterbichler:2011tt}%
  \BibitemOpen
  \bibfield  {author} {\bibinfo {author} {\bibfnamefont {K.}~\bibnamefont {Hinterbichler}},\ }\href {https://doi.org/10.1103/RevModPhys.84.671} {\bibfield  {journal} {\bibinfo  {journal} {Rev. Mod. Phys.}\ }\textbf {\bibinfo {volume} {84}},\ \bibinfo {pages} {671} (\bibinfo {year} {2012})},\ \Eprint {https://arxiv.org/abs/1105.3735} {arXiv:1105.3735 [hep-th]} \BibitemShut {NoStop}%
\bibitem [{\citenamefont {de~Rham}(2014)}]{deRham:2014zqa}%
  \BibitemOpen
  \bibfield  {author} {\bibinfo {author} {\bibfnamefont {C.}~\bibnamefont {de~Rham}},\ }\href {https://doi.org/10.12942/lrr-2014-7} {\bibfield  {journal} {\bibinfo  {journal} {Living Rev. Rel.}\ }\textbf {\bibinfo {volume} {17}},\ \bibinfo {pages} {7} (\bibinfo {year} {2014})},\ \Eprint {https://arxiv.org/abs/1401.4173} {arXiv:1401.4173 [hep-th]} \BibitemShut {NoStop}%
\bibitem [{\citenamefont {Boulware}\ and\ \citenamefont {Deser}(1972)}]{Boulware:1972yco}%
  \BibitemOpen
  \bibfield  {author} {\bibinfo {author} {\bibfnamefont {D.~G.}\ \bibnamefont {Boulware}}\ and\ \bibinfo {author} {\bibfnamefont {S.}~\bibnamefont {Deser}},\ }\href {https://doi.org/10.1103/PhysRevD.6.3368} {\bibfield  {journal} {\bibinfo  {journal} {Phys. Rev. D}\ }\textbf {\bibinfo {volume} {6}},\ \bibinfo {pages} {3368} (\bibinfo {year} {1972})}\BibitemShut {NoStop}%
\bibitem [{\citenamefont {de~Rham}\ \emph {et~al.}(2011)\citenamefont {de~Rham}, \citenamefont {Gabadadze},\ and\ \citenamefont {Tolley}}]{deRham:2010kj}%
  \BibitemOpen
  \bibfield  {author} {\bibinfo {author} {\bibfnamefont {C.}~\bibnamefont {de~Rham}}, \bibinfo {author} {\bibfnamefont {G.}~\bibnamefont {Gabadadze}},\ and\ \bibinfo {author} {\bibfnamefont {A.~J.}\ \bibnamefont {Tolley}},\ }\href {https://doi.org/10.1103/PhysRevLett.106.231101} {\bibfield  {journal} {\bibinfo  {journal} {Phys. Rev. Lett.}\ }\textbf {\bibinfo {volume} {106}},\ \bibinfo {pages} {231101} (\bibinfo {year} {2011})},\ \Eprint {https://arxiv.org/abs/1011.1232} {arXiv:1011.1232 [hep-th]} \BibitemShut {NoStop}%
\bibitem [{\citenamefont {Hassan}\ and\ \citenamefont {Rosen}(2012)}]{Hassan:2011zd}%
  \BibitemOpen
  \bibfield  {author} {\bibinfo {author} {\bibfnamefont {S.~F.}\ \bibnamefont {Hassan}}\ and\ \bibinfo {author} {\bibfnamefont {R.~A.}\ \bibnamefont {Rosen}},\ }\href {https://doi.org/10.1007/JHEP02(2012)126} {\bibfield  {journal} {\bibinfo  {journal} {J. High Energy Phys.}\ }\textbf {\bibinfo {volume} {02}}\bibfield  {number} {\bibinfo  {number} { (2012)},\ \bibinfo {pages} {126}},\ }\Eprint {https://arxiv.org/abs/1109.3515} {arXiv:1109.3515 [hep-th]} \BibitemShut {NoStop}%
\bibitem [{\citenamefont {Hinterbichler}\ and\ \citenamefont {Rosen}(2012)}]{Hinterbichler:2012cn}%
  \BibitemOpen
  \bibfield  {author} {\bibinfo {author} {\bibfnamefont {K.}~\bibnamefont {Hinterbichler}}\ and\ \bibinfo {author} {\bibfnamefont {R.~A.}\ \bibnamefont {Rosen}},\ }\href {https://doi.org/10.1007/JHEP07(2012)047} {\bibfield  {journal} {\bibinfo  {journal} {J. High Energy Phys.}\ }\textbf {\bibinfo {volume} {07}}\bibfield  {number} {\bibinfo  {number} { (2012)},\ \bibinfo {pages} {047}},\ }\Eprint {https://arxiv.org/abs/1203.5783} {arXiv:1203.5783 [hep-th]} \BibitemShut {NoStop}%
\bibitem [{\citenamefont {Li}(2016)}]{Li:2015iwc}%
  \BibitemOpen
  \bibfield  {author} {\bibinfo {author} {\bibfnamefont {W.}~\bibnamefont {Li}},\ }\href {https://doi.org/10.1103/PhysRevD.94.064079} {\bibfield  {journal} {\bibinfo  {journal} {Phys. Rev. D}\ }\textbf {\bibinfo {volume} {94}},\ \bibinfo {pages} {064079} (\bibinfo {year} {2016})},\ \Eprint {https://arxiv.org/abs/1512.06386} {arXiv:1512.06386 [hep-th]} \BibitemShut {NoStop}%
\bibitem [{\citenamefont {Flinckman}\ and\ \citenamefont {Hassan}(2026)}]{Flinckman:2025bje}%
  \BibitemOpen
  \bibfield  {author} {\bibinfo {author} {\bibfnamefont {J.}~\bibnamefont {Flinckman}}\ and\ \bibinfo {author} {\bibfnamefont {S.~F.}\ \bibnamefont {Hassan}},\ }\href {https://doi.org/10.1007/JHEP07(2026)137} {\bibfield  {journal} {\bibinfo  {journal} {J. High Energy Phys.}\ }\textbf {\bibinfo {volume} {07}}\bibfield  {number} {\bibinfo  {number} { (2026)},\ \bibinfo {pages} {137}},\ }\Eprint {https://arxiv.org/abs/2510.03014} {arXiv:2510.03014 [hep-th]} \BibitemShut {NoStop}%
\bibitem [{\citenamefont {Afshar}\ \emph {et~al.}(2015)\citenamefont {Afshar}, \citenamefont {Bergshoeff},\ and\ \citenamefont {Merbis}}]{Afshar:2014dta}%
  \BibitemOpen
  \bibfield  {author} {\bibinfo {author} {\bibfnamefont {H.~R.}\ \bibnamefont {Afshar}}, \bibinfo {author} {\bibfnamefont {E.~A.}\ \bibnamefont {Bergshoeff}},\ and\ \bibinfo {author} {\bibfnamefont {W.}~\bibnamefont {Merbis}},\ }\href {https://doi.org/10.1007/JHEP01(2015)040} {\bibfield  {journal} {\bibinfo  {journal} {J. High Energy Phys.}\ }\textbf {\bibinfo {volume} {01}}\bibfield  {number} {\bibinfo  {number} { (2015)},\ \bibinfo {pages} {040}},\ }\Eprint {https://arxiv.org/abs/1410.6164} {arXiv:1410.6164 [hep-th]} \BibitemShut {NoStop}%
\bibitem [{\citenamefont {Hassan}\ \emph {et~al.}(2015)\citenamefont {Hassan}, \citenamefont {Schmidt-May},\ and\ \citenamefont {von Strauss}}]{Hassan:2013pca}%
  \BibitemOpen
  \bibfield  {author} {\bibinfo {author} {\bibfnamefont {S.~F.}\ \bibnamefont {Hassan}}, \bibinfo {author} {\bibfnamefont {A.}~\bibnamefont {Schmidt-May}},\ and\ \bibinfo {author} {\bibfnamefont {M.}~\bibnamefont {von Strauss}},\ }\href {https://doi.org/10.3390/universe1020092} {\bibfield  {journal} {\bibinfo  {journal} {Universe}\ }\textbf {\bibinfo {volume} {1}},\ \bibinfo {pages} {92} (\bibinfo {year} {2015})},\ \Eprint {https://arxiv.org/abs/1303.6940} {arXiv:1303.6940 [hep-th]} \BibitemShut {NoStop}%
\bibitem [{\citenamefont {Bergshoeff}\ \emph {et~al.}(2014)\citenamefont {Bergshoeff}, \citenamefont {Goya}, \citenamefont {Merbis},\ and\ \citenamefont {Rosseel}}]{Bergshoeff:2014eca}%
  \BibitemOpen
  \bibfield  {author} {\bibinfo {author} {\bibfnamefont {E.~A.}\ \bibnamefont {Bergshoeff}}, \bibinfo {author} {\bibfnamefont {A.~F.}\ \bibnamefont {Goya}}, \bibinfo {author} {\bibfnamefont {W.}~\bibnamefont {Merbis}},\ and\ \bibinfo {author} {\bibfnamefont {J.}~\bibnamefont {Rosseel}},\ }\href {https://doi.org/10.1007/JHEP04(2014)012} {\bibfield  {journal} {\bibinfo  {journal} {J. High Energy Phys.}\ }\textbf {\bibinfo {volume} {04}}\bibfield  {number} {\bibinfo  {number} { (2014)},\ \bibinfo {pages} {012}},\ }\Eprint {https://arxiv.org/abs/1401.5386} {arXiv:1401.5386 [hep-th]} \BibitemShut {NoStop}%
\bibitem [{\citenamefont {Bergshoeff}\ \emph {et~al.}(2013)\citenamefont {Bergshoeff}, \citenamefont {de~Haan}, \citenamefont {Hohm}, \citenamefont {Merbis},\ and\ \citenamefont {Townsend}}]{Bergshoeff:2013xma}%
  \BibitemOpen
  \bibfield  {author} {\bibinfo {author} {\bibfnamefont {E.~A.}\ \bibnamefont {Bergshoeff}}, \bibinfo {author} {\bibfnamefont {S.}~\bibnamefont {de~Haan}}, \bibinfo {author} {\bibfnamefont {O.}~\bibnamefont {Hohm}}, \bibinfo {author} {\bibfnamefont {W.}~\bibnamefont {Merbis}},\ and\ \bibinfo {author} {\bibfnamefont {P.~K.}\ \bibnamefont {Townsend}},\ }\href {https://doi.org/10.1103/PhysRevLett.111.111102} {\bibfield  {journal} {\bibinfo  {journal} {Phys. Rev. Lett.}\ }\textbf {\bibinfo {volume} {111}},\ \bibinfo {pages} {111102} (\bibinfo {year} {2013})},\ \bibinfo {note} {[Erratum: Phys.Rev.Lett. 111, 259902 (2013)]},\ \Eprint {https://arxiv.org/abs/1307.2774} {arXiv:1307.2774 [hep-th]} \BibitemShut {NoStop}%
\bibitem [{\citenamefont {Deser}\ \emph {et~al.}(1982{\natexlab{a}})\citenamefont {Deser}, \citenamefont {Jackiw},\ and\ \citenamefont {Templeton}}]{Deser:1981wh}%
  \BibitemOpen
  \bibfield  {author} {\bibinfo {author} {\bibfnamefont {S.}~\bibnamefont {Deser}}, \bibinfo {author} {\bibfnamefont {R.}~\bibnamefont {Jackiw}},\ and\ \bibinfo {author} {\bibfnamefont {S.}~\bibnamefont {Templeton}},\ }\href {https://doi.org/10.1016/0003-4916(82)90164-6} {\bibfield  {journal} {\bibinfo  {journal} {Annals Phys.}\ }\textbf {\bibinfo {volume} {140}},\ \bibinfo {pages} {372} (\bibinfo {year} {1982}{\natexlab{a}})},\ \bibinfo {note} {[Erratum: Annals Phys. 185, 406 (1988)]}\BibitemShut {NoStop}%
\bibitem [{\citenamefont {Deser}\ \emph {et~al.}(1982{\natexlab{b}})\citenamefont {Deser}, \citenamefont {Jackiw},\ and\ \citenamefont {Templeton}}]{Deser:1982vy}%
  \BibitemOpen
  \bibfield  {author} {\bibinfo {author} {\bibfnamefont {S.}~\bibnamefont {Deser}}, \bibinfo {author} {\bibfnamefont {R.}~\bibnamefont {Jackiw}},\ and\ \bibinfo {author} {\bibfnamefont {S.}~\bibnamefont {Templeton}},\ }\href {https://doi.org/10.1103/PhysRevLett.48.975} {\bibfield  {journal} {\bibinfo  {journal} {Phys. Rev. Lett.}\ }\textbf {\bibinfo {volume} {48}},\ \bibinfo {pages} {975} (\bibinfo {year} {1982}{\natexlab{b}})}\BibitemShut {NoStop}%
\bibitem [{\citenamefont {Ayon-Beato}\ and\ \citenamefont {Hassaine}(2005{\natexlab{a}})}]{Ayon-Beato:2004nrg}%
  \BibitemOpen
  \bibfield  {author} {\bibinfo {author} {\bibfnamefont {E.}~\bibnamefont {Ayon-Beato}}\ and\ \bibinfo {author} {\bibfnamefont {M.}~\bibnamefont {Hassaine}},\ }\href {https://doi.org/10.1016/j.aop.2004.11.006} {\bibfield  {journal} {\bibinfo  {journal} {Annals Phys.}\ }\textbf {\bibinfo {volume} {317}},\ \bibinfo {pages} {175} (\bibinfo {year} {2005}{\natexlab{a}})},\ \Eprint {https://arxiv.org/abs/hep-th/0409150} {arXiv:hep-th/0409150} \BibitemShut {NoStop}%
\bibitem [{\citenamefont {Ayon-Beato}\ and\ \citenamefont {Hassaine}(2005{\natexlab{b}})}]{Ayon-Beato:2005pnc}%
  \BibitemOpen
  \bibfield  {author} {\bibinfo {author} {\bibfnamefont {E.}~\bibnamefont {Ayon-Beato}}\ and\ \bibinfo {author} {\bibfnamefont {M.}~\bibnamefont {Hassaine}},\ }\href {https://doi.org/10.1103/PhysRevD.71.084004} {\bibfield  {journal} {\bibinfo  {journal} {Phys. Rev. D}\ }\textbf {\bibinfo {volume} {71}},\ \bibinfo {pages} {084004} (\bibinfo {year} {2005}{\natexlab{b}})},\ \Eprint {https://arxiv.org/abs/hep-th/0501040} {arXiv:hep-th/0501040} \BibitemShut {NoStop}%
\bibitem [{\citenamefont {Ayon-Beato}\ and\ \citenamefont {Hassaine}(2006)}]{Ayon-Beato:2005gdo}%
  \BibitemOpen
  \bibfield  {author} {\bibinfo {author} {\bibfnamefont {E.}~\bibnamefont {Ayon-Beato}}\ and\ \bibinfo {author} {\bibfnamefont {M.}~\bibnamefont {Hassaine}},\ }\href {https://doi.org/10.1103/PhysRevD.73.104001} {\bibfield  {journal} {\bibinfo  {journal} {Phys. Rev. D}\ }\textbf {\bibinfo {volume} {73}},\ \bibinfo {pages} {104001} (\bibinfo {year} {2006})},\ \Eprint {https://arxiv.org/abs/hep-th/0512074} {arXiv:hep-th/0512074} \BibitemShut {NoStop}%
\bibitem [{\citenamefont {Bergshoeff}\ \emph {et~al.}(2009{\natexlab{a}})\citenamefont {Bergshoeff}, \citenamefont {Hohm},\ and\ \citenamefont {Townsend}}]{Bergshoeff:2009hq}%
  \BibitemOpen
  \bibfield  {author} {\bibinfo {author} {\bibfnamefont {E.~A.}\ \bibnamefont {Bergshoeff}}, \bibinfo {author} {\bibfnamefont {O.}~\bibnamefont {Hohm}},\ and\ \bibinfo {author} {\bibfnamefont {P.~K.}\ \bibnamefont {Townsend}},\ }\href {https://doi.org/10.1103/PhysRevLett.102.201301} {\bibfield  {journal} {\bibinfo  {journal} {Phys. Rev. Lett.}\ }\textbf {\bibinfo {volume} {102}},\ \bibinfo {pages} {201301} (\bibinfo {year} {2009}{\natexlab{a}})},\ \Eprint {https://arxiv.org/abs/0901.1766} {arXiv:0901.1766 [hep-th]} \BibitemShut {NoStop}%
\bibitem [{\citenamefont {Ayon-Beato}\ \emph {et~al.}(2009{\natexlab{a}})\citenamefont {Ayon-Beato}, \citenamefont {Giribet},\ and\ \citenamefont {Hassaine}}]{Ayon-Beato:2009cgh}%
  \BibitemOpen
  \bibfield  {author} {\bibinfo {author} {\bibfnamefont {E.}~\bibnamefont {Ayon-Beato}}, \bibinfo {author} {\bibfnamefont {G.}~\bibnamefont {Giribet}},\ and\ \bibinfo {author} {\bibfnamefont {M.}~\bibnamefont {Hassaine}},\ }\href {https://doi.org/10.1088/1126-6708/2009/05/029} {\bibfield  {journal} {\bibinfo  {journal} {J. High Energy Phys.}\ }\textbf {\bibinfo {volume} {05}}\bibfield  {number} {\bibinfo  {number} { (2009)},\ \bibinfo {pages} {029}},\ }\Eprint {https://arxiv.org/abs/0904.0668} {arXiv:0904.0668 [hep-th]} \BibitemShut {NoStop}%
\bibitem [{\citenamefont {Ay{\'o}n-Beato}\ \emph {et~al.}(2018)\citenamefont {Ay{\'o}n-Beato}, \citenamefont {Higuita-Borja}, \citenamefont {M{\'e}ndez-Zavaleta},\ and\ \citenamefont {Vel{\'a}zquez-Rodr{\'\i}guez}}]{Ayon-Beato:2018hxz}%
  \BibitemOpen
  \bibfield  {author} {\bibinfo {author} {\bibfnamefont {E.}~\bibnamefont {Ay{\'o}n-Beato}}, \bibinfo {author} {\bibfnamefont {D.}~\bibnamefont {Higuita-Borja}}, \bibinfo {author} {\bibfnamefont {J.~A.}\ \bibnamefont {M{\'e}ndez-Zavaleta}},\ and\ \bibinfo {author} {\bibfnamefont {G.}~\bibnamefont {Vel{\'a}zquez-Rodr{\'\i}guez}},\ }\href {https://doi.org/10.1103/PhysRevD.97.084045} {\bibfield  {journal} {\bibinfo  {journal} {Phys. Rev. D}\ }\textbf {\bibinfo {volume} {97}},\ \bibinfo {pages} {084045} (\bibinfo {year} {2018})},\ \Eprint {https://arxiv.org/abs/1801.06764} {arXiv:1801.06764 [hep-th]} \BibitemShut {NoStop}%
\bibitem [{\citenamefont {Andrews}\ \emph {et~al.}(1999)\citenamefont {Andrews}, \citenamefont {Askey},\ and\ \citenamefont {Roy}}]{Andrews_Askey_Roy:1999}%
  \BibitemOpen
  \bibfield  {author} {\bibinfo {author} {\bibfnamefont {G.~E.}\ \bibnamefont {Andrews}}, \bibinfo {author} {\bibfnamefont {R.}~\bibnamefont {Askey}},\ and\ \bibinfo {author} {\bibfnamefont {R.}~\bibnamefont {Roy}},\ }\href {https://doi.org/10.1017/CBO9781107325937} {\emph {\bibinfo {title} {Special Functions}}}\ (\bibinfo  {publisher} {Cambridge University Press},\ \bibinfo {year} {1999})\BibitemShut {NoStop}%
\bibitem [{\citenamefont {Jacobson}(1985)}]{Jacobson:1985}%
  \BibitemOpen
  \bibfield  {author} {\bibinfo {author} {\bibfnamefont {N.}~\bibnamefont {Jacobson}},\ }\href {https://store.doverpublications.com/products/9780486135229} {\emph {\bibinfo {title} {Basic Algebra I}}}\ (\bibinfo  {publisher} {W.H. Freeman \& Co},\ \bibinfo {year} {1985})\BibitemShut {NoStop}%
\bibitem [{\citenamefont {Sturmfels}(2000)}]{Sturmfels:2000}%
  \BibitemOpen
  \bibfield  {author} {\bibinfo {author} {\bibfnamefont {B.}~\bibnamefont {Sturmfels}},\ }\href {https://doi.org/10.1016/S0012-365X(99)00126-0} {\bibfield  {journal} {\bibinfo  {journal} {Discrete Mathematics}\ }\textbf {\bibinfo {volume} {210}},\ \bibinfo {pages} {171} (\bibinfo {year} {2000})}\BibitemShut {NoStop}%
\bibitem [{\citenamefont {Flinckman}\ and\ \citenamefont {Hassan}(2025)}]{Flinckman:2024zpb}%
  \BibitemOpen
  \bibfield  {author} {\bibinfo {author} {\bibfnamefont {J.}~\bibnamefont {Flinckman}}\ and\ \bibinfo {author} {\bibfnamefont {S.~F.}\ \bibnamefont {Hassan}},\ }\href {https://doi.org/10.1007/JHEP02(2025)176} {\bibfield  {journal} {\bibinfo  {journal} {J. High Energy Phys.}\ }\textbf {\bibinfo {volume} {02}}\bibfield  {number} {\bibinfo  {number} { (2025)},\ \bibinfo {pages} {176}},\ }\Eprint {https://arxiv.org/abs/2410.09439} {arXiv:2410.09439 [hep-th]} \BibitemShut {NoStop}%
\bibitem [{\citenamefont {Birkeland}(1927)}]{Birkeland:1927}%
  \BibitemOpen
  \bibfield  {author} {\bibinfo {author} {\bibfnamefont {R.}~\bibnamefont {Birkeland}},\ }\href {https://doi.org/10.1007/BF01475474} {\bibfield  {journal} {\bibinfo  {journal} {Math. Z.}\ }\textbf {\bibinfo {volume} {26}},\ \bibinfo {pages} {566} (\bibinfo {year} {1927})}\BibitemShut {NoStop}%
\bibitem [{\citenamefont {Mayr}(1936)}]{Mayr:1936}%
  \BibitemOpen
  \bibfield  {author} {\bibinfo {author} {\bibfnamefont {K.}~\bibnamefont {Mayr}},\ }\href {https://doi.org/10.1007/BF01707992} {\bibfield  {journal} {\bibinfo  {journal} {Monatsh. Math. Phy.}\ }\textbf {\bibinfo {volume} {45}},\ \bibinfo {pages} {280} (\bibinfo {year} {1936})}\BibitemShut {NoStop}%
\bibitem [{\citenamefont {Akbar}\ and\ \citenamefont {Gibbons}(2003)}]{Akbar:2003gf}%
  \BibitemOpen
  \bibfield  {author} {\bibinfo {author} {\bibfnamefont {M.~M.}\ \bibnamefont {Akbar}}\ and\ \bibinfo {author} {\bibfnamefont {G.~W.}\ \bibnamefont {Gibbons}},\ }\href {https://doi.org/10.1088/0264-9381/20/9/314} {\bibfield  {journal} {\bibinfo  {journal} {Class. Quant. Grav.}\ }\textbf {\bibinfo {volume} {20}},\ \bibinfo {pages} {1787} (\bibinfo {year} {2003})},\ \Eprint {https://arxiv.org/abs/hep-th/0301026} {arXiv:hep-th/0301026} \BibitemShut {NoStop}%
\bibitem [{\citenamefont {Passare}\ and\ \citenamefont {Tsikh}(2004)}]{Passare:2004}%
  \BibitemOpen
  \bibfield  {author} {\bibinfo {author} {\bibfnamefont {M.}~\bibnamefont {Passare}}\ and\ \bibinfo {author} {\bibfnamefont {A.}~\bibnamefont {Tsikh}},\ }in\ \href {https://doi.org/10.1007/978-3-642-18908-1_21} {\emph {\bibinfo {booktitle} {The Legacy of Niels Henrik Abel}}},\ \bibinfo {editor} {edited by\ \bibinfo {editor} {\bibfnamefont {O.~A.}\ \bibnamefont {Laudal}}\ and\ \bibinfo {editor} {\bibfnamefont {R.}~\bibnamefont {Piene}}}\ (\bibinfo  {publisher} {Springer},\ \bibinfo {year} {2004})\ pp.\ \bibinfo {pages} {653--672}\BibitemShut {NoStop}%
\bibitem [{\citenamefont {Gel'fand}\ \emph {et~al.}(1989)\citenamefont {Gel'fand}, \citenamefont {Kapranov},\ and\ \citenamefont {Zelevinsky}}]{Gelfand:1989}%
  \BibitemOpen
  \bibfield  {author} {\bibinfo {author} {\bibfnamefont {I.~M.}\ \bibnamefont {Gel'fand}}, \bibinfo {author} {\bibfnamefont {M.~M.}\ \bibnamefont {Kapranov}},\ and\ \bibinfo {author} {\bibfnamefont {A.~V.}\ \bibnamefont {Zelevinsky}},\ }\href {https://doi.org/10.1007/BF01078777} {\bibfield  {journal} {\bibinfo  {journal} {Funct. Anal. Appl.}\ }\textbf {\bibinfo {volume} {23}},\ \bibinfo {pages} {94} (\bibinfo {year} {1989})}\BibitemShut {NoStop}%
\bibitem [{\citenamefont {Gel'fand}\ \emph {et~al.}(1990)\citenamefont {Gel'fand}, \citenamefont {Kapranov},\ and\ \citenamefont {Zelevinsky}}]{Gelfand:1990bua}%
  \BibitemOpen
  \bibfield  {author} {\bibinfo {author} {\bibfnamefont {I.~M.}\ \bibnamefont {Gel'fand}}, \bibinfo {author} {\bibfnamefont {M.~M.}\ \bibnamefont {Kapranov}},\ and\ \bibinfo {author} {\bibfnamefont {A.~V.}\ \bibnamefont {Zelevinsky}},\ }\href {https://doi.org/10.1016/0001-8708(90)90048-R} {\bibfield  {journal} {\bibinfo  {journal} {Adv. Math.}\ }\textbf {\bibinfo {volume} {84}},\ \bibinfo {pages} {255} (\bibinfo {year} {1990})}\BibitemShut {NoStop}%
\bibitem [{\citenamefont {de~la Cruz}(2019)}]{delaCruz:2019skx}%
  \BibitemOpen
  \bibfield  {author} {\bibinfo {author} {\bibfnamefont {L.}~\bibnamefont {de~la Cruz}},\ }\href {https://doi.org/10.1007/JHEP12(2019)123} {\bibfield  {journal} {\bibinfo  {journal} {J. High Energy Phys.}\ }\textbf {\bibinfo {volume} {12}}\bibfield  {number} {\bibinfo  {number} { (2019)},\ \bibinfo {pages} {123}},\ }\Eprint {https://arxiv.org/abs/1907.00507} {arXiv:1907.00507 [math-ph]} \BibitemShut {NoStop}%
\bibitem [{\citenamefont {Candelas}\ \emph {et~al.}(1991)\citenamefont {Candelas}, \citenamefont {De~La~Ossa}, \citenamefont {Green},\ and\ \citenamefont {Parkes}}]{Candelas:1990rm}%
  \BibitemOpen
  \bibfield  {author} {\bibinfo {author} {\bibfnamefont {P.}~\bibnamefont {Candelas}}, \bibinfo {author} {\bibfnamefont {X.~C.}\ \bibnamefont {De~La~Ossa}}, \bibinfo {author} {\bibfnamefont {P.~S.}\ \bibnamefont {Green}},\ and\ \bibinfo {author} {\bibfnamefont {L.}~\bibnamefont {Parkes}},\ }\href {https://doi.org/10.1016/0550-3213(91)90292-6} {\bibfield  {journal} {\bibinfo  {journal} {Nucl. Phys. B}\ }\textbf {\bibinfo {volume} {359}},\ \bibinfo {pages} {21} (\bibinfo {year} {1991})}\BibitemShut {NoStop}%
\bibitem [{\citenamefont {Batyrev}(1993)}]{Batyrev:1993}%
  \BibitemOpen
  \bibfield  {author} {\bibinfo {author} {\bibfnamefont {V.~V.}\ \bibnamefont {Batyrev}},\ }\href {https://doi.org/10.1215/S0012-7094-93-06917-7} {\bibfield  {journal} {\bibinfo  {journal} {Duke Math. J.}\ }\textbf {\bibinfo {volume} {69}},\ \bibinfo {pages} {349} (\bibinfo {year} {1993})}\BibitemShut {NoStop}%
\bibitem [{\citenamefont {Stienstra}(2007)}]{Stienstra:2007}%
  \BibitemOpen
  \bibfield  {author} {\bibinfo {author} {\bibfnamefont {J.}~\bibnamefont {Stienstra}},\ }in\ \href {https://doi.org/10.1007/978-3-7643-8284-1_12} {\emph {\bibinfo {booktitle} {{Arithmetic and Geometry Around Hypergeometric Functions}}}},\ \bibinfo {editor} {edited by\ \bibinfo {editor} {\bibfnamefont {R.-P.}\ \bibnamefont {Holzapfel}}, \bibinfo {editor} {\bibfnamefont {A.~M.}\ \bibnamefont {Uluda{\u{g}}}},\ and\ \bibinfo {editor} {\bibfnamefont {M.}~\bibnamefont {Yoshida}}}\ (\bibinfo  {publisher} {Birkh{\"a}user Basel},\ \bibinfo {year} {2007})\ pp.\ \bibinfo {pages} {313--371},\ \Eprint {https://arxiv.org/abs/math/0511351} {arXiv:math/0511351} \BibitemShut {NoStop}%
\bibitem [{\citenamefont {Backdahl}\ and\ \citenamefont {Herberthson}(2005)}]{Backdahl:2005uz}%
  \BibitemOpen
  \bibfield  {author} {\bibinfo {author} {\bibfnamefont {T.}~\bibnamefont {Backdahl}}\ and\ \bibinfo {author} {\bibfnamefont {M.}~\bibnamefont {Herberthson}},\ }\href {https://doi.org/10.1088/0264-9381/22/9/009} {\bibfield  {journal} {\bibinfo  {journal} {Class. Quant. Grav.}\ }\textbf {\bibinfo {volume} {22}},\ \bibinfo {pages} {1607} (\bibinfo {year} {2005})},\ \Eprint {https://arxiv.org/abs/gr-qc/0502012} {arXiv:gr-qc/0502012} \BibitemShut {NoStop}%
\bibitem [{\citenamefont {Wood}\ \emph {et~al.}(2024)\citenamefont {Wood}, \citenamefont {Saffin},\ and\ \citenamefont {Avgoustidis}}]{Wood:2024acv}%
  \BibitemOpen
  \bibfield  {author} {\bibinfo {author} {\bibfnamefont {K.}~\bibnamefont {Wood}}, \bibinfo {author} {\bibfnamefont {P.~M.}\ \bibnamefont {Saffin}},\ and\ \bibinfo {author} {\bibfnamefont {A.}~\bibnamefont {Avgoustidis}},\ }\href {https://doi.org/10.1103/PhysRevD.109.124006} {\bibfield  {journal} {\bibinfo  {journal} {Phys. Rev. D}\ }\textbf {\bibinfo {volume} {109}},\ \bibinfo {pages} {124006} (\bibinfo {year} {2024})},\ \Eprint {https://arxiv.org/abs/2402.17835} {arXiv:2402.17835 [gr-qc]} \BibitemShut {NoStop}%
\bibitem [{\citenamefont {Wood}\ \emph {et~al.}(2025)\citenamefont {Wood}, \citenamefont {Saffin},\ and\ \citenamefont {Avgoustidis}}]{Wood:2024eol}%
  \BibitemOpen
  \bibfield  {author} {\bibinfo {author} {\bibfnamefont {K.}~\bibnamefont {Wood}}, \bibinfo {author} {\bibfnamefont {P.~M.}\ \bibnamefont {Saffin}},\ and\ \bibinfo {author} {\bibfnamefont {A.}~\bibnamefont {Avgoustidis}},\ }\href {https://doi.org/10.1103/PhysRevD.111.024057} {\bibfield  {journal} {\bibinfo  {journal} {Phys. Rev. D}\ }\textbf {\bibinfo {volume} {111}},\ \bibinfo {pages} {024057} (\bibinfo {year} {2025})},\ \Eprint {https://arxiv.org/abs/2410.10976} {arXiv:2410.10976 [gr-qc]} \BibitemShut {NoStop}%
\bibitem [{\citenamefont {Garc{\'\i}a-Compe{\'a}n}\ and\ \citenamefont {Rivera-Oliva}(2026)}]{Garcia-Compean:2026cnq}%
  \BibitemOpen
  \bibfield  {author} {\bibinfo {author} {\bibfnamefont {H.}~\bibnamefont {Garc{\'\i}a-Compe{\'a}n}}\ and\ \bibinfo {author} {\bibfnamefont {E.}~\bibnamefont {Rivera-Oliva}},\ }\href@noop {} {\bibinfo {title} {{Kerr-Schild solutions in Multigravity and the Classical Double Copy}}} (\bibinfo {year} {2026}),\ \Eprint {https://arxiv.org/abs/2602.16905} {arXiv:2602.16905 [gr-qc]} \BibitemShut {NoStop}%
\bibitem [{\citenamefont {Ay\'on-Beato}\ \emph {et~al.}(2016)\citenamefont {Ay\'on-Beato}, \citenamefont {Higuita-Borja},\ and\ \citenamefont {M\'endez-Zavaleta}}]{Ayon-Beato:2015qtt}%
  \BibitemOpen
  \bibfield  {author} {\bibinfo {author} {\bibfnamefont {E.}~\bibnamefont {Ay\'on-Beato}}, \bibinfo {author} {\bibfnamefont {D.}~\bibnamefont {Higuita-Borja}},\ and\ \bibinfo {author} {\bibfnamefont {J.~A.}\ \bibnamefont {M\'endez-Zavaleta}},\ }\href {https://doi.org/10.1103/PhysRevD.93.024049} {\bibfield  {journal} {\bibinfo  {journal} {Phys. Rev. D}\ }\textbf {\bibinfo {volume} {93}},\ \bibinfo {pages} {024049} (\bibinfo {year} {2016})},\ \Eprint {https://arxiv.org/abs/1511.01108} {arXiv:1511.01108 [hep-th]} \BibitemShut {NoStop}%
\bibitem [{\citenamefont {Ay{\'o}n-Beato}\ \emph {et~al.}(2016)\citenamefont {Ay{\'o}n-Beato}, \citenamefont {Hassa{\"\i}ne},\ and\ \citenamefont {Higuita-Borja}}]{Ayon-Beato:2015nvz}%
  \BibitemOpen
  \bibfield  {author} {\bibinfo {author} {\bibfnamefont {E.}~\bibnamefont {Ay{\'o}n-Beato}}, \bibinfo {author} {\bibfnamefont {M.}~\bibnamefont {Hassa{\"\i}ne}},\ and\ \bibinfo {author} {\bibfnamefont {D.}~\bibnamefont {Higuita-Borja}},\ }\href {https://doi.org/10.1103/PhysRevD.94.064073} {\bibfield  {journal} {\bibinfo  {journal} {Phys. Rev. D}\ }\textbf {\bibinfo {volume} {94}},\ \bibinfo {pages} {064073} (\bibinfo {year} {2016})},\ \Eprint {https://arxiv.org/abs/1512.06870} {arXiv:1512.06870 [hep-th]} \BibitemShut {NoStop}%
\bibitem [{\citenamefont {Ay{\'o}n-Beato}\ \emph {et~al.}(2025)\citenamefont {Ay{\'o}n-Beato}, \citenamefont {Flores-Alfonso}, \citenamefont {Hassaine},\ and\ \citenamefont {Higuita-Borja}}]{Ayon-Beato:2025ahb}%
  \BibitemOpen
  \bibfield  {author} {\bibinfo {author} {\bibfnamefont {E.}~\bibnamefont {Ay{\'o}n-Beato}}, \bibinfo {author} {\bibfnamefont {D.}~\bibnamefont {Flores-Alfonso}}, \bibinfo {author} {\bibfnamefont {M.}~\bibnamefont {Hassaine}},\ and\ \bibinfo {author} {\bibfnamefont {D.~F.}\ \bibnamefont {Higuita-Borja}},\ }\href {https://doi.org/10.1103/l1f2-vxlc} {\bibfield  {journal} {\bibinfo  {journal} {Phys. Rev. D}\ }\textbf {\bibinfo {volume} {112}},\ \bibinfo {pages} {104020} (\bibinfo {year} {2025})},\ \Eprint {https://arxiv.org/abs/2508.02986} {arXiv:2508.02986 [hep-th]} \BibitemShut {NoStop}%
\bibitem [{\citenamefont {Babichev}\ and\ \citenamefont {Fabbri}(2014)}]{Babichev:2014tfa}%
  \BibitemOpen
  \bibfield  {author} {\bibinfo {author} {\bibfnamefont {E.}~\bibnamefont {Babichev}}\ and\ \bibinfo {author} {\bibfnamefont {A.}~\bibnamefont {Fabbri}},\ }\href {https://doi.org/10.1103/PhysRevD.90.084019} {\bibfield  {journal} {\bibinfo  {journal} {Phys. Rev. D}\ }\textbf {\bibinfo {volume} {90}},\ \bibinfo {pages} {084019} (\bibinfo {year} {2014})},\ \Eprint {https://arxiv.org/abs/1406.6096} {arXiv:1406.6096 [gr-qc]} \BibitemShut {NoStop}%
\bibitem [{\citenamefont {Garc{\'\i}a-Compe{\'a}n}\ and\ \citenamefont {Ramos}(2025)}]{Garcia-Compean:2025wkj}%
  \BibitemOpen
  \bibfield  {author} {\bibinfo {author} {\bibfnamefont {H.}~\bibnamefont {Garc{\'\i}a-Compe{\'a}n}}\ and\ \bibinfo {author} {\bibfnamefont {C.~I.}\ \bibnamefont {Ramos}},\ }\href {https://doi.org/10.1007/JHEP11(2025)154} {\bibfield  {journal} {\bibinfo  {journal} {J. High Energy Phys.}\ }\textbf {\bibinfo {volume} {11}}\bibfield  {number} {\bibinfo  {number} { (2025)},\ \bibinfo {pages} {154}},\ }\Eprint {https://arxiv.org/abs/2510.01550} {arXiv:2510.01550 [gr-qc]} \BibitemShut {NoStop}%
\bibitem [{\citenamefont {Garc{\'{i}}a~D.}\ and\ \citenamefont {Pleba{\'{n}}ski}(1981)}]{Garcia:1981}%
  \BibitemOpen
  \bibfield  {author} {\bibinfo {author} {\bibfnamefont {A.}~\bibnamefont {Garc{\'{i}}a~D.}}\ and\ \bibinfo {author} {\bibfnamefont {J.~F.}\ \bibnamefont {Pleba{\'{n}}ski}},\ }\href {https://doi.org/10.1063/1.524843} {\bibfield  {journal} {\bibinfo  {journal} {J. Math. Phys.}\ }\textbf {\bibinfo {volume} {22}},\ \bibinfo {pages} {2655} (\bibinfo {year} {1981})}\BibitemShut {NoStop}%
\bibitem [{\citenamefont {Salazar~I.}\ \emph {et~al.}(1983)\citenamefont {Salazar~I.}, \citenamefont {Garc{\'{i}}a~D.},\ and\ \citenamefont {Pleba{\'{n}}ski}}]{Salazar:1983}%
  \BibitemOpen
  \bibfield  {author} {\bibinfo {author} {\bibfnamefont {H.}~\bibnamefont {Salazar~I.}}, \bibinfo {author} {\bibfnamefont {A.}~\bibnamefont {Garc{\'{i}}a~D.}},\ and\ \bibinfo {author} {\bibfnamefont {J.~F.}\ \bibnamefont {Pleba{\'{n}}ski}},\ }\href {https://doi.org/10.1063/1.525930} {\bibfield  {journal} {\bibinfo  {journal} {J. Math. Phys.}\ }\textbf {\bibinfo {volume} {24}},\ \bibinfo {pages} {2191} (\bibinfo {year} {1983})}\BibitemShut {NoStop}%
\bibitem [{\citenamefont {Garc{\'{i}}a~D.}(1983)}]{Garcia:1983}%
  \BibitemOpen
  \bibfield  {author} {\bibinfo {author} {\bibfnamefont {A.}~\bibnamefont {Garc{\'{i}}a~D.}},\ }\href {https://doi.org/10.1007/BF02721101} {\bibfield  {journal} {\bibinfo  {journal} {Il Nuovo Cimento B}\ }\textbf {\bibinfo {volume} {78}},\ \bibinfo {pages} {255–262} (\bibinfo {year} {1983})}\BibitemShut {NoStop}%
\bibitem [{\citenamefont {Ozsvath}\ \emph {et~al.}(1985)\citenamefont {Ozsvath}, \citenamefont {Robinson},\ and\ \citenamefont {Rozga}}]{Ozsvath:1985qn}%
  \BibitemOpen
  \bibfield  {author} {\bibinfo {author} {\bibfnamefont {I.}~\bibnamefont {Ozsvath}}, \bibinfo {author} {\bibfnamefont {I.}~\bibnamefont {Robinson}},\ and\ \bibinfo {author} {\bibfnamefont {K.}~\bibnamefont {Rozga}},\ }\href {https://doi.org/10.1063/1.526887} {\bibfield  {journal} {\bibinfo  {journal} {J. Math. Phys.}\ }\textbf {\bibinfo {volume} {26}},\ \bibinfo {pages} {1755} (\bibinfo {year} {1985})}\BibitemShut {NoStop}%
\bibitem [{\citenamefont {Bicak}\ and\ \citenamefont {Podolsky}(1999{\natexlab{a}})}]{Bicak:1999ha}%
  \BibitemOpen
  \bibfield  {author} {\bibinfo {author} {\bibfnamefont {J.}~\bibnamefont {Bicak}}\ and\ \bibinfo {author} {\bibfnamefont {J.}~\bibnamefont {Podolsky}},\ }\href {https://doi.org/10.1063/1.532981} {\bibfield  {journal} {\bibinfo  {journal} {J. Math. Phys.}\ }\textbf {\bibinfo {volume} {40}},\ \bibinfo {pages} {4495} (\bibinfo {year} {1999}{\natexlab{a}})},\ \Eprint {https://arxiv.org/abs/gr-qc/9907048} {arXiv:gr-qc/9907048} \BibitemShut {NoStop}%
\bibitem [{\citenamefont {Bicak}\ and\ \citenamefont {Podolsky}(1999{\natexlab{b}})}]{Bicak:1999hb}%
  \BibitemOpen
  \bibfield  {author} {\bibinfo {author} {\bibfnamefont {J.}~\bibnamefont {Bicak}}\ and\ \bibinfo {author} {\bibfnamefont {J.}~\bibnamefont {Podolsky}},\ }\href {https://doi.org/10.1063/1.532982} {\bibfield  {journal} {\bibinfo  {journal} {J. Math. Phys.}\ }\textbf {\bibinfo {volume} {40}},\ \bibinfo {pages} {4506} (\bibinfo {year} {1999}{\natexlab{b}})},\ \Eprint {https://arxiv.org/abs/gr-qc/9907049} {arXiv:gr-qc/9907049} \BibitemShut {NoStop}%
\bibitem [{\citenamefont {Kundt}(1961)}]{Kundt:1961}%
  \BibitemOpen
  \bibfield  {author} {\bibinfo {author} {\bibfnamefont {W.}~\bibnamefont {Kundt}},\ }\href {https://doi.org/10.1007/BF01328918} {\bibfield  {journal} {\bibinfo  {journal} {Zeitschrift für Physik}\ }\textbf {\bibinfo {volume} {163}},\ \bibinfo {pages} {77–86} (\bibinfo {year} {1961})}\BibitemShut {NoStop}%
\bibitem [{\citenamefont {Robinson}\ and\ \citenamefont {Trautman}(1962)}]{Robinson:1962zz}%
  \BibitemOpen
  \bibfield  {author} {\bibinfo {author} {\bibfnamefont {I.}~\bibnamefont {Robinson}}\ and\ \bibinfo {author} {\bibfnamefont {A.}~\bibnamefont {Trautman}},\ }\href {https://doi.org/10.1098/rspa.1962.0036} {\bibfield  {journal} {\bibinfo  {journal} {Proc. Roy. Soc. Lond. A}\ }\textbf {\bibinfo {volume} {265}},\ \bibinfo {pages} {463} (\bibinfo {year} {1962})}\BibitemShut {NoStop}%
\bibitem [{\citenamefont {Siklos}(1985)}]{Siklos:1985}%
  \BibitemOpen
  \bibfield  {author} {\bibinfo {author} {\bibfnamefont {S.}~\bibnamefont {Siklos}},\ }in\ \href {https://www.cambridge.org/us/universitypress/subjects/physics/astrophysics/galaxies-axisymmetric-systems-and-relativity-essays-presented-w-b-bonnor-his-65th-birthday} {\emph {\bibinfo {booktitle} {Galaxies, Axisymmetric Systems and Relativity}}},\ \bibinfo {editor} {edited by\ \bibinfo {editor} {\bibfnamefont {M.}~\bibnamefont {MacCallum}}}\ (\bibinfo  {publisher} {Cambridge University Press},\ \bibinfo {year} {1985})\ pp.\ \bibinfo {pages} {247--274}\BibitemShut {NoStop}%
\bibitem [{\citenamefont {Podolsky}(1998)}]{Podolsky:1997ik}%
  \BibitemOpen
  \bibfield  {author} {\bibinfo {author} {\bibfnamefont {J.}~\bibnamefont {Podolsky}},\ }\href {https://doi.org/10.1088/0264-9381/15/3/019} {\bibfield  {journal} {\bibinfo  {journal} {Class. Quant. Grav.}\ }\textbf {\bibinfo {volume} {15}},\ \bibinfo {pages} {719} (\bibinfo {year} {1998})},\ \Eprint {https://arxiv.org/abs/gr-qc/9801052} {arXiv:gr-qc/9801052} \BibitemShut {NoStop}%
\bibitem [{\citenamefont {Bergshoeff}\ \emph {et~al.}(2015)\citenamefont {Bergshoeff}, \citenamefont {Hohm}, \citenamefont {Merbis}, \citenamefont {Routh},\ and\ \citenamefont {Townsend}}]{Bergshoeff:2014bia}%
  \BibitemOpen
  \bibfield  {author} {\bibinfo {author} {\bibfnamefont {E.~A.}\ \bibnamefont {Bergshoeff}}, \bibinfo {author} {\bibfnamefont {O.}~\bibnamefont {Hohm}}, \bibinfo {author} {\bibfnamefont {W.}~\bibnamefont {Merbis}}, \bibinfo {author} {\bibfnamefont {A.~J.}\ \bibnamefont {Routh}},\ and\ \bibinfo {author} {\bibfnamefont {P.~K.}\ \bibnamefont {Townsend}},\ }\href {https://doi.org/10.1007/978-3-319-10070-8_7} {\bibfield  {journal} {\bibinfo  {journal} {Lect. Notes Phys.}\ }\textbf {\bibinfo {volume} {892}},\ \bibinfo {pages} {181} (\bibinfo {year} {2015})},\ \Eprint {https://arxiv.org/abs/1402.1688} {arXiv:1402.1688 [hep-th]} \BibitemShut {NoStop}%
\bibitem [{\citenamefont {Scargill}\ \emph {et~al.}(2014)\citenamefont {Scargill}, \citenamefont {Noller},\ and\ \citenamefont {Ferreira}}]{Scargill:2014wya}%
  \BibitemOpen
  \bibfield  {author} {\bibinfo {author} {\bibfnamefont {J.~H.~C.}\ \bibnamefont {Scargill}}, \bibinfo {author} {\bibfnamefont {J.}~\bibnamefont {Noller}},\ and\ \bibinfo {author} {\bibfnamefont {P.~G.}\ \bibnamefont {Ferreira}},\ }\href {https://doi.org/10.1007/JHEP12(2014)160} {\bibfield  {journal} {\bibinfo  {journal} {J. High Energy Phys.}\ }\textbf {\bibinfo {volume} {12}}\bibfield  {number} {\bibinfo  {number} { (2014)},\ \bibinfo {pages} {160}},\ }\Eprint {https://arxiv.org/abs/1410.7774} {arXiv:1410.7774 [hep-th]} \BibitemShut {NoStop}%
\bibitem [{\citenamefont {de~Rham}\ and\ \citenamefont {Tolley}(2015)}]{deRham:2015cha}%
  \BibitemOpen
  \bibfield  {author} {\bibinfo {author} {\bibfnamefont {C.}~\bibnamefont {de~Rham}}\ and\ \bibinfo {author} {\bibfnamefont {A.~J.}\ \bibnamefont {Tolley}},\ }\href {https://doi.org/10.1103/PhysRevD.92.024024} {\bibfield  {journal} {\bibinfo  {journal} {Phys. Rev. D}\ }\textbf {\bibinfo {volume} {92}},\ \bibinfo {pages} {024024} (\bibinfo {year} {2015})},\ \Eprint {https://arxiv.org/abs/1505.01450} {arXiv:1505.01450 [hep-th]} \BibitemShut {NoStop}%
\bibitem [{\citenamefont {Goon}\ \emph {et~al.}(2015)\citenamefont {Goon}, \citenamefont {Hinterbichler}, \citenamefont {Joyce},\ and\ \citenamefont {Trodden}}]{Goon:2014paa}%
  \BibitemOpen
  \bibfield  {author} {\bibinfo {author} {\bibfnamefont {G.}~\bibnamefont {Goon}}, \bibinfo {author} {\bibfnamefont {K.}~\bibnamefont {Hinterbichler}}, \bibinfo {author} {\bibfnamefont {A.}~\bibnamefont {Joyce}},\ and\ \bibinfo {author} {\bibfnamefont {M.}~\bibnamefont {Trodden}},\ }\href {https://doi.org/10.1007/JHEP07(2015)101} {\bibfield  {journal} {\bibinfo  {journal} {J. High Energy Phys.}\ }\textbf {\bibinfo {volume} {07}}\bibfield  {number} {\bibinfo  {number} { (2015)},\ \bibinfo {pages} {101}},\ }\Eprint {https://arxiv.org/abs/1412.6098} {arXiv:1412.6098 [hep-th]} \BibitemShut {NoStop}%
\bibitem [{\citenamefont {Deffayet}\ \emph {et~al.}(2013)\citenamefont {Deffayet}, \citenamefont {Mourad},\ and\ \citenamefont {Zahariade}}]{Deffayet:2012zc}%
  \BibitemOpen
  \bibfield  {author} {\bibinfo {author} {\bibfnamefont {C.}~\bibnamefont {Deffayet}}, \bibinfo {author} {\bibfnamefont {J.}~\bibnamefont {Mourad}},\ and\ \bibinfo {author} {\bibfnamefont {G.}~\bibnamefont {Zahariade}},\ }\href {https://doi.org/10.1007/JHEP03(2013)086} {\bibfield  {journal} {\bibinfo  {journal} {J. High Energy Phys.}\ }\textbf {\bibinfo {volume} {03}}\bibfield  {number} {\bibinfo  {number} { (2013)},\ \bibinfo {pages} {086}},\ }\Eprint {https://arxiv.org/abs/1208.4493} {arXiv:1208.4493 [gr-qc]} \BibitemShut {NoStop}%
\bibitem [{\citenamefont {Breitenlohner}\ and\ \citenamefont {Freedman}(1982)}]{Breitenlohner:1982jf}%
  \BibitemOpen
  \bibfield  {author} {\bibinfo {author} {\bibfnamefont {P.}~\bibnamefont {Breitenlohner}}\ and\ \bibinfo {author} {\bibfnamefont {D.~Z.}\ \bibnamefont {Freedman}},\ }\href {https://doi.org/10.1016/0003-4916(82)90116-6} {\bibfield  {journal} {\bibinfo  {journal} {Annals Phys.}\ }\textbf {\bibinfo {volume} {144}},\ \bibinfo {pages} {249} (\bibinfo {year} {1982})}\BibitemShut {NoStop}%
\bibitem [{\citenamefont {Mezincescu}\ and\ \citenamefont {Townsend}(1985)}]{Mezincescu:1984ev}%
  \BibitemOpen
  \bibfield  {author} {\bibinfo {author} {\bibfnamefont {L.}~\bibnamefont {Mezincescu}}\ and\ \bibinfo {author} {\bibfnamefont {P.~K.}\ \bibnamefont {Townsend}},\ }\href {https://doi.org/10.1016/0003-4916(85)90150-2} {\bibfield  {journal} {\bibinfo  {journal} {Annals Phys.}\ }\textbf {\bibinfo {volume} {160}},\ \bibinfo {pages} {406} (\bibinfo {year} {1985})}\BibitemShut {NoStop}%
\bibitem [{\citenamefont {Ay{\'o}n-Beato}\ and\ \citenamefont {Vel{\'a}zquez-Rodr{\'\i}guez}(2016)}]{Ayon-Beato:2015xsz}%
  \BibitemOpen
  \bibfield  {author} {\bibinfo {author} {\bibfnamefont {E.}~\bibnamefont {Ay{\'o}n-Beato}}\ and\ \bibinfo {author} {\bibfnamefont {G.}~\bibnamefont {Vel{\'a}zquez-Rodr{\'\i}guez}},\ }\href {https://doi.org/10.1103/PhysRevD.93.044040} {\bibfield  {journal} {\bibinfo  {journal} {Phys. Rev. D}\ }\textbf {\bibinfo {volume} {93}},\ \bibinfo {pages} {044040} (\bibinfo {year} {2016})},\ \Eprint {https://arxiv.org/abs/1511.07461} {arXiv:1511.07461 [gr-qc]} \BibitemShut {NoStop}%
\bibitem [{\citenamefont {Bergshoeff}\ \emph {et~al.}(2012)\citenamefont {Bergshoeff}, \citenamefont {de~Haan}, \citenamefont {Merbis}, \citenamefont {Rosseel},\ and\ \citenamefont {Zojer}}]{Bergshoeff:2012ev}%
  \BibitemOpen
  \bibfield  {author} {\bibinfo {author} {\bibfnamefont {E.~A.}\ \bibnamefont {Bergshoeff}}, \bibinfo {author} {\bibfnamefont {S.}~\bibnamefont {de~Haan}}, \bibinfo {author} {\bibfnamefont {W.}~\bibnamefont {Merbis}}, \bibinfo {author} {\bibfnamefont {J.}~\bibnamefont {Rosseel}},\ and\ \bibinfo {author} {\bibfnamefont {T.}~\bibnamefont {Zojer}},\ }\href {https://doi.org/10.1103/PhysRevD.86.064037} {\bibfield  {journal} {\bibinfo  {journal} {Phys. Rev. D}\ }\textbf {\bibinfo {volume} {86}},\ \bibinfo {pages} {064037} (\bibinfo {year} {2012})},\ \Eprint {https://arxiv.org/abs/1206.3089} {arXiv:1206.3089 [hep-th]} \BibitemShut {NoStop}%
\bibitem [{\citenamefont {Brown}\ and\ \citenamefont {Henneaux}(1986)}]{Brown:1986nw}%
  \BibitemOpen
  \bibfield  {author} {\bibinfo {author} {\bibfnamefont {J.~D.}\ \bibnamefont {Brown}}\ and\ \bibinfo {author} {\bibfnamefont {M.}~\bibnamefont {Henneaux}},\ }\href {https://doi.org/10.1007/BF01211590} {\bibfield  {journal} {\bibinfo  {journal} {Commun. Math. Phys.}\ }\textbf {\bibinfo {volume} {104}},\ \bibinfo {pages} {207} (\bibinfo {year} {1986})}\BibitemShut {NoStop}%
\bibitem [{\citenamefont {Chernicoff}\ \emph {et~al.}(2024)\citenamefont {Chernicoff}, \citenamefont {Giribet}, \citenamefont {Moreno}, \citenamefont {Oliva}, \citenamefont {Rojas},\ and\ \citenamefont {Torres}}]{Chernicoff:2024dll}%
  \BibitemOpen
  \bibfield  {author} {\bibinfo {author} {\bibfnamefont {M.}~\bibnamefont {Chernicoff}}, \bibinfo {author} {\bibfnamefont {G.}~\bibnamefont {Giribet}}, \bibinfo {author} {\bibfnamefont {J.}~\bibnamefont {Moreno}}, \bibinfo {author} {\bibfnamefont {J.}~\bibnamefont {Oliva}}, \bibinfo {author} {\bibfnamefont {R.}~\bibnamefont {Rojas}},\ and\ \bibinfo {author} {\bibfnamefont {C.~R. d.~A.}\ \bibnamefont {Torres}},\ }\href {https://doi.org/10.1103/PhysRevD.110.044021} {\bibfield  {journal} {\bibinfo  {journal} {Phys. Rev. D}\ }\textbf {\bibinfo {volume} {110}},\ \bibinfo {pages} {044021} (\bibinfo {year} {2024})},\ \Eprint {https://arxiv.org/abs/2404.10127} {arXiv:2404.10127 [hep-th]} \BibitemShut {NoStop}%
\bibitem [{\citenamefont {Setare}\ and\ \citenamefont {Hatami}(2013)}]{Setare:2013fza}%
  \BibitemOpen
  \bibfield  {author} {\bibinfo {author} {\bibfnamefont {M.~R.}\ \bibnamefont {Setare}}\ and\ \bibinfo {author} {\bibfnamefont {N.}~\bibnamefont {Hatami}},\ }\href {https://doi.org/10.1007/JHEP04(2013)142} {\bibfield  {journal} {\bibinfo  {journal} {J. High Energy Phys.}\ }\textbf {\bibinfo {volume} {04}}\bibinfo  {number} { (2013)},\ \bibinfo {pages} {142}}\BibitemShut {NoStop}%
\bibitem [{\citenamefont {Setare}(2015)}]{Setare:2014zea}%
  \BibitemOpen
\bibfield  {number} {  }\bibfield  {author} {\bibinfo {author} {\bibfnamefont {M.~R.}\ \bibnamefont {Setare}},\ }\href {https://doi.org/10.1016/j.nuclphysb.2015.07.006} {\bibfield  {journal} {\bibinfo  {journal} {Nucl. Phys. B}\ }\textbf {\bibinfo {volume} {898}},\ \bibinfo {pages} {259} (\bibinfo {year} {2015})},\ \Eprint {https://arxiv.org/abs/1412.2151} {arXiv:1412.2151 [hep-th]} \BibitemShut {NoStop}%
\bibitem [{\citenamefont {Gurses}\ \emph {et~al.}(2015)\citenamefont {Gurses}, \citenamefont {Sisman},\ and\ \citenamefont {Tekin}}]{Gurses:2015zia}%
  \BibitemOpen
  \bibfield  {author} {\bibinfo {author} {\bibfnamefont {M.}~\bibnamefont {Gurses}}, \bibinfo {author} {\bibfnamefont {T.~C.}\ \bibnamefont {Sisman}},\ and\ \bibinfo {author} {\bibfnamefont {B.}~\bibnamefont {Tekin}},\ }\href {https://doi.org/10.1103/PhysRevD.92.084016} {\bibfield  {journal} {\bibinfo  {journal} {Phys. Rev. D}\ }\textbf {\bibinfo {volume} {92}},\ \bibinfo {pages} {084016} (\bibinfo {year} {2015})},\ \Eprint {https://arxiv.org/abs/1509.03167} {arXiv:1509.03167 [hep-th]} \BibitemShut {NoStop}%
\bibitem [{\citenamefont {Grumiller}\ and\ \citenamefont {Johansson}(2008)}]{Grumiller:2008qz}%
  \BibitemOpen
  \bibfield  {author} {\bibinfo {author} {\bibfnamefont {D.}~\bibnamefont {Grumiller}}\ and\ \bibinfo {author} {\bibfnamefont {N.}~\bibnamefont {Johansson}},\ }\href {https://doi.org/10.1088/1126-6708/2008/07/134} {\bibfield  {journal} {\bibinfo  {journal} {J. High Energy Phys.}\ }\textbf {\bibinfo {volume} {07}}\bibfield  {number} {\bibinfo  {number} { (2008)},\ \bibinfo {pages} {134}},\ }\Eprint {https://arxiv.org/abs/0805.2610} {arXiv:0805.2610 [hep-th]} \BibitemShut {NoStop}%
\bibitem [{\citenamefont {Liu}\ and\ \citenamefont {Sun}(2009{\natexlab{a}})}]{Liu:2009bk}%
  \BibitemOpen
  \bibfield  {author} {\bibinfo {author} {\bibfnamefont {Y.}~\bibnamefont {Liu}}\ and\ \bibinfo {author} {\bibfnamefont {Y.-W.}\ \bibnamefont {Sun}},\ }\href {https://doi.org/10.1088/1126-6708/2009/04/106} {\bibfield  {journal} {\bibinfo  {journal} {J. High Energy Phys.}\ }\textbf {\bibinfo {volume} {04}}\bibfield  {number} {\bibinfo  {number} { (2009)},\ \bibinfo {pages} {106}},\ }\Eprint {https://arxiv.org/abs/0903.0536} {arXiv:0903.0536 [hep-th]} \BibitemShut {NoStop}%
\bibitem [{\citenamefont {Liu}\ and\ \citenamefont {Sun}(2009{\natexlab{b}})}]{Liu:2009kc}%
  \BibitemOpen
  \bibfield  {author} {\bibinfo {author} {\bibfnamefont {Y.}~\bibnamefont {Liu}}\ and\ \bibinfo {author} {\bibfnamefont {Y.-W.}\ \bibnamefont {Sun}},\ }\href {https://doi.org/10.1088/1126-6708/2009/05/039} {\bibfield  {journal} {\bibinfo  {journal} {J. High Energy Phys.}\ }\textbf {\bibinfo {volume} {05}}\bibfield  {number} {\bibinfo  {number} { (2009)},\ \bibinfo {pages} {039}},\ }\Eprint {https://arxiv.org/abs/0903.2933} {arXiv:0903.2933 [hep-th]} \BibitemShut {NoStop}%
\bibitem [{\citenamefont {Zucker}(2008)}]{Zucker:2008}%
  \BibitemOpen
  \bibfield  {author} {\bibinfo {author} {\bibfnamefont {I.}~\bibnamefont {Zucker}},\ }\href {http://www.jstor.org/stable/27821778} {\bibfield  {journal} {\bibinfo  {journal} {Math. Gaz.}\ }\textbf {\bibinfo {volume} {92}},\ \bibinfo {pages} {264} (\bibinfo {year} {2008})}\BibitemShut {NoStop}%
\bibitem [{\citenamefont {Gording}\ and\ \citenamefont {Schmidt-May}(2018)}]{Gording:2018not}%
  \BibitemOpen
  \bibfield  {author} {\bibinfo {author} {\bibfnamefont {B.}~\bibnamefont {Gording}}\ and\ \bibinfo {author} {\bibfnamefont {A.}~\bibnamefont {Schmidt-May}},\ }\href {https://doi.org/10.1007/JHEP09(2018)044} {\bibfield  {journal} {\bibinfo  {journal} {J. High Energy Phys.}\ }\textbf {\bibinfo {volume} {09}}\bibfield  {number} {\bibinfo  {number} { (2018)},\ \bibinfo {pages} {044}},\ }\bibinfo {note} {[Erratum: JHEP 10, 115 (2018)]},\ \Eprint {https://arxiv.org/abs/1807.05011} {arXiv:1807.05011 [gr-qc]} \BibitemShut {NoStop}%
\bibitem [{\citenamefont {Ay{\'o}n-Beato}\ \emph {et~al.}(2014)\citenamefont {Ay{\'o}n-Beato}, \citenamefont {Hassa{\"\i}ne},\ and\ \citenamefont {Ju{\'a}rez-Aubry}}]{Ayon-Beato:2014wla}%
  \BibitemOpen
  \bibfield  {author} {\bibinfo {author} {\bibfnamefont {E.}~\bibnamefont {Ay{\'o}n-Beato}}, \bibinfo {author} {\bibfnamefont {M.}~\bibnamefont {Hassa{\"\i}ne}},\ and\ \bibinfo {author} {\bibfnamefont {M.~M.}\ \bibnamefont {Ju{\'a}rez-Aubry}},\ }\href {https://doi.org/10.1103/PhysRevD.90.044026} {\bibfield  {journal} {\bibinfo  {journal} {Phys. Rev. D}\ }\textbf {\bibinfo {volume} {90}},\ \bibinfo {pages} {044026} (\bibinfo {year} {2014})},\ \Eprint {https://arxiv.org/abs/1406.1588} {arXiv:1406.1588 [hep-th]} \BibitemShut {NoStop}%
\bibitem [{\citenamefont {Bergshoeff}\ \emph {et~al.}(2009{\natexlab{b}})\citenamefont {Bergshoeff}, \citenamefont {Hohm},\ and\ \citenamefont {Townsend}}]{Bergshoeff:2009aq}%
  \BibitemOpen
  \bibfield  {author} {\bibinfo {author} {\bibfnamefont {E.~A.}\ \bibnamefont {Bergshoeff}}, \bibinfo {author} {\bibfnamefont {O.}~\bibnamefont {Hohm}},\ and\ \bibinfo {author} {\bibfnamefont {P.~K.}\ \bibnamefont {Townsend}},\ }\href {https://doi.org/10.1103/PhysRevD.79.124042} {\bibfield  {journal} {\bibinfo  {journal} {Phys. Rev. D}\ }\textbf {\bibinfo {volume} {79}},\ \bibinfo {pages} {124042} (\bibinfo {year} {2009}{\natexlab{b}})},\ \Eprint {https://arxiv.org/abs/0905.1259} {arXiv:0905.1259 [hep-th]} \BibitemShut {NoStop}%
\bibitem [{\citenamefont {Oliva}\ \emph {et~al.}(2009)\citenamefont {Oliva}, \citenamefont {Tempo},\ and\ \citenamefont {Troncoso}}]{Oliva:2009ip}%
  \BibitemOpen
  \bibfield  {author} {\bibinfo {author} {\bibfnamefont {J.}~\bibnamefont {Oliva}}, \bibinfo {author} {\bibfnamefont {D.}~\bibnamefont {Tempo}},\ and\ \bibinfo {author} {\bibfnamefont {R.}~\bibnamefont {Troncoso}},\ }\href {https://doi.org/10.1088/1126-6708/2009/07/011} {\bibfield  {journal} {\bibinfo  {journal} {J. High Energy Phys.}\ }\textbf {\bibinfo {volume} {07}}\bibfield  {number} {\bibinfo  {number} { (2009)},\ \bibinfo {pages} {011}},\ }\Eprint {https://arxiv.org/abs/0905.1545} {arXiv:0905.1545 [hep-th]} \BibitemShut {NoStop}%
\bibitem [{\citenamefont {Ayon-Beato}\ \emph {et~al.}(2009{\natexlab{b}})\citenamefont {Ayon-Beato}, \citenamefont {Garbarz}, \citenamefont {Giribet},\ and\ \citenamefont {Hassaine}}]{Ayon-Beato:2009rgu}%
  \BibitemOpen
  \bibfield  {author} {\bibinfo {author} {\bibfnamefont {E.}~\bibnamefont {Ayon-Beato}}, \bibinfo {author} {\bibfnamefont {A.}~\bibnamefont {Garbarz}}, \bibinfo {author} {\bibfnamefont {G.}~\bibnamefont {Giribet}},\ and\ \bibinfo {author} {\bibfnamefont {M.}~\bibnamefont {Hassaine}},\ }\href {https://doi.org/10.1103/PhysRevD.80.104029} {\bibfield  {journal} {\bibinfo  {journal} {Phys. Rev. D}\ }\textbf {\bibinfo {volume} {80}},\ \bibinfo {pages} {104029} (\bibinfo {year} {2009}{\natexlab{b}})},\ \Eprint {https://arxiv.org/abs/0909.1347} {arXiv:0909.1347 [hep-th]} \BibitemShut {NoStop}%
\bibitem [{\citenamefont {Beukers}(2014)}]{Beukers:2014}%
  \BibitemOpen
  \bibfield  {author} {\bibinfo {author} {\bibfnamefont {F.}~\bibnamefont {Beukers}},\ }\href {https://www.ams.org/journals/notices/201401/rnoti-p48.pdf?adat=January%202014&trk=20140148&cat=none&type=.pdf} {\bibfield  {journal} {\bibinfo  {journal} {Notices of the AMS}\ }\textbf {\bibinfo {volume} {61}},\ \bibinfo {pages} {48} (\bibinfo {year} {2014})}\BibitemShut {NoStop}%
\bibitem [{\citenamefont {Beukers}(2011)}]{Beukers:2011}%
  \BibitemOpen
  \bibfield  {author} {\bibinfo {author} {\bibfnamefont {F.}~\bibnamefont {Beukers}},\ }\href {https://smf.emath.fr/publications/notes-sur-les-fonctions-hypergeometriques} {\bibfield  {journal} {\bibinfo  {journal} {Séminaires et Congrès}\ }\textbf {\bibinfo {volume} {23}},\ \bibinfo {pages} {25} (\bibinfo {year} {2011})}\BibitemShut {NoStop}%
\end{thebibliography}%

\end{document}